\documentclass[pdflatex,sn-mathphys-ay]{sn-jnl}% Math and Physical Sciences Author Year Reference Style
\usepackage{amssymb}
\usepackage{amsmath}
\usepackage{enumitem}
\def\T{{\mathrm{\scriptscriptstyle T}}}
\newcommand{\suppref}[1]{\hyperref[#1]{\ref*{#1}}}
\newcounter{suppsec}
\renewcommand{\thesuppsec}{section S\arabic{suppsec}}
\newcommand{\suppsection}[2]{%
    \refstepcounter{suppsec}%
    \hypertarget{#1}{}\textbf{\thesuppsec:} #2\label{#1}%
}
\setcitestyle{aysep={,}}
\usepackage{pdfpages}

\begin{document}
	
\title{A stable method for direct estimation of single-index terms in GAMs with an application to bike-sharing demand}

%%=============================================================%%
%% Prefix	-> \pfx{Dr}
%% GivenName	-> \fnm{Joergen W.}
%% Particle	-> \spfx{van der} -> surname prefix
%% FamilyName	-> \sur{Ploeg}
%% Suffix	-> \sfx{IV}
%% NatureName	-> \tanm{Poet Laureate} -> Title after name
%% Degrees	-> \dgr{MSc, PhD}
%% \author*[1,2]{\pfx{Dr} \fnm{Joergen W.} \spfx{van der} \sur{Ploeg} \sfx{IV} \tanm{Poet Laureate} 
%%                 \dgr{MSc, PhD}}\email{iauthor@gmail.com}
%%=============================================================%%
	
\author[1]{\fnm{Danilo V.} \sur{Silva}}
\email{danilo.silva@ime.usp.br}

\author*[1]{\fnm{Gilberto A.} \sur{Paula}}
\email{giapaula@ime.usp.br}	
	
\affil[1]{\orgdiv{Department of Statistics, Institute of Mathematics, Statistics and Computer Science}, \orgname{Universidade de S\~ao Paulo}, \orgaddress{\city{S\~ao Paulo}, \country{Brazil}}}
	
%% sample for unstructured abstract %%
%%==================================%%
	
\abstract{Single-index terms have been increasingly applied in generalized additive models (GAMs) across diverse areas due to their versatility in integrating functional flexibility with parametric dimension reduction while maintaining interpretability. However, the estimation presents severe computational challenges. This paper introduces a stable method based on Fisher scoring that dynamically updates the boundaries of the single-index covariates within a unified iterative framework using P-splines. Simulation studies with moderate sample sizes under non-Gaussian distributions confirm the empirical consistency of the estimation across multiple scenarios. Notably, the proposed method remains stable where state-of-the-art competitive methods fail to recover true single-index coefficients and their associated nonlinear functions, and is 80.13 times faster than the usual two-step method in the most computationally intensive scenario. The modeling advantage is illustrated through an application to Capital Bike Sharing data, where we deal with a single-index interaction effect for each year, with distinct single-index coefficients, a complex structure that makes competitive methods inapplicable. The proposed method is implemented in R, with all functions available at \url{https://github.com/silva-danilo/sipgam} for reproducibility and transparency in comparisons.} 
% The derived model matrices enable the fast computation of the estimated effective degrees of freedom and pointwise confidence bands for the single-index terms. The smoothing parameter updates are integrated into the iterative process via the generalized Fellner-Schall method, which recycles the derived matrix decompositions, thereby providing an efficient approximation to the global penalized optimization problem. 

\maketitle 

\section{Introduction}
Generalized additive models (GAMs) \citep{wood_gam} are a powerful class of models that have the flexibility of the exponential family and deal with nonlinear effects. These effects are elegantly modeled by terms of one or multiple covariates via smooth penalised functions. A single-index term combines similar continuous covariates, based on their importance, into the same linear index. Thus, one can improve efficiency by reducing the number of coefficients rather than using a smooth function for each covariate, and by interpreting the resulting index covariates. Also, it is simpler than using multidimensional smooth terms as tensor product P-splines \citep{eilers_joys}. In addition to their advantages, these models exhibit substantial application potential in situations involving multiple covariates of a similar nature, possibly correlated.

Various studies examine the estimation of single-index terms in GAMs. A relevant sequence begins with \citet{yu_2002}, initially formalising estimation of a unique single-index term alongside simultaneous linear terms in the least squares context, and \citet{yu_2017} extend it to the exponential family context. \citet{zu_2023} implement the two-step estimation method in \texttt{gplsim} \citep{gplsim_2023}, an efficient R package for \citet{yu_2017} especification with a unique single-index term. In these works, the authors note that estimation convergence is not guaranteed and that a larger sample size is required to obtain a satisfactory estimate under non-Gaussian distributions. Recently, \citet{li_2025} explore the same two-step method for estimation of a unique single-index term alongside simultaneous linear and one-covariate smooth terms. 

In the two-step estimation approach, the single-index coefficients are initially fixed (the single-index term becomes a one-covariate smooth term) and then optimized alternately with the complete GAM estimation. This provides implementation advantages by avoiding the need to manually specify smooth bases, identifiability constraints, and to estimate smoothing parameters, since the inner model can be fitted, for instance, using the consolidated \texttt{mgcv} \citep{mgcv_2025} R package. However, this nested approach makes the estimation process unstable and sensitive to the starting values of the single-index coefficients, and incurs the high computational cost of computing a complete GAM fit at each optimization step. To illustrate this, consider the model from  \citet{li_2025} such that $y_i | {\bf z}_i \sim \text{poi}(\mu_i)$ independent with $\mu_i$ denoting the conditional mean of the $i$-th observation and the relation $\log (\mu_i) = 0.25 + f(u_i)$, with
\begin{align*}
f(u_i) & = \dfrac{0.2(t_i)^{11} \{10(1-t_i)\}^6 + 10(10t_i)^3 (1-t_i)^{10}}{4} \quad \text{with}  \quad t_i = \dfrac{u_i + 0.41}{1.4},
\end{align*}
in which $u_i = (z_{i1} - z_{i2} - 0.5 z_{i3})/1.5$, the single-index coefficients for this model are $(0.66, -0.66, 0.33)^\T$. Starting with $(1,0,0)^\T$ is obtained $(0.6517, -0.6719, 0.3519)^\T$ needind to compute $115$ complete GAMs via \texttt{mgcv}. However, starting with $(1,2,1)^\T / \sqrt{6}$ is obtained $(0, 0.8803, -0.4744)^\T$ needind to compute $195$ complete GAMs. Furthermore, while these methods under this approach can be extended to simultaneous single-index terms and interaction effects, treating the index as a fixed covariate in the inner loop restricts the model to maintaining the same single-index coefficients across interaction effects, severely limiting its flexibility.

Alternatively, directly maximizing all coefficients may be robust to the indicated stability issues. This approach, however, requires computing the inner derivatives of smooth functions and dealing with the boundaries of single-index covariate bases. Since the domain of these smooth functions depends on the single-index coefficients, which are updated during model fitting, one solution is to fix the boundaries such that the probability of any single-index covariate falling outside is low for any plausible single-index coefficient estimate. Following this line, \citet{collarin_2025b} integrate simultaneous single-index terms and other complex covariate transformations with \texttt{mgcv} through the method available in the under-development R package \texttt{gamFactory} \citep{gamFactory_2026}. This user-friendly implementation estimates the index transformation coefficients simultaneously with all other model coefficients via a Bayesian approach with multivariate Gaussian priors for the single-index coefficients. While the theoretical framework is general, the stability analyses for the complete exponential family have not yet been provided, and the current package implementation supports only the Gaussian, binomial, generalized Pareto, Poisson, and Sinh-Arsinh families. This method still does not accommodate a single-index interaction effect, unlike the \texttt{by} argument in \texttt{mgcv}.

This paper proposes an alternative method for directly estimating all coefficients of GAMs with simultaneous single-index terms (alongside simultaneous linear and one-covariate smooth terms) using P-splines with dynamically updated boundaries. The use of P-splines is motivated by their renowned flexibility in regularizing B-spline bases via simple difference penalties (see, for instance, \citeauthor{durban_2015}, \citeyear{durban_2015}). The model matrices for single-index terms are defined to implement an estimation method that uses the penalized complete Fisher information matrix and a fast approximation update to optimize smoothing via the generalized Fellner-Schall method \citep{wood_2017} in the single-index additive case, recycling the derived matrix decompositions. The convergence of the proposed method is verified in simulation studies with moderate sample sizes under non-Gaussian distributions, demonstrating substantial improvements in stability and computational efficiency over state-of-the-art competitive methods.

The novelty lies in combining, based on Fisher scoring, the strengths of competitive methods: the joint estimation of all coefficients, characteristic of \texttt{gamFactory}, and the dynamic updates of the single-index boundaries, characteristic of two-step estimation. Furthermore, we demonstrate the modeling flexibility of our method in the application to Capital Bike Sharing data by modeling a single-index interaction effect for each year, with distinct single-index coefficients, a complex structure that makes competitive methods inapplicable.

The paper is organized as follows: the GAMs with single-index terms definition and notation are defined in \autoref{sec_model}, a PGAM-type iterative process is derived in \autoref{sec_pgam}, simulation results are given in \autoref{sec_simulation}, and an application to bike-sharing demand is presented in \autoref{sec_bike}. \autoref{sec_conclusion} deals with concluding remarks, and some derivations are given in appendices \autoref{append_eknots} -- \autoref{append_vcov_f}. Technical details are available as supplementary materials, and the R functions for replication and transparency are at \url{https://github.com/silva-danilo/sipgam}.

\section{The model} \label{sec_model}
The GAM with $m$ single-index terms is defined as
\[
\text{(i)} \ y_i | (\mathbf{x}_i, \mathbf{z}_i^1, \cdots, \mathbf{z}_i^m) \sim \text{EF}(\mu_i, \phi) \quad \text{and} \quad \text{(ii)} \ g(\mu_i) = \eta_i =  {\bf x}_i^\T {\boldsymbol \beta} + \sum_{j}^m f_j(u_i^j),
\]
where $\text{EF}(\mu_i, \phi)$ denotes exponential family distribution of mean $\mu_i$ and precision parameter $\phi$ of the $i$-th observation, $g$ is the link function that maps the linear predictor $\eta_i$ to the parameter space, ${\bf x}_i^\T {\boldsymbol \beta}$ is the linear term, $u_i^j = \{{\bf z}_i^j\}^\T {\boldsymbol \alpha}^j$ is the $j$-th single-index covariate (unobserved) and $f_j$ is its smooth function with ${\bf x}_i = (x_{i1}, \cdots, x_{ip})^\T$ and ${\bf z}_i^j = (z_{i1}^j, \cdots, z_{is_j+1}^j)^\T$ containing values of observed covariates, whereas ${\boldsymbol \beta} = (\beta_1, \cdots, \beta_p)^\T$ and ${\boldsymbol \alpha}^j = (\alpha_1^j, \cdots, \alpha_{s_j+1}^j)^\T$ are the single-index coefficients to be estimated. The joint specification alongside simultaneous one-covariate smooth terms, $u_i^j = z_i^j$, is easily particularized.

\subsection{Identifiability}
% In this paper ${\bf z}_i^j$ and ${\bf z}_i^{j'}$ do not share any covariates for all $j \neq j'$.
Usually $x_{i1} = 1$ for all observations, then $\beta_1$ is the model intercept. Single-index coefficients identifiability requires the constraints $||{\boldsymbol \alpha}^j|| = 1$ with $\alpha_1^j > 0$ for all $j$, and a usual reparameterization (see, for instance, \citeauthor{yu_2002}, \citeyear{yu_2002}) is 
$$
{\boldsymbol \alpha}^j = [1, \{\tilde{\boldsymbol \alpha}{}^j\}^\T]^\T/\sqrt{1 + ||\tilde{\boldsymbol \alpha}{}^j||^2},
$$
in which $\tilde{\boldsymbol \alpha}{}^j = (\tilde{\alpha}_1^j, \cdots, \tilde{\alpha}_{s_j}^j)^\T$ does not have constraints. In addition, we consider P-spline smoothings \citep{eilers_1996, durban_2015} such that $f_j(u^j) = \sum_{i}^{q_j+1} \text{N}_i^{k_j}(u^j)\gamma_i^j$ with $\text{N}_i^{k_j}(u^j)$ denoting the $j$-th B-spline basis of degree $k_j$ (order $d_j = k_j+1$) and adopted basis using equally-spaced knots (see \autoref{append_eknots}). 

Let ${\bf N}^j$ be the B-spline basis with rows $\{{\bf N}_i^j\}^\T$ with ${\bf N}_i^j = [\text{N}_1^{k_j}(u_i^j), \cdots, \text{N}_{q_j+1}^{k_j}(u_i^j)]^\T$ for each observation. These bases have no zero mean, which introduces another identifiability problem, now with respect to $\beta_1$. For the interpretation of $g^{-1}(\beta_1)$ as being the mean value of the reference observation with the additive functions having zero mean, the constraints $\sum_{i} f_j(u_i^j) = {\bf 1}^\T{\bf N}^j {\boldsymbol \gamma}^j = 0$ for all possible ${\boldsymbol \gamma}^j = (\gamma_1^j, \cdots, \gamma_{q_j+1}^j)^\T$ \citep{wood_gam} are required. That is equivalent to ${\bf 1}^\T{\bf N}^j = {\bf 0}_{1 \times q_j+1}$. Although this identifiability issue could be alternatively addressed by adding an extra quadratic penalty to the penalized log-likelihood during fitting \citep{wood_2020}, we adopt a reparameterization approach. Specifically, we take the column-centered basis matrix and drop the last column ($\tilde \gamma_{q_j+1}^j = 0$), yielding the basis $\tilde {\bf N}^j$ with $q_j$ columns. 

\subsection{Penalty structure}
The penalized log-likelihood function may be expressed as
\begin{equation} \label{eq_loglike}
\text{L}_p({\boldsymbol \psi}, \phi, {\boldsymbol \lambda}) = \text{L}({\boldsymbol \psi}, \phi) - \frac{1}{2} {\boldsymbol \psi}^\T {\bf P}_{\lambda} {\boldsymbol \psi} = \text{L}({\boldsymbol \psi}, \phi) -\frac{1}{2} \sum_j^m \lambda_j \{ \tilde{\boldsymbol \gamma}^j \}^\T \tilde{\bf P}^j \tilde{\boldsymbol \gamma}^j,
\end{equation}
where $\text{L}({\boldsymbol \psi}, \phi) = \sum_{i}^n \phi\{y_i\tau_i - b(\tau_i)\} + c(y_i,\phi)$ is the regular exponential family log-likelihood function with ${\boldsymbol \psi} = [{\boldsymbol \beta}^\T, \{\tilde {\boldsymbol \gamma}^1\}^\T,
\{\tilde{\boldsymbol \alpha}^1\}^\T, \cdots, \{\tilde{\boldsymbol \gamma}^m\}^\T,  \{\tilde{\boldsymbol \alpha}^m\}^\T]^\T$ (this order affects the model penalty and matrix), $\tau_i = \tau(\mu_i)$ denotes the canonical parameter, $b$ and $c$ are twice differentiable functions. The model penalization is ${\bf P}_{\lambda} = \sum_j \lambda_j {\bf P}^j$ \citep{wood_gam}, whereas $\lambda_j > 0$ is the smoothing parameter and 
\begin{equation} \label{eq_penalty_matrix}
{\bf P}^j = \text{blockdiag} \left( {\bf 0}_{p \times p}, \ {\bf 0}_{q_{1} \times q_{1}}, \ {\bf 0}_{s_{1} \times s_{1}}, \ \cdots, \ \tilde {\bf P}^j, \ {\bf 0}_{s_{j} \times s_{j}} \ \cdots, \ {\bf 0}_{s_{m} \times s_{m}} \right)
\end{equation}
% \begin{equation} \label{eq_penalty_matrix}
% {\bf P}^j = \left[ \begin{array}{ccccccccc} 
% {\bf 0}_{p \times p} & {\bf 0}_{p \times q_{1}} & {\bf 0}_{p \times s_{1}} & \cdots & {\bf 0}_{p \times q_{j}} & {\bf 0}_{p \times s_{j}} & \cdots & {\bf 0}_{p \times q_{m}} & {\bf 0}_{p \times s_{m}} \\
% {\bf 0}_{q_{1} \times p} & {\bf 0}_{q_{1} \times q_{1}} & {\bf 0}_{q_{1} \times s_{1}} & \cdots & {\bf 0}_{q_{1} \times q_{j}} & {\bf 0}_{q_{1} \times s_{j}} & \cdots & {\bf 0}_{q_{1} \times q_{m}} & {\bf 0}_{q_{1} \times s_{m}} \\
% \vdots \\ 
% {\bf 0}_{q_{j} \times p} & {\bf 0}_{q_{j} \times q_{1}} & {\bf 0}_{q_{j} \times s_{1}} & \cdots & \tilde {\bf P}^j & {\bf 0}_{q_{j} \times s_{j}} & \cdots & {\bf 0}_{q_{j} \times q_{m}} & {\bf 0}_{q_{j} \times s_{m}} \\
% \vdots \\ 
% {\bf 0}_{s_{m} \times p} & {\bf 0}_{s_{m} \times q_{1}} & {\bf 0}_{s_{m} \times s_{1}} & \cdots & {\bf 0}_{s_{m} \times q_{j}} & {\bf 0}_{s_{m} \times s_{j}} & \cdots & {\bf 0}_{s_{m} \times q_{m}} & {\bf 0}_{s_{m} \times s_{m}} 
% \end{array} \right],
% \end{equation}
in which $\tilde {\bf P}^j = \{\tilde{\bf D}_{\textit{dif}_j}^j\}^\T \tilde{\bf D}_{\textit{dif}_j}^j$ with $\tilde{\bf D}_{\textit{dif}_j}^j$ being the difference of order $\textit{dif}_j$ having the last column dropped, particularly for $q_j = 4$:
\[
\tilde{\bf D}_{1}^j = \left[ \begin{array}{rrrr} 
-1 & 1 & 0 & 0 \\
0 & -1 & 1 & 0 \\
0 & 0 & -1 & 1  \\
0 & 0 &  0 & -1
\end{array} \right] 
\ \text{and} \
\tilde{\bf D}_{2}^j =  \left[ \begin{array}{rrrr} 
1 & -2 & 1 & 0 \\
0 & 1 & -2 & 1 \\
0 & 0 & 1 & -2 
\end{array} \right]. 
\]
Note that the last row is deliberately retained to ensure the second-order differences are computed correctly, since $\tilde \gamma_{q_j+1}^j = 0$, preventing underpenalization at the right boundary. 
% This is correct because $\tilde \gamma_{q_j+1}^j = 0$.

Maximizing \eqref{eq_loglike}, for ${\boldsymbol \lambda} = (\lambda_1, \cdots, \lambda_m)^\T$ fixed, results in the maximum penalized log-likelihood estimate (MPLE) of $({\boldsymbol \psi}^\T, \phi)^\T$. In addition to the already well-explored problem of selecting the smoothing parameter ${\boldsymbol \lambda}$ \citep{wood_2011}, the direct estimation of single-index terms introduces a profound computational difficulty: estimating the single-index coefficients $\tilde{\boldsymbol \alpha}^1, \cdots, \tilde{\boldsymbol \alpha}^m$ whose updates change their smooth bases design.
%In the next section, we derive a PGAM-type iterative process for direct estimation of all coefficients, integrating the smoothing parameters.
%The precision parameter is considered in the second term of the penalized log-likelihood function to eliminate its need in the iterative process and smoothing optimization. In the usual approach, the smoothing parameters are given by $\phi \lambda_j$. 

\section{PGAM-type iterative process} \label{sec_pgam}
The penalized score of $\tilde{\alpha}_1^j$ is given by
\begin{align*}
\text{U}_{\tilde{\alpha}_1^j} & = \frac{\partial}{\partial \tilde{\alpha}_1^j} \left \{ \sum_{i}^n \phi\{y_i\tau_i - b(\tau_i)\} + c(y_i,\phi) -\frac{1}{2} \sum_j^m \lambda_j \{ \tilde{\boldsymbol \gamma}^j \}^\T \tilde{\bf P}^j \tilde{\boldsymbol \gamma}^j \right \} \\
& = \phi \sum_{i}^n y_i \frac{d \tau_i}{d \mu_i} \frac{d \mu_i}{d \eta_i}  \tilde{f}'_j(u_i^j) \frac{d }{d \tilde{\alpha}_1^j} \{ u_i^j \} - b'(\tau_i) \frac{d \tau_i}{d \mu_i} \frac{d \mu_i}{d \eta_i} \tilde{f}'_j(u_i^j) \frac{d }{d \tilde{\alpha}_1^j} \{ u_i^j \} \\
& = \phi \sum_{i}^n \tilde{f}'_j(u_i^j) \frac{d }{d \tilde{\alpha}_1^j} \left\{ \alpha^j_1 z_{i1}^j + \cdots + \alpha^j_{s_j+1} z_{is_j+1}^j \right\} \frac{1}{\text{V}_i} \frac{d \mu_i}{d \eta_i} (y_i - \mu_i) \\
%& = \phi \sum_{i}^n \sqrt{ \left( \frac{1}{V_i} \frac{d \mu_i}{d \eta_i} \right)^2} (y_i - \mu_i)  x_{i1} \\
& = \phi \sum_{i}^n \left\{ \frac{d \alpha^j_1}{d \tilde{\alpha}_1^j} z_{i1}^j + \cdots + \frac{d \alpha^j_{s_j+1}}{d \tilde{\alpha}_1^j} z_{is_j+1}^j \right\} \tilde{f}'_j(u_i^j) \sqrt{ \frac{\omega_i}{\text{V}_i} } (y_i - \mu_i), 
\end{align*}
with $\omega_i = \{g'(\mu_i)\}^{-2}\text{V}_i^{-1}$, $\text{V}_i = d\mu_i/d\tau_i$. Let  ${\bf Z}^j$ be the matrix with rows $\{{\bf z}_i^j\}^\T$, ${\bf W} = \text{diag}\{\omega_1, \cdots, \omega_n \}$, ${\bf V} = \text{diag}\{\text{V}_1, \cdots, \text{V}_n\}$, ${\bf y} = (y_1, \cdots, y_n)^\T$ and ${\boldsymbol \mu} = (\mu_1, \cdots, \mu_n)^\T$. Also, let ${\bf J}^j$ be the $j$-th Jacobian matrix of ${\boldsymbol \alpha}^j$ with respect to $\tilde{\boldsymbol \alpha}^j$ (see \autoref{append_jacobian}) and $\{\tilde{\bf F}^j\}^{(1)} = \text{diag}\{\tilde{f}'_j(u_1^j), \cdots, \tilde{f}'_j(u_n^j)\}$ in which $\tilde{f}'_j$ is another B-spline of degree $k_j - 1$ and can be expressed using the reparametrized coefficients $\tilde{\boldsymbol \gamma}^j$ from the original curve $\tilde{f}_j$ (see \autoref{append_df}). Then, the penalized score of $\tilde{\boldsymbol \alpha}^j$ takes the form
\begin{align*}
\text{U}_{\tilde{\alpha}^j} & = \frac{\partial \text{L}_p({\boldsymbol \psi}, \phi, {\boldsymbol \lambda})}{\partial \tilde{\boldsymbol \alpha}^j} = \phi \{ \{\tilde{\bf F}^j\}^{(1)} {\bf Z}^j {\bf J}^j \}^\T {\bf W}^{\frac{1}{2}} {\bf V}^{-\frac{1}{2}}({\bf y} - {\boldsymbol \mu}).
\end{align*}
Defining $\tilde{\bf T}^j = \{\tilde{\bf F}^j\}^{(1)} {\bf Z}^j {\bf J}^j$ as the $j$-th single-index term model matrix, ${\bf X}$ with rows ${\bf x}_i^\T$ as linear term model matrix and $\tilde{\bf N}^j$ with rows $\{\tilde{\bf N}_i^j\}^\T$ as reparameterized B-spline basis, therefore the penalized score function of $({\boldsymbol \psi}^\T, \phi)^\T$ (see supplementary \suppref{supp_pscore}) is expressed as
\[
\begin{bmatrix}
{\bf U}_{\beta} \\ 
{\bf U}_{\tilde{\gamma}_1} \\
{\bf U}_{\tilde{\alpha}_1} \\
\vdots \\
{\bf U}_{\tilde{\gamma}_m} \\
{\bf U}_{\tilde{\alpha}_m} \\
\text{U}_\phi 
\end{bmatrix} = 
\begin{bmatrix}
\phi {\bf X}^\T {\bf W}^{\frac{1}{2}}{\bf V}^{-\frac{1}{2}}({\bf y} - {\boldsymbol \mu}) \\ 
\phi \{\tilde{\bf N}^1\}^\T {\bf W}^{\frac{1}{2}}{\bf V}^{-\frac{1}{2}}({\bf y} - {\boldsymbol \mu}) - \lambda_1 \tilde{\bf P}^1 \tilde{\boldsymbol \gamma}^1 \\
\phi \{\tilde{\bf T}^1\}^\T {\bf W}^{\frac{1}{2}} {\bf V}^{-\frac{1}{2}}({\bf y} - {\boldsymbol \mu}) \\
\vdots \\
\phi \{\tilde{\bf N}^m\}^\T {\bf W}^{\frac{1}{2}}{\bf V}^{-\frac{1}{2}}({\bf y} - {\boldsymbol \mu}) - \lambda_m \tilde{\bf P}^m \tilde{\boldsymbol \gamma}^m \\
\phi \{\tilde{\bf T}^m\}^\T {\bf W}^{\frac{1}{2}} {\bf V}^{-\frac{1}{2}}({\bf y} - {\boldsymbol \mu}) \\
\sum_{i} y_i\tau_i - b(\tau_i) + c'(y_i,\phi)
\end{bmatrix}.
\]
The penalized Fisher information of $(\tilde{\alpha}_1^j, \tilde{\alpha}_2^j)$ given by
\begin{align*}
\text{K}_{\tilde{\alpha}_1^j \tilde{\alpha}_2^j} & = -\text{E} \left[ \frac{\partial}{\partial \tilde{\alpha}_2^j} \left \{ \phi \sum_{i}^n \left\{ \frac{d \alpha^j_1}{d \tilde{\alpha}_1^j} z_{i1}^j + \cdots + \frac{d \alpha^j_{s_j+1}}{d \tilde{\alpha}_1^j} z_{is_j+1}^j \right\} \tilde{f}'_j(u_i^j) \sqrt{ \frac{\omega_i}{\text{V}_i} } (y_i - \mu_i) \right \} \right] \\
& = \phi \sum_{i}^n \left\{ \sum_k^{s_j+1} \frac{d \alpha^j_k}{d \tilde{\alpha}_1^j} z_{ik}^j \right\} \tilde{f}'_j(u_i^j) \sqrt{ \frac{\omega_i}{\text{V}_i} } \frac{\text{V}_i}{\text{V}_i} \frac{d \mu_i}{d \eta_i}  \tilde{f}'_j(u_i^j) \left\{ \sum_k^{s_j+1} \frac{d \alpha^j_k}{d \tilde{\alpha}_2^j} z_{ik}^j \right\} \\
& = \phi \sum_{i}^n \left\{ \sum_k^{s_j+1} \frac{d \alpha^j_k}{d \tilde{\alpha}_1^j} z_{ik}^j \right\} \tilde{f}'_j(u_i^j) \omega_i  \tilde{f}'_j(u_i^j) \left\{ \sum_k^{s_j+1} \frac{d \alpha^j_k}{d \tilde{\alpha}_2^j} z_{ik}^j \right\},
\end{align*}
so that the penalized Fisher information matrix of $(\tilde{\boldsymbol \alpha}^j, \tilde{\boldsymbol \alpha}^j)$ takes the form
\begin{align*}
{\bf K}_{\tilde{\alpha}^j \tilde{\alpha}^j} & = \text{E} \left[ -\frac{\partial^2 \text{L}_p({\boldsymbol \psi}, \phi, {\boldsymbol \lambda})}{\partial \tilde{\boldsymbol \alpha}^j \partial \{\tilde{\boldsymbol \alpha}^j\}^\T} \right] =  \phi \{ \{\tilde{\bf F}^j\}^{(1)} {\bf Z}^j {\bf J}^j \}^\T {\bf W} \{\tilde{\bf F}^j\}^{(1)} {\bf Z}^j {\bf J}^j,
\end{align*}
and the penalized complete Fisher information (see supplementary \suppref{supp_pfisher}) may be expressed as  
\[
\begin{bmatrix}
{\bf K}_{\psi \psi} & {\bf K}_{\psi \phi} \\
{\bf K}_{\phi \psi} & \text{K}_{\phi \phi} 
\end{bmatrix}, \ \text{where} \quad
\begin{aligned}
{\bf K}_{\psi \psi} & = \phi{\bf M}^\T {\bf W} {\bf M} + {\bf P}_\lambda \\
{\bf K}_{\psi \phi} & = \{ {\bf K}_{\phi \psi} \}^\T = {\bf 0}_{p + \sum_j q_j + s_j \times 1}, \end{aligned}
\]
%The penalized Fisher information matrix (see \ref{append_fisher}) is ${\bf K}_{\psi \psi} = \phi \{ {\bf M}^\T {\bf W} {\bf M} + {\bf P}_\lambda \} $, and with ${\boldsymbol \rho} = \phi^{-1}{\boldsymbol \lambda}$ as the reparametrized smoothing parameters (${\bf P}_\rho = \phi^{-1}{\bf P}_\lambda$),
$\text{K}_{\phi \phi} = \text{E}\{-\sum_{i}^n c''(y_i,\phi)\}$ in which ${\bf P}_{\lambda}$ is the model penalization from the previous section and ${\bf M} = [{\bf X}, \tilde{\bf N}^1, \tilde{\bf T}^1, \cdots, \tilde{\bf N}^m, \tilde{\bf T}^m]$ is the model matrix. Due to the orthogonality between ${\boldsymbol \psi}$ and $\phi$, the precision parameter can be estimated separately from ${\boldsymbol \psi}$. In particular, for the canonical link function ($\tau_i = \eta_i$), it follows that ${\bf W} = {\bf V}$ and the penalized score function assumes a reduced form. 

\subsection{Estimation framework}
The Fisher scoring procedure for obtaining the MPLE of ${\boldsymbol \psi}$, fixed ${\boldsymbol \lambda}$, leads to the following PGAM-type \citep{marx_1998} iterative process (see \autoref{append_pgam}):
\begin{equation} \label{eq_pgam}
{\boldsymbol \psi}^{(t+1)} = [ \{{\bf M}^{(t)}\}^\T {\bf W}^{(t)} {\bf M}^{(t)} + \phi^{-1} {\bf P}_\lambda]^{-1} \{{\bf M}^{(t)}\}^\T {\bf W}^{(t)} {\tilde{\bf y}}^{(t)},
\end{equation}
%in which  ${\tilde{\bf y}} - \sum_j^m \tilde{\bf u}^j$ 
with ${\tilde{\bf y}} = {\bf M}{\boldsymbol \psi} + {\bf W}^{-\frac{1}{2}}{\bf V}^{-\frac{1}{2}}({\bf y} - {\boldsymbol \mu})$ denotes the dependent modified response. The precision parameter can not be incorporated into the penalty because it is explicitly needed in the smoothing optimization. Actually, the Fisher inverse is not computed directly to improve numerical precision and efficiency. Instead, the Cholesky decomposition can be used (since the matrix is symmetric and positive definite, ensured by model identifiability) to solve two triangular systems. 
% as in \citet{cardozo_2022}

Let ${\bf L}$ be the upper triangular square matrix such that $ {\bf L}^\T {\bf L} = \{ {\bf M}^\T {\bf W} {\bf M} \} + \phi^{-1} {\bf P}_\lambda $, then we derive the following procedure that integrates the smoothing parameter updates for direct estimation of all model coefficients:
\begin{align} \label{eq_chol}
\text{(i)} \ \{{\bf L}^{(t)}\}^\T {\bf b}^{(t)} = \{{\bf M}^{(t)}\}^\T {\bf W}^{(t)} {\tilde{\bf y}}^{(t)} \quad \text{and} \quad \text{(ii)} \ {\bf L}^{(t)} {\boldsymbol \psi}^{(t+1)} = {\bf b}^{(t)}, 
\end{align}
%\begin{align} \label{eq_chol}
%\text{(i)} \ {\bf L}^{(t)} \{{\bf L}^{(t)}\}^{-1} = {\bf I}_{p + \sum_j q_j + s_j} \quad \text{and} \quad \text{(ii)} \ \{{\bf L}^{(t)}\}^{-1} \{{\bf L}^{(t)}\}^{-\T} \{\tilde{\bf M}^{(t)}\}^\T {\tilde{\bf y}}^{(t)}, 
%\end{align}
%Note that {\bf W} and {\bf M} change for each step, then the order cost of always computing the Fisher inverse is irreducible. 
solving via forward and backward substitutions, respectively, with a simple update for smoothing optimization via the generalized Fellner-Schall method \citep{wood_2017} for the additive single-index case (see \autoref{append_smooth}) given by 
\begin{equation} \label{eq_lambda_up}
\lambda_j^{(t+1)} = \frac{\text{tr} \left\{ {\bf Q}^{(t)}  {\bf P}^j \right\} }{ \{ \tilde{\boldsymbol \gamma}^{j(t+1)}  \} ^\T \tilde{\bf P}^j \tilde{\boldsymbol \gamma}^{j(t+1)} } \lambda_j^{(t)},
\end{equation}
in which ${\bf Q} = {\bf P}_\lambda^{-} - \phi^{-1} \text{cp} \{ {\bf B} \}$ with ${\bf B}$ denoting the solution of the triangular system ${\bf L}^\T {\bf B} = {\bf I}_{p + \sum_j q_j + s_j}$. 

The order, keeping the coefficient dimensions, of floating-point operations (flops) required for the update in step \eqref{eq_chol}, given ${\bf M}$, ${\boldsymbol \psi}$, ${\boldsymbol \mu}$ and $\phi^{-1} {\bf P}_\lambda$, can be detailed as follows. The computation of both $\tilde{\bf M} = {{\bf W}}^{\frac{1}{2}} {\bf M}$ and ${\bf M}^\T {\bf W} {\tilde{\bf y}}$ is $O[n(p + \sum_j q_j + s_j)]$ flops combined, exploiting the diagonal structure of {\bf W} and {\bf V}. The evaluation of $\text{cp} \{ \tilde{\bf M} \} + \phi^{-1} {\bf P}_\lambda$ is $O[n(p + \sum_j q_j + s_j)^2]$ flops, with the number of operations reduced by exploiting the symmetry of the cross-product matrix $\text{cp}$. Obtaining ${\bf L}$ via Cholesky decomposition is $O[(p + \sum_j q_j + s_j)^3]$ flops, requiring approximately one-third of the operations of a general matrix inversion. Solving systems with forward and backward substitution requires $O[(p + \sum_j q_j + s_j)^2]$ flops in total, with the number of operations reduced by exploiting the triangular structure. As $n \gg p, q_j, s_j$, the total cost of each step in the iterative process given by the dominant terms is $O[n(p + \sum_j q_j + s_j)^2]$. We propose the implementation in \autoref{append_implem} for the algorithm \eqref{eq_chol} and \eqref{eq_lambda_up}. 

Dynamically updated boundaries of the single-index covariate are obtained by recomputing the domain of the respective smooth functions at each optimization step, so that the model matrix design depends continuously on the single-index coefficients. Statistically, this prevents the scaling problem discussed by \citet{collarin_2025b}, in which the single-index covariates cover only a small part of their fixed domain, wasting basis resolution. However, this dynamic updating introduces a profound form of nonlinearity, rendering existing optimal algorithms (see, for instance, \citeauthor{wood_gam}, \citeyear{wood_gam}) that use penalized iterative reweighted least squares (PIRLS)-type updates with locally fixed model matrices structurally incompatible. 

\subsection{Inference}
The single-index term variability estimator must also account for variability in the single-index coefficients. The direct approach allows us to incorporate this detail easily. An efficient, with $O[n(q_j + s_j)^2]$ flops, 95\% asymptotic pointwise confidence band for a single-index term $\tilde{\bf f}_j$ can be obtained by using the delta method (see \autoref{append_vcov_f}) and interpolating the intervals:
$$
{\hat{\tilde{\bf f}}}_j \pm 1.96 \sqrt{\dfrac{1}{\phi} \text{rowSums} \left\{
\left( \left[ \begin{array}{cc}
\tilde{\bf N}^j & \tilde{\bf T}^j \\
\end{array} \right]
{\bf B}^\T_{[p+1+\sum_i^{j-1}q_i+s_i:p+\sum_i^{j}q_i+s_i]} \right)^{\circ 2}
\right\}},
$$
in which the subscript $[p+1+\sum_i^{j-1}q_i+s_i:p+\sum_i^{j}q_i+s_i]$ indicates the block of the covariance matrix decomposition corresponding to the parameters $\tilde{\boldsymbol \gamma}^j$, $\tilde{\boldsymbol \alpha}^j$, and $\circ$ is the Hadamard product. 

Let $\textit{edf}$ be $\text{tr} \left \{ [ \text{cp} \{ \tilde{\bf M} \} + \phi^{-1} {\bf P}_\lambda]^{-1} \tilde{\bf M}^\T\tilde{\bf M} \right \} = \text{tr} \left \{ \text{cp} \{ {\bf B} \} \text{cp} \{ \tilde{\bf M} \} \right \}$. This quantity represents the effective degrees of freedom of the model, providing a measure of its complexity. The $\textit{edf}_{\tilde \alpha^j} = s_j$ is the effective degrees of freedom specifically to coefficients $\tilde{\boldsymbol \alpha}^j$ that coincides with the sum of the principal diagonal elements of the matrix $\text{cp} \{ {\bf B} \} \text{cp} \{ \tilde{\bf M} \}$ at the corresponding positions.

\section{Simulation} \label{sec_simulation}
We consider scenarios with $n = 200, 800, 3200$ observations and compare the proposed, two-step, and \texttt{gamFactory} methods across $500$ simulation replicates to assess their stability and empirical consistency. For the same $n$, the observed covariates are fixed uniformly on the unit interval, and only the response is simulated. We specify $q_j = 9$, $d_j = 4$, and $\textit{dif}_j = 2$ for all terms, as is usually done, and apply the true link function in all methods. The simulations are performed on a personal laptop with a 13th Gen Intel(R) Core(TM) i5-13450HX (8x 2.40 GHz) processor and 16.0 GB of RAM.

The two-step method uses the \texttt{optim()} function to obtain initial values from a non-penalized model, as in \citet{li_2025}, and we extend this to simultaneous single-index terms. The \texttt{gamFactory} method employs constraints that differ from those of the other considered methods and prevent immediate recovery of the single-index coefficients. For the single-index transformation, the covariate that enters the smooth function is a centered and scaled version of the linear combination, where the scaling factor ($a_0$) is estimated alongside the coefficients. This method imposes a penalty on the empirical variance of the linear combination to scale it to a constant (by default $c=1$), leaving the norm only weakly identified. Then, we normalize the estimated single-index coefficients to unit norm, set the sign of the first coefficient to positive, and set the smooth functions of covariates to zero mean. After this postprocessing, the resulting single-index coefficients and fitted smooth functions are directly comparable across all methods. Furthermore, following a recommendation from the package developers for these simulation studies, we use random initialization of the single-index coefficients and the generalized Fellner-Schall iteration as the optimizer.

We consider models with simultaneous single-index terms under non-Gaussian distributions, organized into three simulation scenarios: Poisson I, Gamma, and Poisson II, covering discrete and continuous responses. The first scenario serves as a computational baseline to validate correct implementation of the methods using simple nonlinear functions, while the last two scenarios incorporate more complex nonlinear functions to assess the methods under more demanding conditions. The \texttt{gamFactory} is not compared in the Gamma scenario because it currently does not support this family. 

To comprehensively evaluate the methods, we consider the average relative error of the estimated single-index coefficients and the instability rate. Because the single-index coefficients are simulated with unit norm, a relative error exceeding 0.5 indicates severe failure of the method, rendering the corresponding smooth function meaningless. Therefore, we adopt this threshold not as an arbitrary precision measure but as a binary criterion for identifying problematic fits. Fits exceeding this threshold are classified as unstable. Furthermore, we saved the outputs of each fitted model in an equal-sized list to standardize and ensure that the memory overhead of any method did not affect the runtime comparisons.
% These coefficients are chosen so that nonlinear functions have a similar range of variation. 

\subsection{Poisson I}
Let $y_i | ({\bf x}_i, {\bf z}_i^1, {\bf z}_i^2) \sim \text{poi}(\mu_i)$ independent with $\mu_i$ denoting the conditional mean of the $i$-th observation. The simulated model is given by the relation $\log (\mu_i) = 2 + 0.7{x}_i + \sum_j^2 \tilde{f}_j(u_i^j)$ in which $\tilde{f}_j(u^j_i) = f_j(u^j_i) - \sum_i f_j(u^j_i) / n$ for all $j$, with
\begin{align*}
f_1(u^1_i) & = \sin(4t^1_i) \quad \text{with} \quad t^1_i = u^1_i/\sqrt{12} - 0.11, \\
f_2(u^2_i) & = \sin(4t^2_i) - \cos(4t^2_i) \quad \text{with} \quad t^2_i = u^2_i/\sqrt{12} + 0.45,
\end{align*}
in which $u^1_i = (\tilde z_{i1}^1 - 1.4 \tilde z_{i2}^1)/1.72$, and $u^2_i = (\tilde z_{i1}^2 + 1.7 \tilde z_{i2}^2 - 0.8 \tilde z_{i3}^2)/2.13$ and $\tilde z_{ij}$ are the centered and scaled transformation of $z_{ij}$, resulting in single-index coefficients ${\boldsymbol \alpha}^1 = (0.58, -0.81)^\T$ and ${\boldsymbol \alpha}^2 = (0.47, 0.80, -0.38)^\T$. Under this configuration, the dimension of ${\boldsymbol \psi}$ is $23$. 

\begin{figure}[!htb]\centering
\includegraphics[width=11cm]{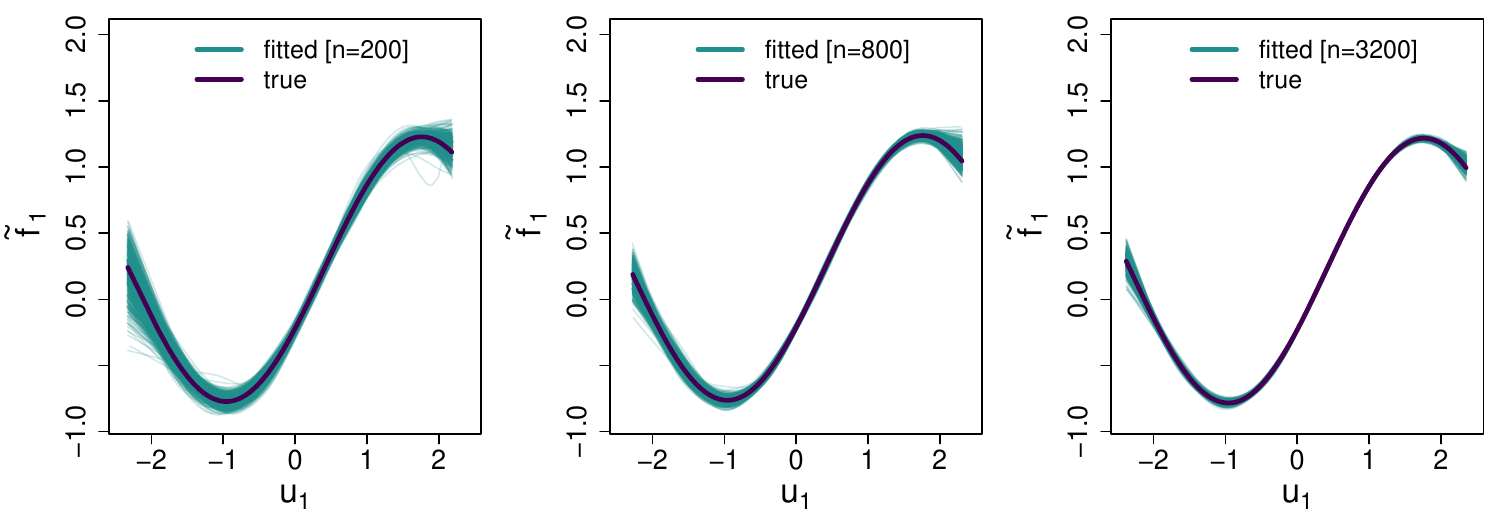}
\includegraphics[width=11cm]{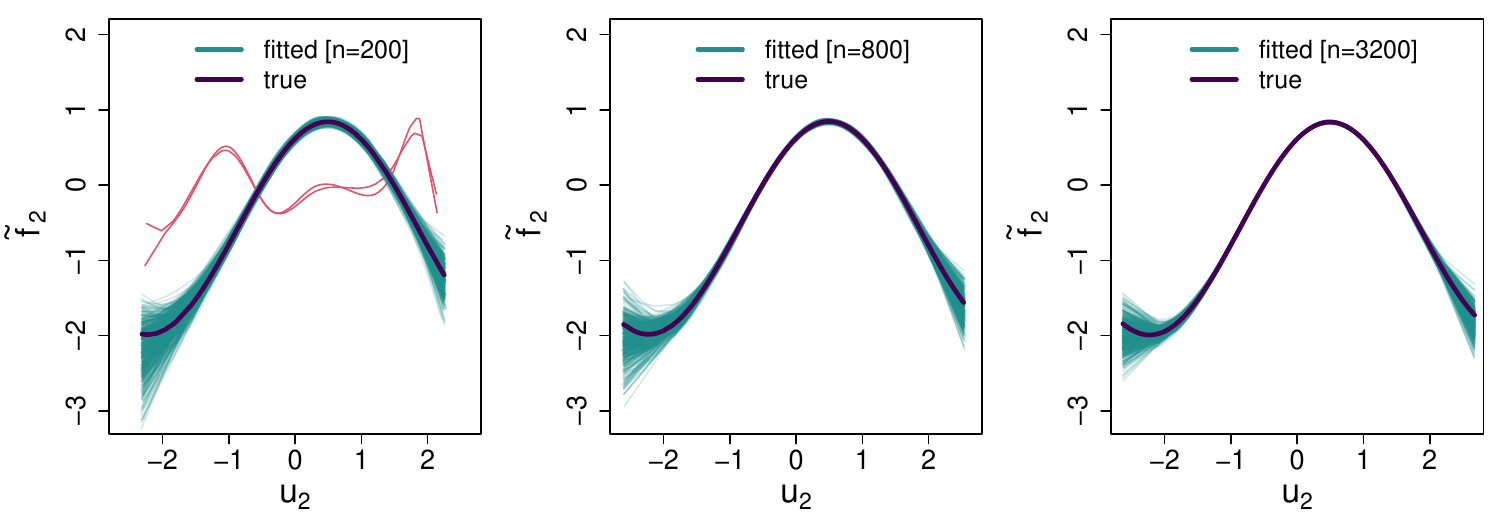}
\caption{\label{simu_poisson1_f_prop} Fitted additive function $\tilde{f}_1$ and single-index covariate ${u}_1$ (top-left, top-middle, top-right), $\tilde{f}_2$ and ${u}_2$ (bottom-left, bottom-middle, bottom-right) from all fitted models with the proposed method in the Poisson I scenario. The red curves are unstable fitted models.}
\end{figure}
\begin{figure}[!htb]\centering
\includegraphics[width=11cm]{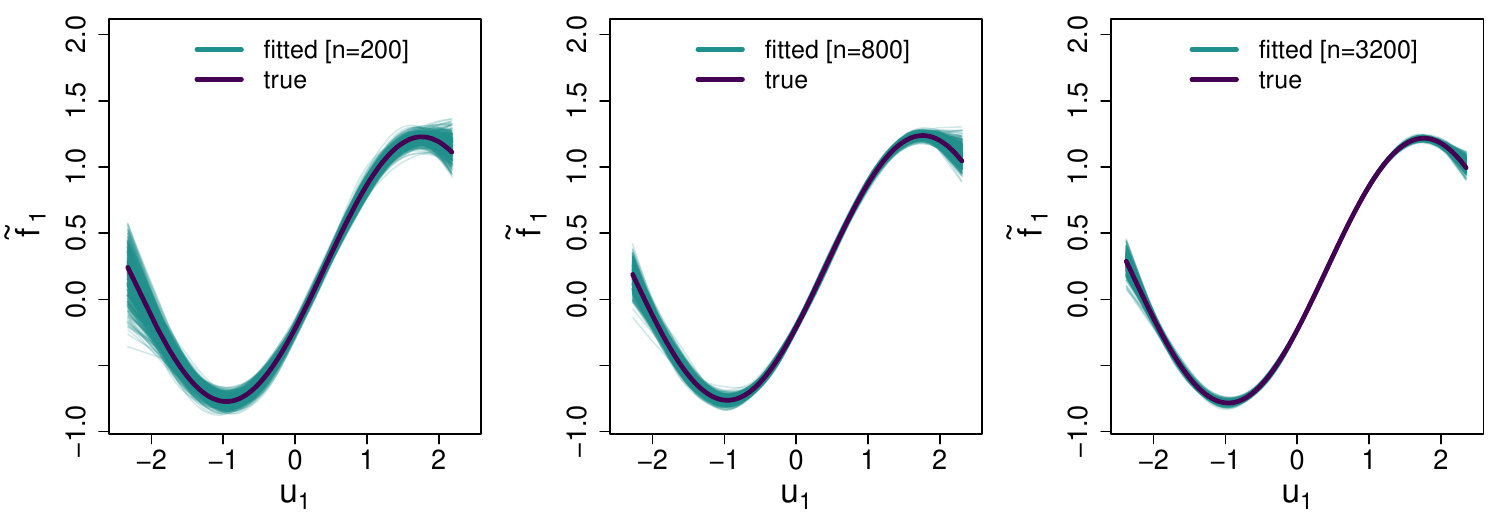}
\includegraphics[width=11cm]{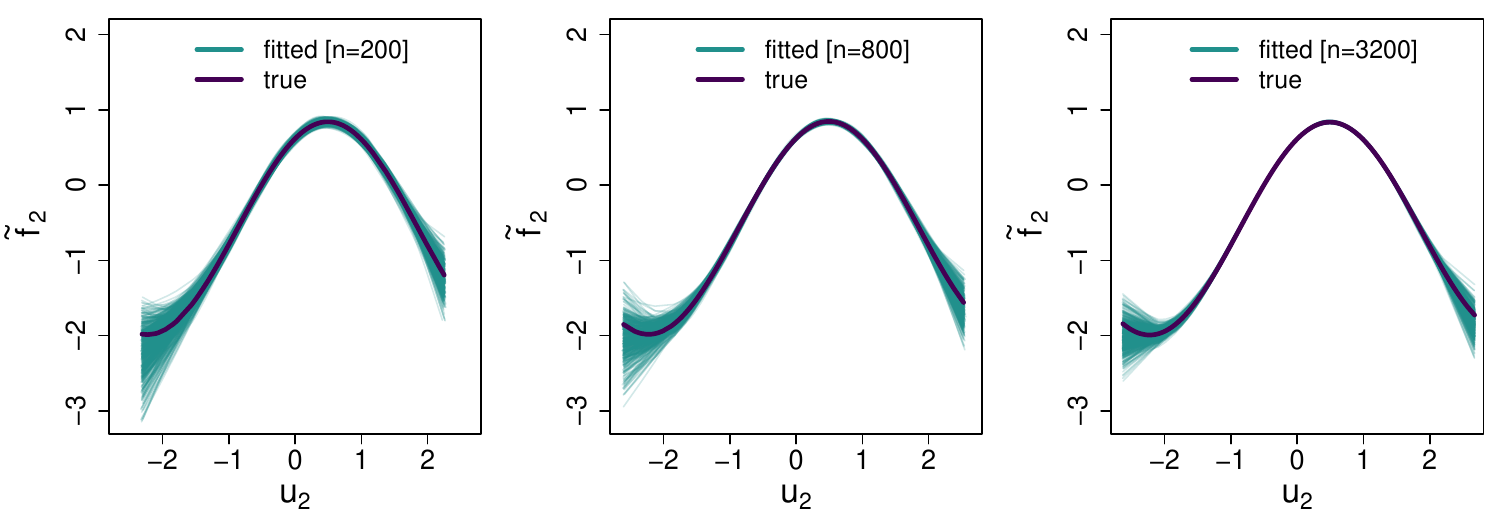}
\caption{\label{simu_poisson1_f_2step} Fitted additive function $\tilde{f}_1$ and single-index covariate ${u}_1$ (top-left, top-middle, top-right), $\tilde{f}_2$ and ${u}_2$ (bottom-left, bottom-middle, bottom-right) from all fitted models with the two-step method in the Poisson I scenario. The red curves are unstable fitted models.}
\end{figure}
\begin{figure}[!htb]\centering
\includegraphics[width=11cm]{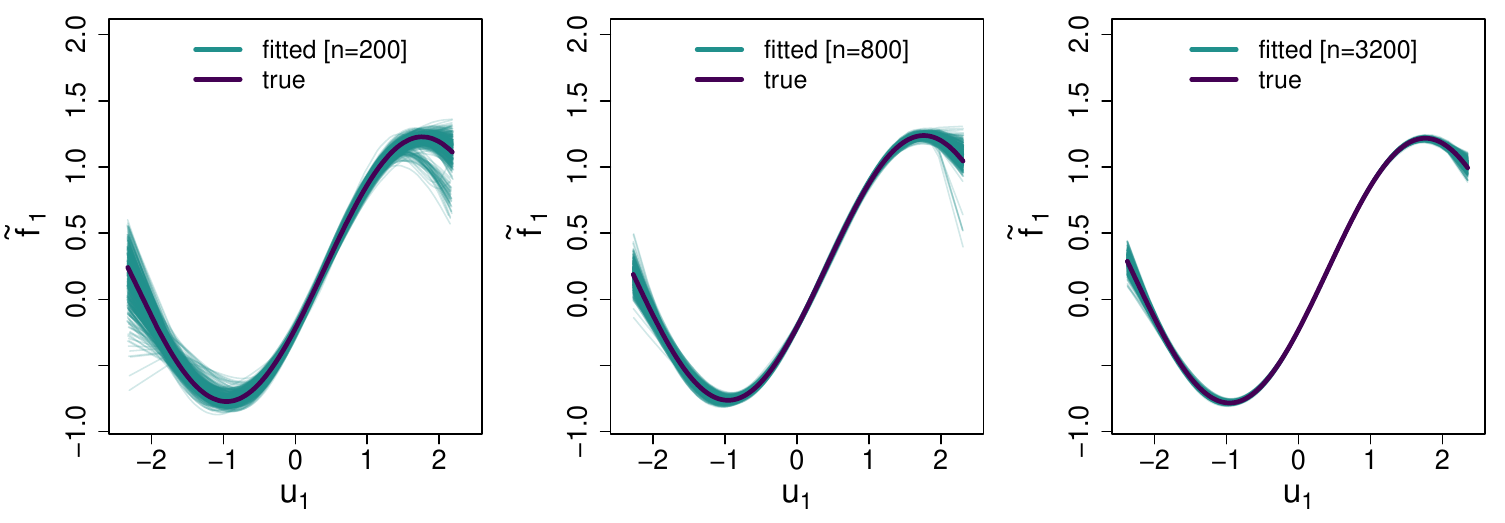}
\includegraphics[width=11cm]{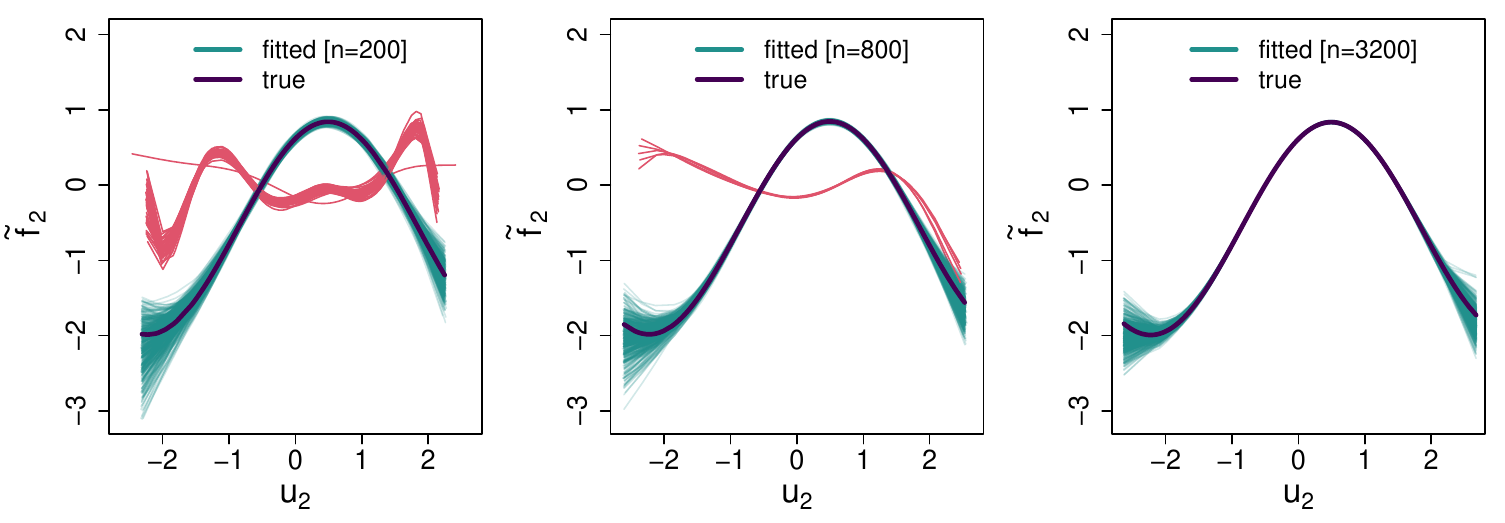}
\caption{\label{simu_poisson1_f_gamfactory} Fitted additive function $\tilde{f}_1$ and single-index covariate ${u}_1$ (top-left, top-middle, top-right), $\tilde{f}_2$ and ${u}_2$ (bottom-left, bottom-middle, bottom-right) from all fitted models with the \texttt{gamFactory} method in the Poisson I scenario. The red curves are unstable fitted models.}
\end{figure}

The proposed method takes a total runtime of $5.91$ minutes, with $90\%$ of all fits concluding in less than $0.37$ seconds. The fitted nonlinear functions and single-index covariates are presented in \autoref{simu_poisson1_f_prop}. For $n=200$, only $0.4\%$ of the fitted models are unstable (highlighted in red). 

The two-step method takes a total runtime of $166.01$ minutes, which is $28.09$ times longer than the proposed method due to the need to compute $341738$ complete GAMs via \texttt{mgcv}. The fitted nonlinear functions and single-index covariates are presented in \autoref{simu_poisson1_f_2step}. None of the fitted models is unstable.

The \texttt{gamFactory} method takes a total runtime of $41.74$ minutes, which is $7.06$ times longer than the proposed method. The fitted nonlinear functions and single-index covariates are presented in \autoref{simu_poisson1_f_gamfactory}. For $n=200$, $9\%$ of the fitted models are unstable (highlighted in red), and $1\%$ for $n=800$.

\subsection{Gamma}
Let $y_i | ({\bf x}_i, {\bf z}_i^1, {\bf z}_i^2, {\bf z}_i^3) \sim \text{gamma}(\mu_i, \phi = 9)$ independent with $\mu_i$ denoting the conditional mean of the $i$-th observation and $\phi^{-\frac{1}{2}}$ the common coefficient of variation. The simulated model is given by the relation $\log (\mu_i) = 2 -1.8{x}_i + \sum_j^3 \tilde{f}_j(u_i^j)$ in which $\tilde{f}_j(u^j_i) = f_j(u^j_i) - \sum_i f_j(u^j_i) / n$ for all $j$, with
\begin{align*}
f_1(u^1_i) & = ( 1.8 u^1_i)^3 -\sin(u^1_i) \quad \text{with} \quad u^1_i = (z_{i1}^1 - 1.4z_{i2}^1)/1.72, \\
f_2(u^2_i) & = \exp({u^2_i}) -3(u^2_i)^3 \quad \text{with} \quad u^2_i = (z_{i1}^2 + 1.7z_{i2}^2 - 0.8z_{i3}^2)/2.13, \\
f_3(u^3_i) & = (u^3_i)^2/6 -\cos(\pi u^3_i) \quad \text{with} \quad u^3_i = (z_{i1}^3 + 3.4z_{i2}^3 - 0.5z_{i3}^3 - 1.6z_{i4}^3)/3.92,
\end{align*}
resulting in single-index coefficients ${\boldsymbol \alpha}^1 = (0.58, -0.81)^\T$, ${\boldsymbol \alpha}^2 = (0.47, 0.80, -0.38)^\T$, and ${\boldsymbol \alpha}^3 = (0.26, 0.87, -0.13, -0.41)^\T$. Under this configuration, the dimension of ${\boldsymbol \psi}$ is $36$.

\begin{figure}[!htb]\centering
\includegraphics[width=11cm]{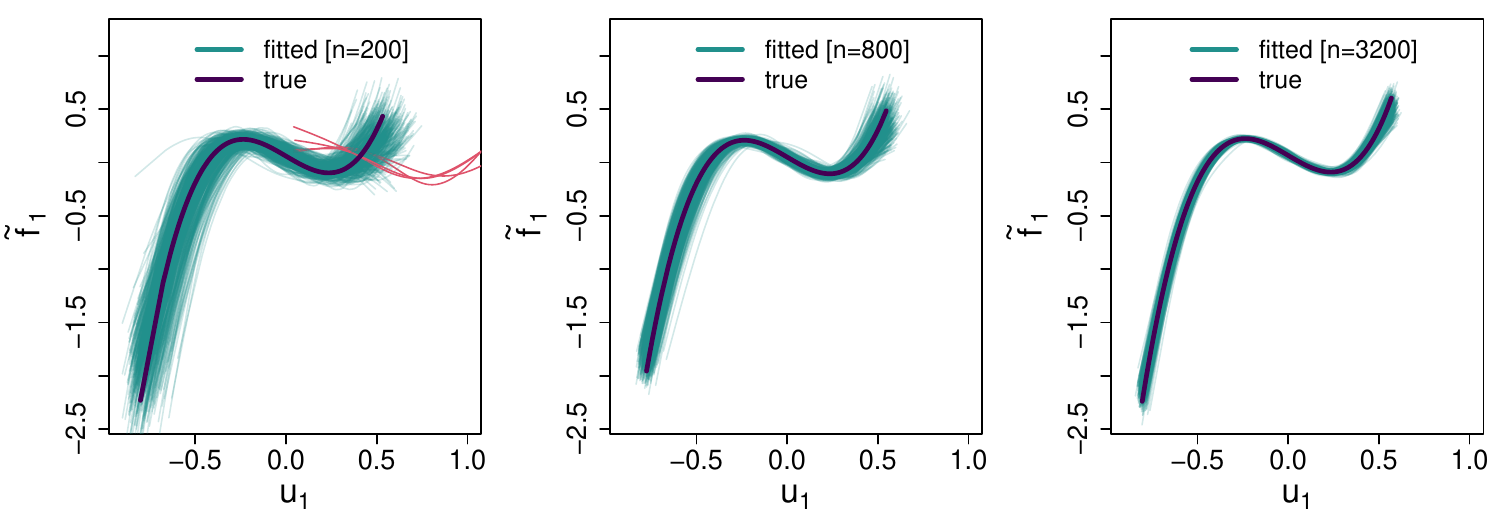}
\includegraphics[width=11cm]{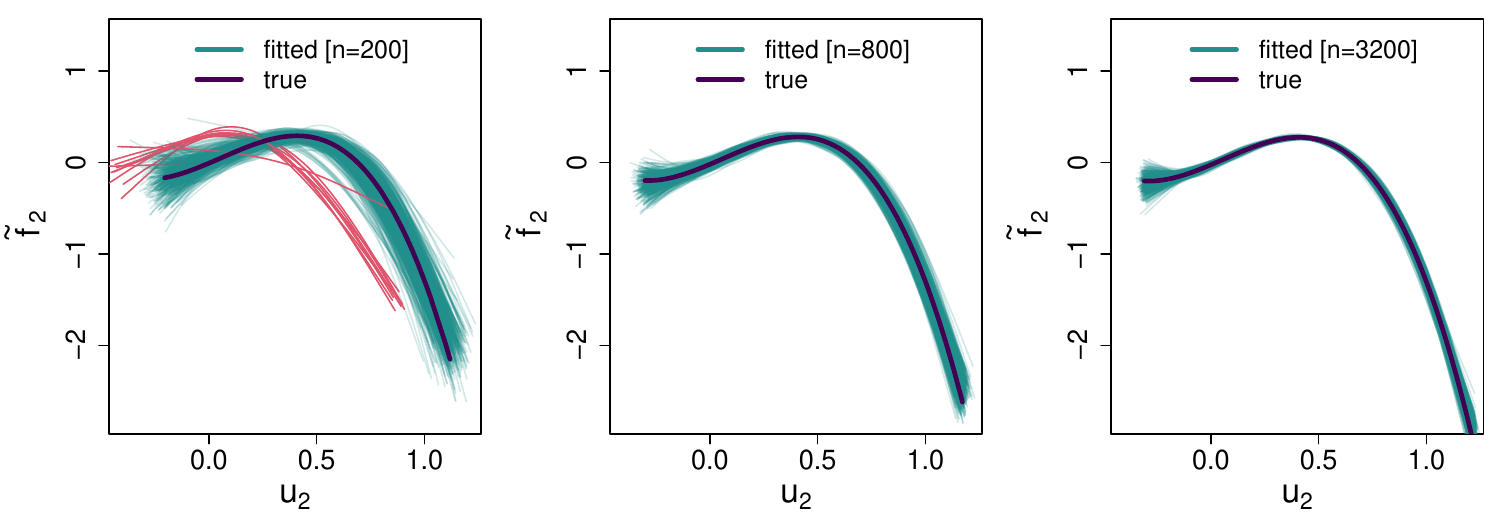}
\includegraphics[width=11cm]{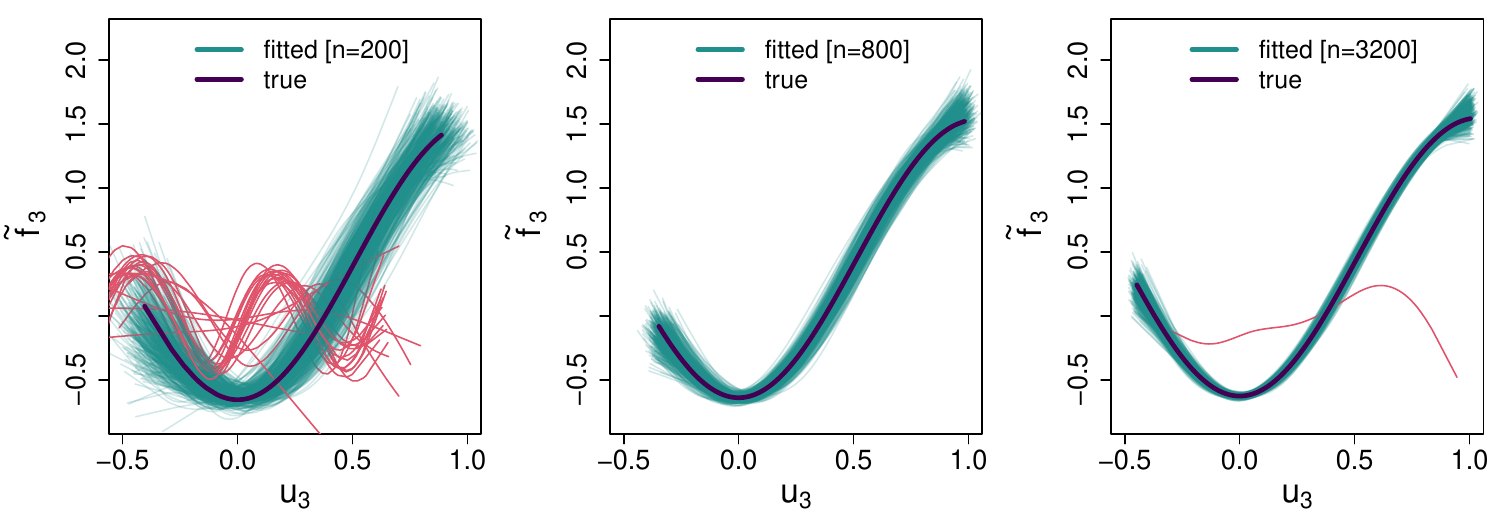}
\caption{\label{simu_gamma1_f_prop} Fitted additive function $\tilde{f}_1$ and single-index covariate ${u}_1$ (top-left, top-middle, top-right), $\tilde{f}_2$ and ${u}_2$ (center-left, center-middle, center-right), $\tilde{f}_3$ and ${u}_3$ (bottom-left, bottom-middle, bottom-right) from all fitted models with the proposed method in the Gamma scenario. The red curves are unstable fitted models.}
\end{figure}
\begin{figure}[!htb]\centering
\includegraphics[width=11cm]{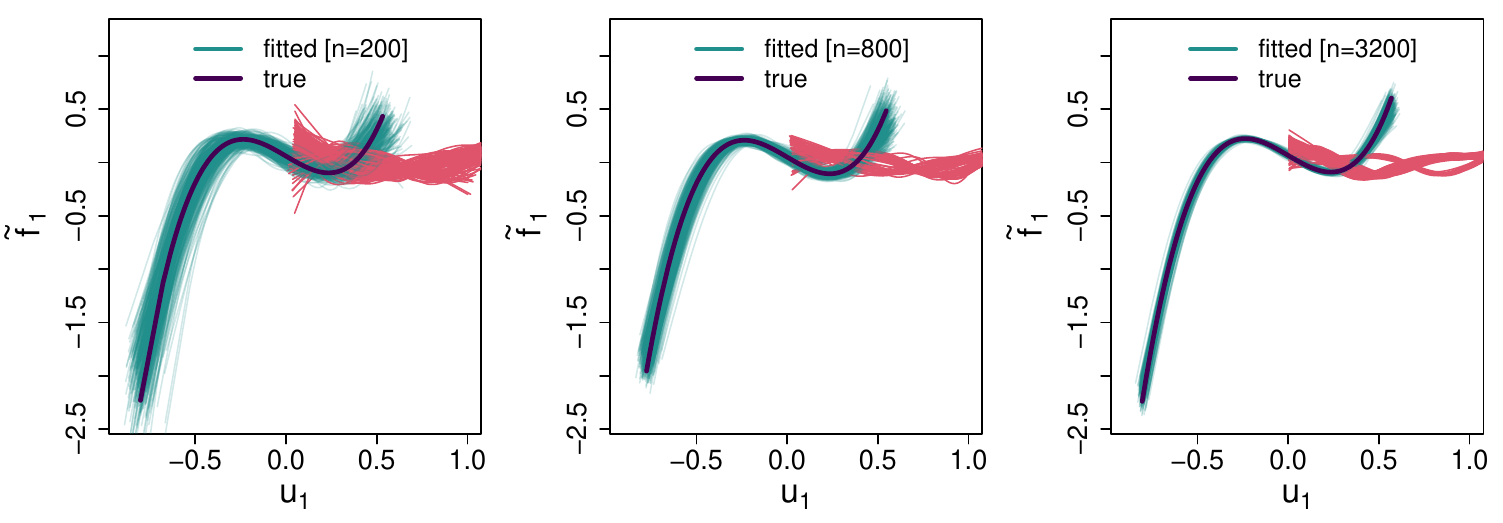}
\includegraphics[width=11cm]{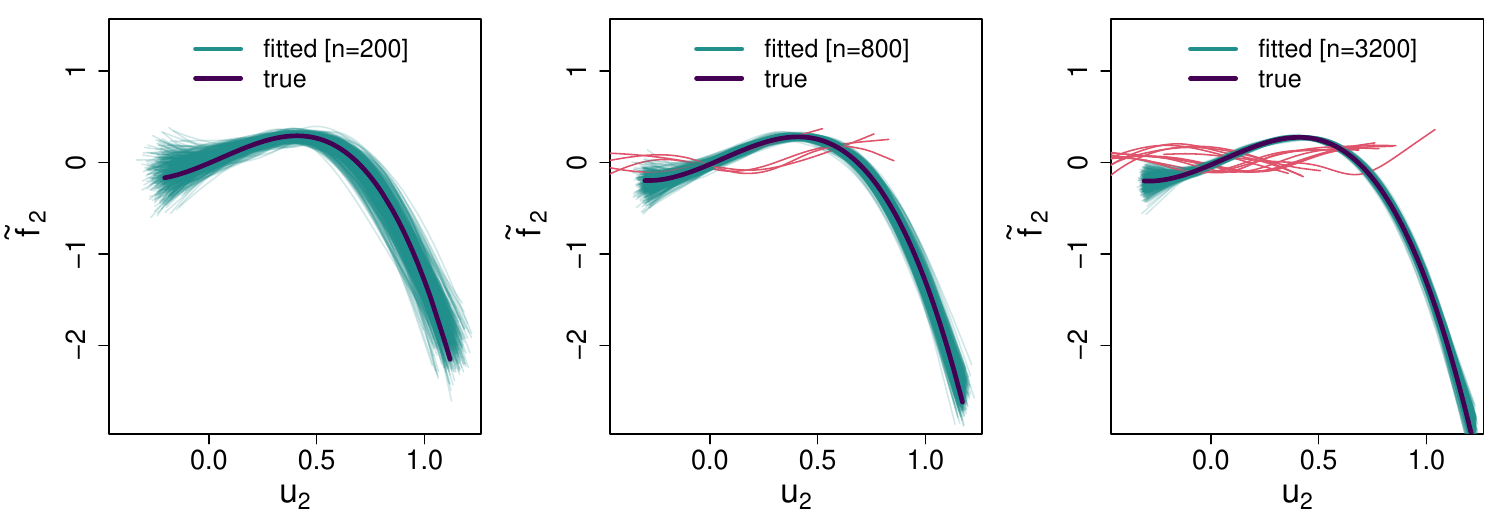}
\includegraphics[width=11cm]{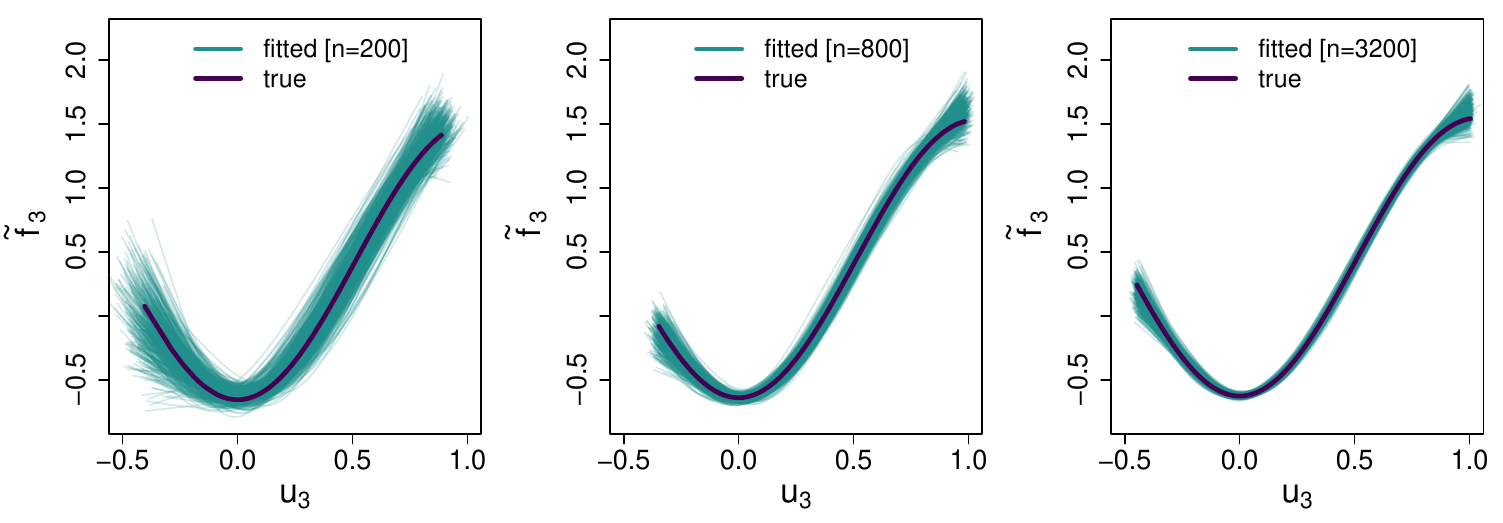}
\caption{\label{simu_gamma1_f_2step} Fitted additive function $\tilde{f}_1$ and single-index covariate ${u}_1$ (top-left, top-middle, top-right), $\tilde{f}_2$ and ${u}_2$ (center-left, center-middle, center-right), $\tilde{f}_3$ and ${u}_3$ (bottom-left, bottom-middle, bottom-right) from all fitted models with the two-step method in the Gamma scenario. The red curves are unstable fitted models.}
\end{figure}

The proposed method takes a total runtime of $20.33$ minutes, with $90\%$ of all fits concluding in less than $1.70$ seconds. The fitted nonlinear functions and single-index covariates are presented in \autoref{simu_gamma1_f_prop}. For $n=200$, only $4.4\%$ of the fitted models are unstable (highlighted in red), and $0.2\%$ for $n=3200$. 

The two-step method takes a total runtime of $1628.95$ minutes, which is $80.13$ times longer than the proposed method due to the need to compute $1412736$ complete GAMs via \texttt{mgcv}. The fitted nonlinear functions and single-index covariates are presented in \autoref{simu_gamma1_f_2step}. For $n=200$, $29.2\%$ of the fitted models are unstable (highlighted in red), $15.2\%$ for $n=800$, and $37\%$ for $n=3200$.

\subsection{Poisson II}
Let $y_i | ({\bf x}_i, {\bf z}_i^1, {\bf z}_i^2) \sim \text{poi}(\mu_i)$ independent with $\mu_i$ denoting the conditional mean of the $i$-th observation. The simulated model is given by the relation $\log (\mu_i) = 2 + 0.7{x}_i + \sum_j^2 \tilde{f}_j(u_i^j)$ in which $\tilde{f}_j(u^j_i) = f_j(u^j_i) - \sum_i f_j(u^j_i) / n$ for all $j$, with
\begin{align*}
f_1(u^1_i) & = ( 1.8 t^1_i)^3 -\sin(t^1_i) \quad \text{with} \quad t^1_i = u^1_i/\sqrt{12} - 0.11, \\
f_2(u^2_i) & = \dfrac{0.2(t^2_i)^{11} \{10(1-t^2_i)\}^6 + 10(10t^2_i)^3 (1-t^2_i)^{10}}{8} \quad \text{with}  \quad t^2_i = \dfrac{u^2_i/\sqrt{12} + 0.57}{1.4},
\end{align*}
in which $u^1_i = (\tilde z_{i1}^1 - 1.4 \tilde z_{i2}^1)/1.72$, and $u^2_i = (\tilde z_{i1}^2 -  \tilde z_{i2}^2 - 0.5 \tilde z_{i3}^2)/1.5$ and $\tilde z_{ij}$ are the centered and scaled transformation of $z_{ij}$, resulting in single-index coefficients ${\boldsymbol \alpha}^1 = (0.58, -0.81)^\T$ and ${\boldsymbol \alpha}^2 = (0.66, -0.66, 0.33)^\T$. Under this configuration, the dimension of ${\boldsymbol \psi}$ is $23$. 

\begin{figure}[!htb]\centering
\includegraphics[width=11cm]{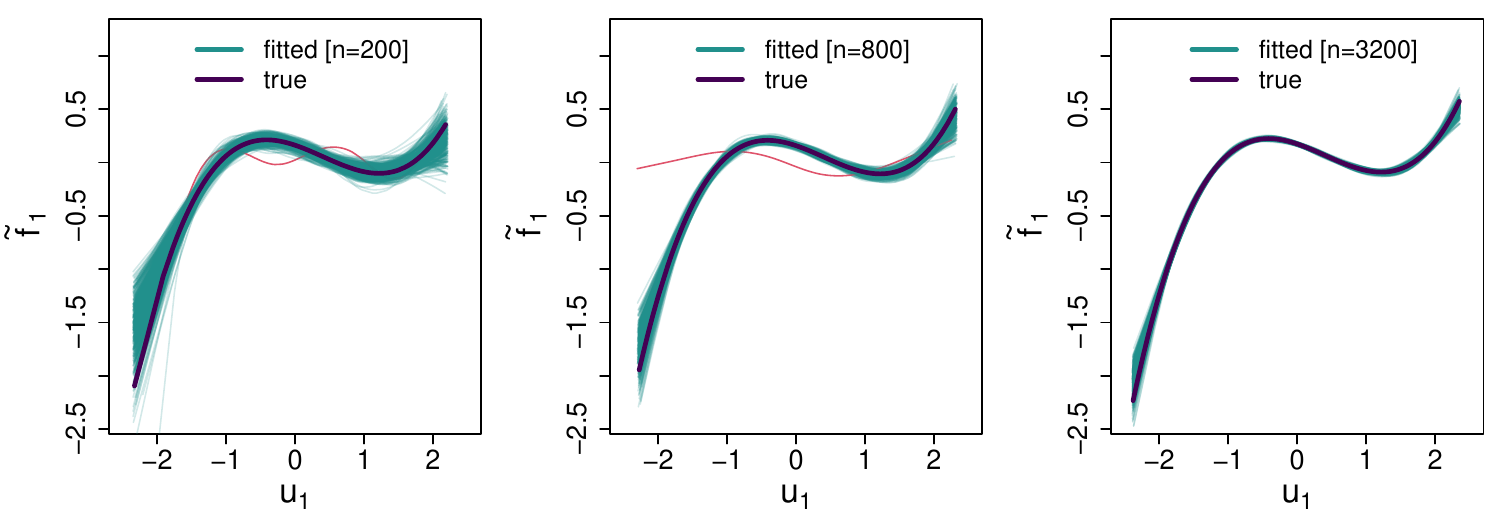}
\includegraphics[width=11cm]{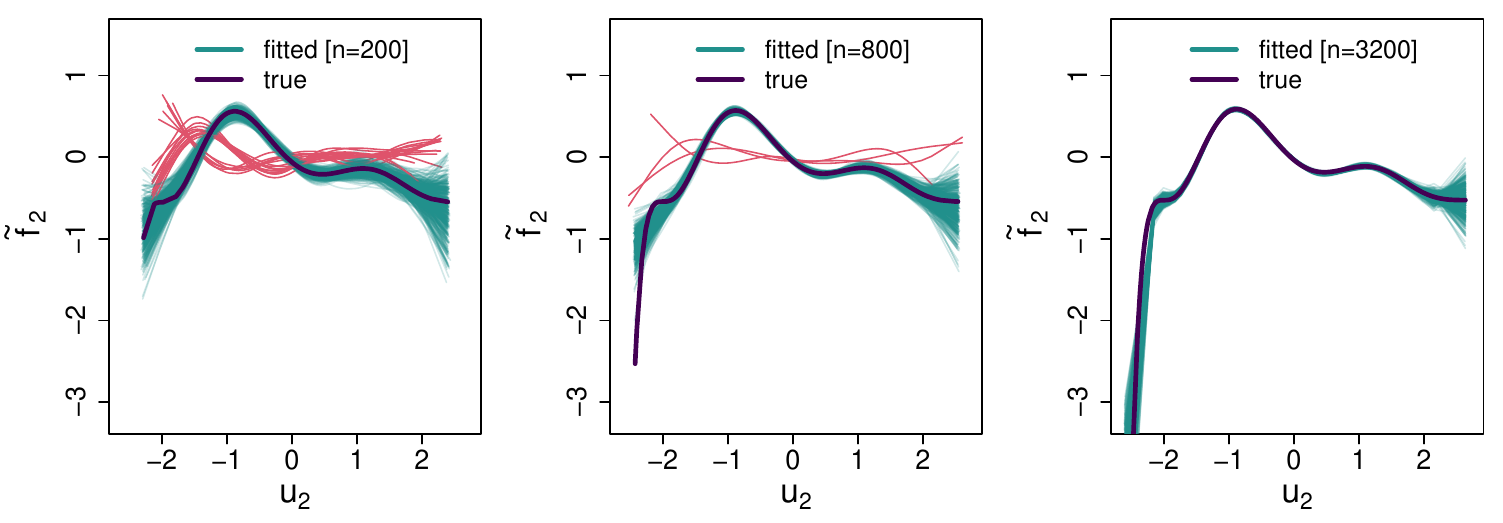}
\caption{\label{simu_poisson2_f_prop} Fitted additive function $\tilde{f}_1$ and single-index covariate ${u}_1$ (top-left, top-middle, top-right), $\tilde{f}_2$ and ${u}_2$ (bottom-left, bottom-middle, bottom-right) from all fitted models with the proposed method in the Poisson II scenario. The red curves are unstable fitted models.}
\end{figure}
\begin{figure}[!htb]\centering
\includegraphics[width=11cm]{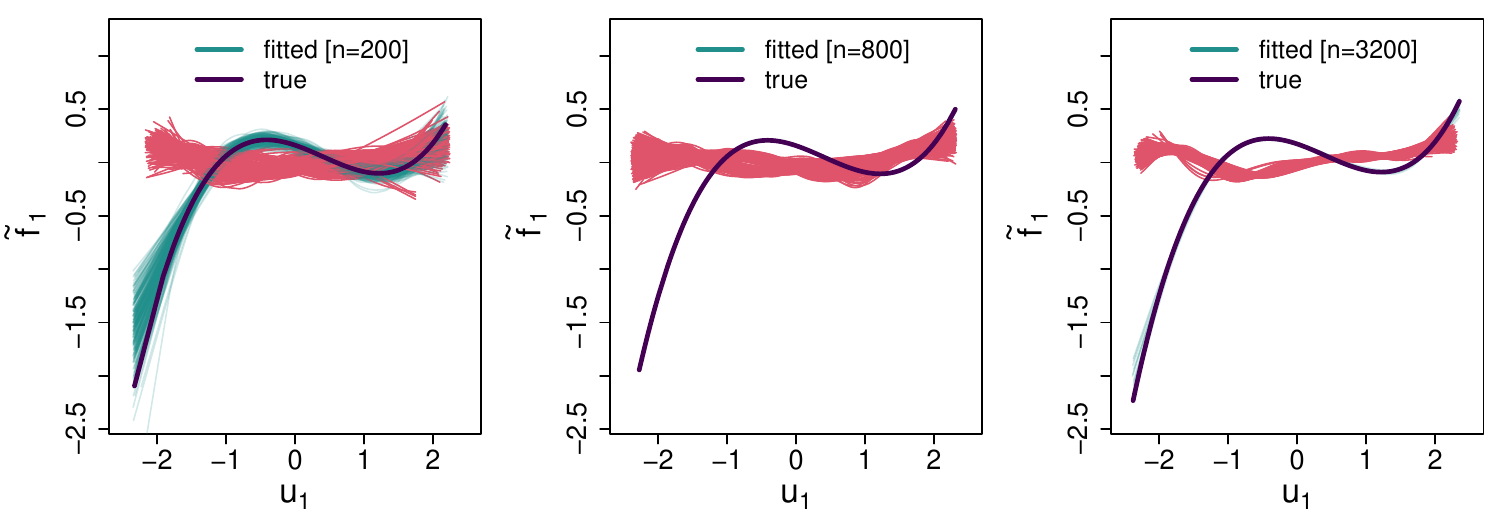}
\includegraphics[width=11cm]{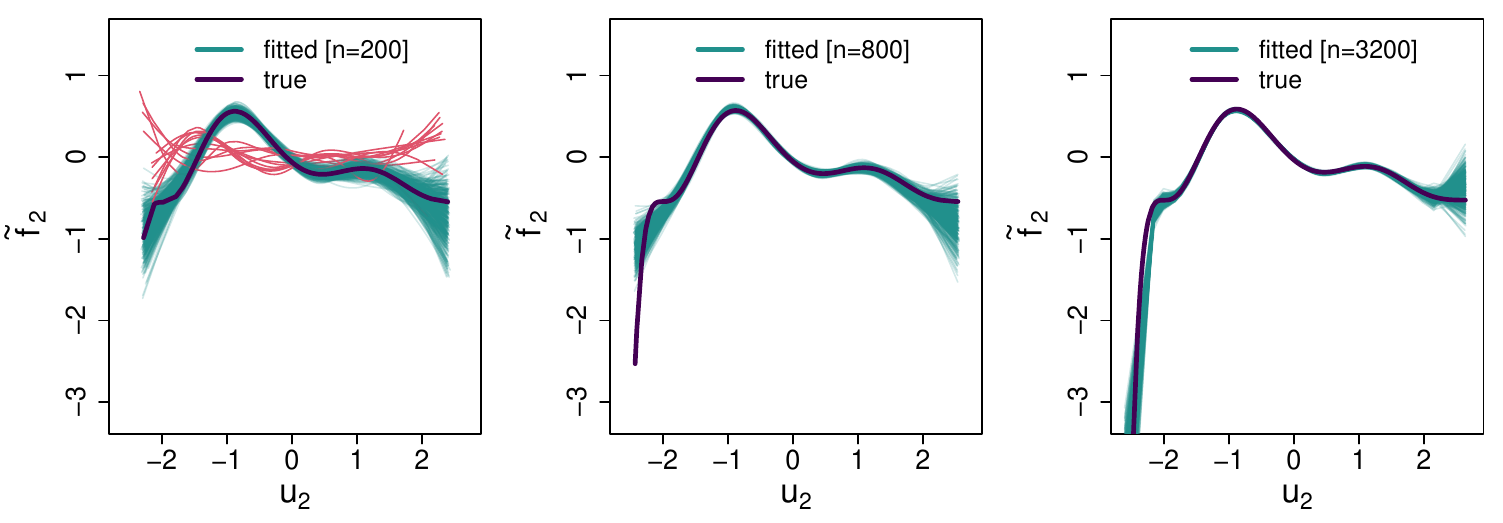}
\caption{\label{simu_poisson2_f_2step} Fitted additive function $\tilde{f}_1$ and single-index covariate ${u}_1$ (top-left, top-middle, top-right), $\tilde{f}_2$ and ${u}_2$ (bottom-left, bottom-middle, bottom-right) from all fitted models with the two-step method in the Poisson II scenario. The red curves are unstable fitted models.}
\end{figure}
\begin{figure}[!htb]\centering
\includegraphics[width=11cm]{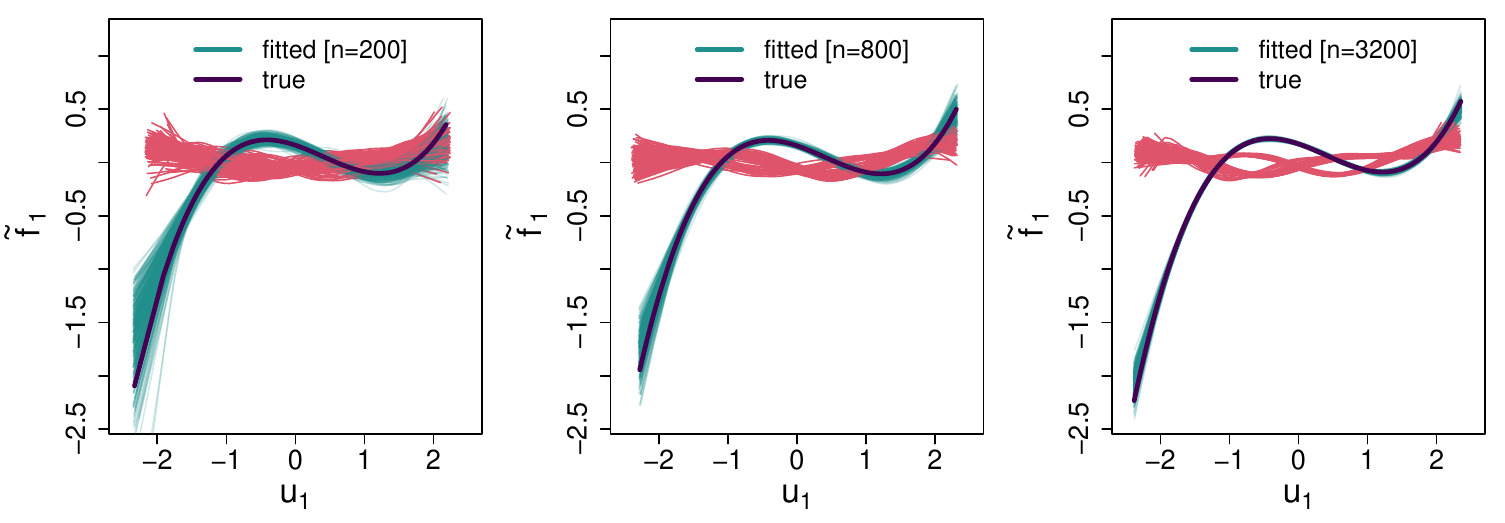}
\includegraphics[width=11cm]{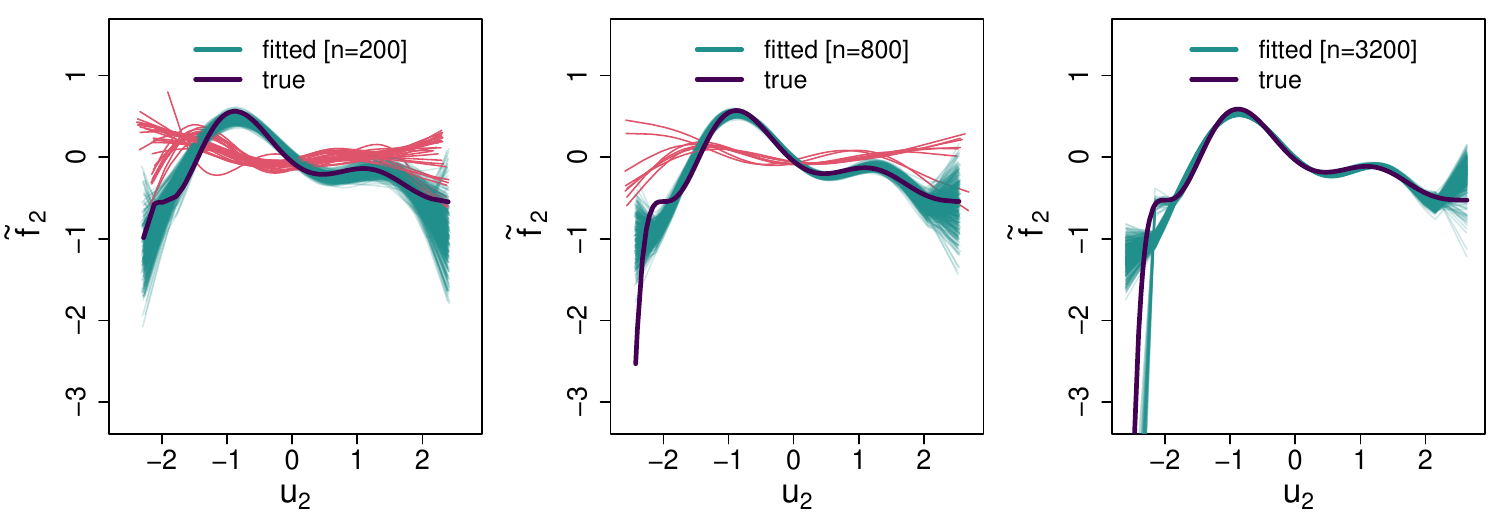}
\caption{\label{simu_poisson2_f_gamfactory} Fitted additive function $\tilde{f}_1$ and single-index covariate ${u}_1$ (top-left, top-middle, top-right), $\tilde{f}_2$ and ${u}_2$ (bottom-left, bottom-middle, bottom-right) from all fitted models with the \texttt{gamFactory} method in the Poisson II scenario. The red curves are unstable fitted models.}
\end{figure}

The proposed method takes a total runtime of $8.80$ minutes, with $90\%$ of all fits concluding in less than $0.64$ seconds. The fitted nonlinear functions and single-index covariates are presented in \autoref{simu_poisson2_f_prop}. For $n=200$, only $3.6\%$ of the fitted models are unstable (highlighted in red), and $0.8\%$ for $n=800$. 

The two-step method takes a total runtime of $251.57$ minutes, which is $28.59$ times longer than the proposed method due to the need to compute $354510$ complete GAMs via \texttt{mgcv}. The fitted nonlinear functions and single-index covariates are presented in \autoref{simu_poisson2_f_2step}. For $n=200$, $55\%$ of the fitted models are unstable (highlighted in red), $100\%$ for $n=800$, and $97.4\%$ for $n=3200$.

The \texttt{gamFactory} method takes a total runtime of $45.15$ minutes, which is $5.13$ times longer than the proposed method. The fitted nonlinear functions and single-index covariates are presented in \autoref{simu_poisson2_f_gamfactory}. For $n=200$, $34\%$ of the fitted models are unstable (highlighted in red), $44.2\%$ for $n=800$, and $41.8\%$ for $n=3200$.

\subsection{Overall performance and stability}
The Poisson I scenario works as expected, demonstrating that all methods are comparable in their ability to recover true single-index coefficients and nonlinear functions with simple conditions. The competitive methods show significant performance degradation in the last two scenarios, confirming the superiority of the proposed method. The Gamma and Poisson II scenarios demonstrate the robustness of the proposed method under non-Gaussian distributions and complex nonlinearity, with all fitted models meeting the convergence criteria without computational issues. The reported runtime includes smoothing optimization and the computation of pointwise confidence bands, ensuring a fair comparison. After removing the few unstable fitted models, the coefficient estimates are generally precise and empirically consistent (see R functions and more details in supplementary \suppref{supp_simu}).

\section{Bike-sharing demand} \label{sec_bike}
Bike-sharing programs have grown in recent years, serving both workers and leisure travelers. This growth is driven by conveniences such as avoiding the need to carry a personal bike (thus reducing theft risk), eliminating maintenance concerns, and seamless integration with public transport. In big cities, biking can be faster than driving, while also providing physical exercise and environmental benefits.

Understanding daily and seasonal demand for bicycles at a bike-sharing station is crucial to avoid under-allocation during high season and peak hours, and over-allocation during low season. The mean total bike rental count can be modeled using environmental and seasonal covariates. To this end, we analyzed real bike-sharing demand in Washington, D.C., USA. The Capital Bike Sharing data \citep{fanaee_2014} includes two years of historical data (2011 and 2012) on hourly total bike rentals, along with corresponding weather information. The dataset has 17379 observations.

The response variable \texttt{hdemand} is binary, indicating high demand when the total count of rental bikes exceeds 150. The normalized covariates are \texttt{hum}: humidity, and \texttt{windspeed}: wind speed. Also, the \texttt{hr}: hour from 0 to 23, \texttt{yday}: day of the year from 1 to 366, \texttt{yr}: 2011 and 2012 years, \texttt{holiday}: holiday indicator, and \texttt{weekday}: day of the week. The other covariates in the dataset are not used because they were already included in the set of covariates under consideration.

There is a nonlinear relationship between the total count of rental bikes and the environmental covariates, with different forms across years, suggesting an interaction. The seasonal covariates do not appear to interact with year. Based on this, the postulate model is $\texttt{hdemand}_i | ({\bf x}_i, \texttt{yday}_i, \texttt{hr}_i, {\bf z}_i) \ {\sim} \ \text{Be}(\mu_i)$ independent with $\mu_i$ denoting the conditional mean of the $i$-th observation, ${\bf x}_i = (\texttt{holiday}_i, \texttt{weekday}_i, \texttt{yr}_i)^\T$, ${\bf z}_i = (\texttt{hum}_i, \texttt{windspeed}_i)^\T$ and the linear predictor $\text{logit} (\mu_i)$ given that
$$
\begin{cases}
{\bf x}_i^\T {\boldsymbol \beta} + \tilde{f}_1(\texttt{yday}_i) + \tilde{f}_2(\texttt{hr}_i) + \tilde{f}_3(\alpha_1^1\texttt{hum}_i + \alpha_2^1 \texttt{windspeed}_i) \ \text{for} \ \texttt{yr}_i = 2011, \\
{\bf x}_i^\T {\boldsymbol \beta} + \tilde{f}_1(\texttt{yday}_i) + \tilde{f}_2(\texttt{hr}_i) + \tilde{f}_4(\alpha_1^2\texttt{hum}_i + \alpha_2^2 \texttt{windspeed}_i) \ \text{for} \ \texttt{yr}_i = 2012.
\end{cases}
$$

To fit this model, minor adaptations of the R functions presented in the simulation are needed. The adaptations include specialization for the Bernoulli distribution, an additional loop for smooth functions of covariates that are not single-index, and an incidence-matrix organization for the single-index interaction effect (see R functions and more details in supplementary \suppref{supp_bike}). We postulate $q_1 = q_2 = 12$ to capture seasonal trends (roughly one knot per month) and daily cycles (one knot per two hours) while avoiding over-parameterization, and set $q_3 = q_4 = 24$ along with $d_j = 4$, $\textit{dif}_j = 2$ for all terms. We only use the simple difference penalties and do not consider modifications for periodic nonlinearity. Under this configuration, the dimension of ${\boldsymbol \psi}$ is $83$. Note that it is inapplicable to perform the competitive methods to fit this model because there is a single-index interaction effect for each year with different coefficients $\tilde{\alpha}_1^1 \neq \tilde{\alpha}_1^2$ (${\alpha}_2^1$ = 0.29 and ${\alpha}_2^2$ = 0.67). The fitted coefficients and their approximate standard errors are presented in \autoref{tab_bike_coef}. All coefficients are highly significant via the asymptotic $z$-tests. The inference aligns with the descriptive analysis, showing that the probability of high demand is lower on holidays and weekends, and in 2011. The form of the environmental index covariate differs between the two years for small index values. Still, it is similar just for large values of the environmental index, highlighting the interaction (bottom-left and bottom-right). The environmental index varies by year, with humidity making the largest contribution (${\alpha}_1^1 = 0.96$ and ${\alpha}_1^2 = 0.75$). The model is fitted in $2.94$ seconds.

\begin{table}[htb]
\centering
\caption{Coefficients estimated and their approximate standard errors from the fitted models with the proposed method for bike-sharing demand.} \label{tab_bike_coef}
\begin{tabular}{lrrr}
\hline
Linear coefficient  & Estimate  & Std. Error & $p$-value\\
\hline
$\beta_1$ (\texttt{intercept})  &  -3.05 & 0.14  & $<$ 0.001\\
$\beta_2$ (\texttt{holiday = yes})  &  -0.89 & 0.15  & $<$ 0.001\\
$\beta_3$ (\texttt{weekday = mon}) &  0.61    &  0.10  & $<$ 0.001\\
$\beta_4$ (\texttt{weekday = tue}) &   0.75 &  0.09  & $<$ 0.001\\
$\beta_5$ (\texttt{weekday = wed})  &   0.70 &  0.09  & $<$ 0.001\\
$\beta_6$ (\texttt{weekday = thu}) &   0.86 &  0.09   & $<$ 0.001\\
$\beta_7$ (\texttt{weekday = fri})  &   1.13 &  0.10   & $<$ 0.001\\
$\beta_8$ (\texttt{weekday = sat}) &   0.41  &  0.09   & $<$ 0.001\\
$\beta_9$ (\texttt{yr = 2012}) &   2.14 &  0.06   & $<$ 0.001\\
$\tilde{\alpha}_1^1$  &   0.31 &  0.06   & $<$ 0.001\\
$\tilde{\alpha}_1^2$  &   0.89 &  0.10  & $<$ 0.001\\
\hline
Smooth term & $\textit{edf}$ & $q$ & $\lambda$ \\
\hline
$\tilde{f}_1$ & 11.604 & 12 & 0.013   \\
$\tilde{f}_2$ & 10.893 & 12 & 0.010  \\
$\tilde{f}_3$ & 6.057 & 24 & 37.191   \\      
$\tilde{f}_4$ & 4.922 & 24 & 45.971   \\     
\hline
\end{tabular}
\end{table}

\begin{figure}[!htb]\centering
\includegraphics[width=13cm]{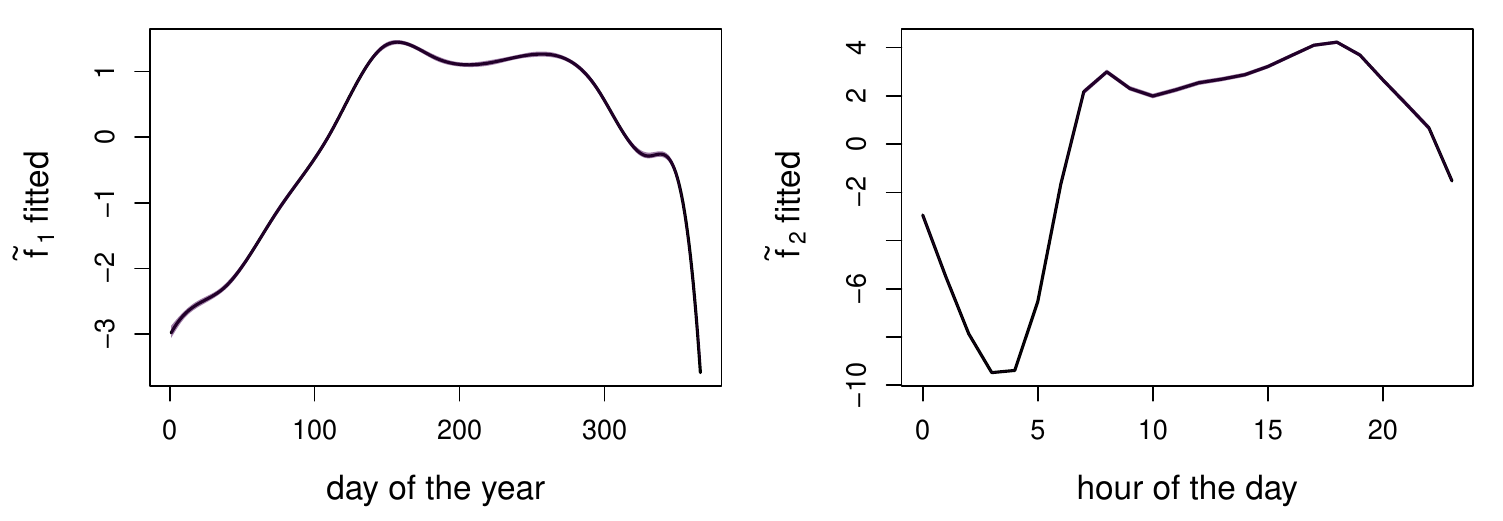}
\includegraphics[width=13cm]{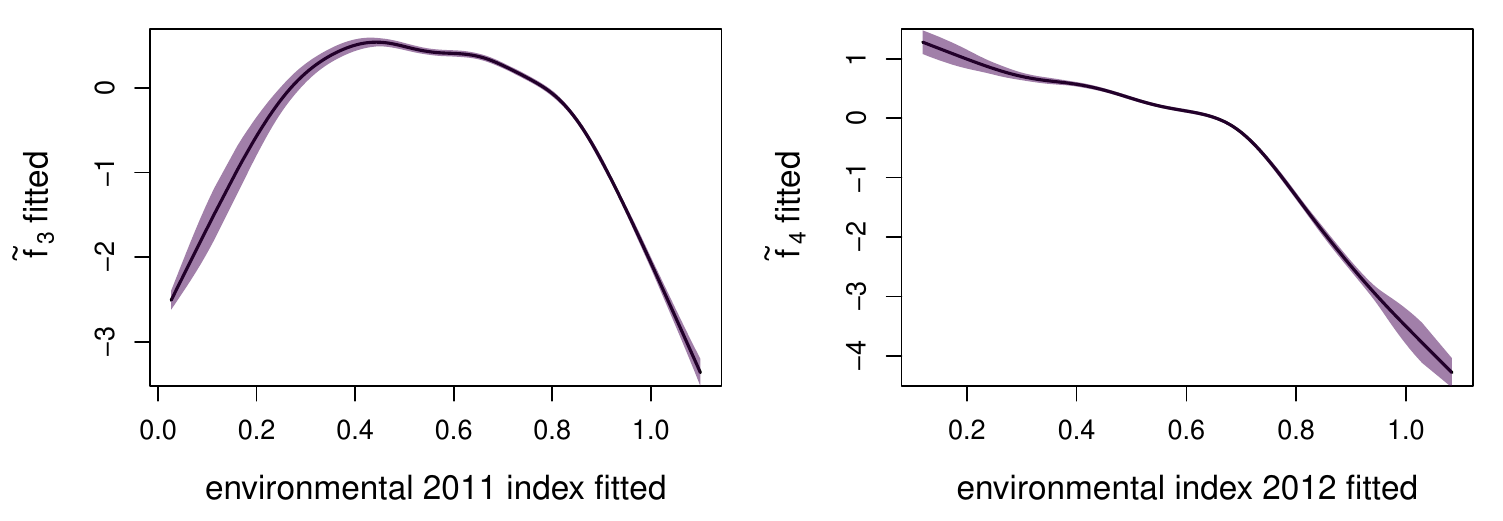}
\caption{\label{bike_f} Fitted additive function $\tilde{f}_1$ for day of the year (top-left), $\tilde{f}_2$ for hour of day (top-right). Fitted additive function $\tilde{f}_3$ and environmental index covariate for 2011 (bottom-left), $\tilde{f}_4$ and environmental index covariate for 2012 (bottom-right), with the 95\% asymptotic pointwise confidence band from the fitted models with the proposed method for bike-sharing demand.}
\end{figure}

\autoref{bike_f} describes the four fitted additive functions and the two fitted environmental index covariates with the 95\% asymptotic pointwise confidence bands (also accounting for the variability in the single-index coefficients). The probability of high demand increases until June and decreases at the end of the year (top-left). Over the first 4 hours of the day, the probability decreases, then increases until around 8:00 AM (top-right). The AUC is $0.9489$, with specificity $0.840$ and sensitivity $0.906$ at a threshold of $0.457$, indicating an excellent fit.
% Comparar predições para dez usando temp2011 e true nos modelos naive, gam e si. Ilustrar maldição da dimensionalidade para predições

\begin{figure}[!htb]\centering
\includegraphics[width=13cm]{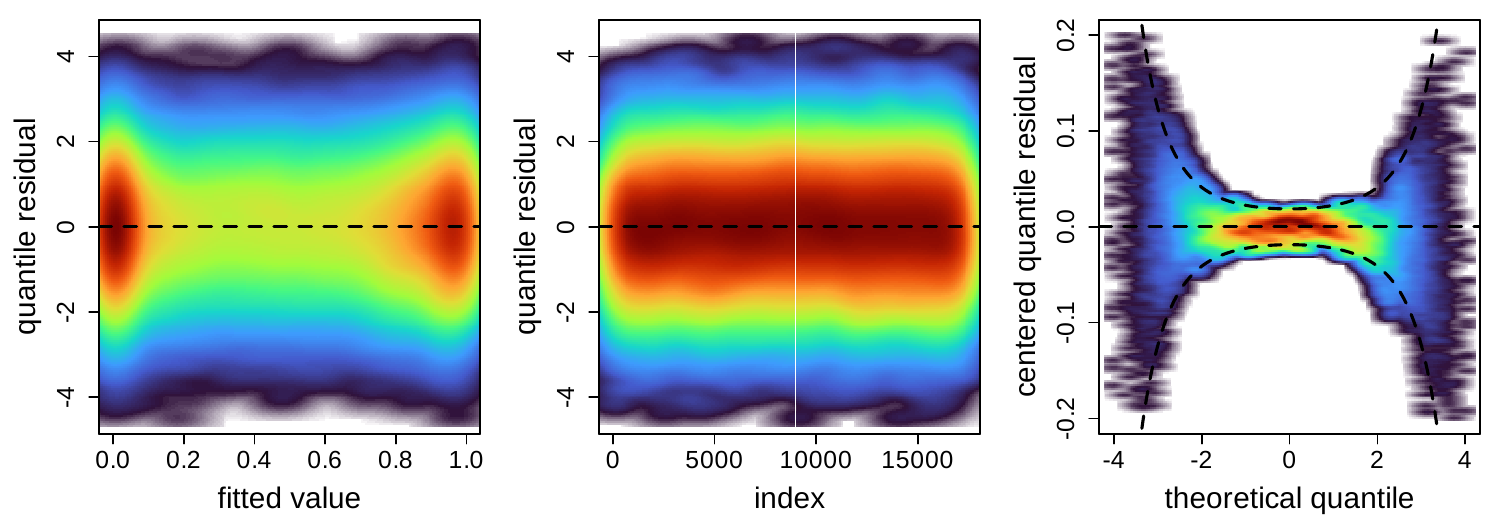}
\caption{\label{bike_res} Scatterplots smoothed for randomized quantile residual versus fitted value (left), observations index (center), and theoretical quantiles of the standard Gaussian distribution (right) from the fitted model with the proposed method for bike-sharing demand.}
\end{figure} 

Additionally, a basic diagnostic tool is needed to assess the adequacy of the postulated model. We base the diagnostic on the quantile residual \citep{dunn_1996} and the corresponding worm plot \citep{buuren_2001}. For discrete responses, the quantile residual is randomized, computed as \texttt{rq = qnorm(runif(n, min=a, max=b))} with \texttt{a = pbinom(y-1, 1, mu)} and \texttt{b = pbinom(y, 1, mu)} in the Bernoulli case. Consequently, the diagnostic plots change each time this residual is computed. To eliminate the need to check multiple plots, we implement a customized diagnostic tool that replicates the randomized residuals 40 times to construct a single smoothed density plot. \autoref{bike_res} presents the residual analysis, which verifies that the postulated model is well-suited to the dataset.

\section{Concluding remarks} \label{sec_conclusion}
This paper contributes a stable method, based on complete Fisher scoring, for fitting GAMs with simultaneous single-index terms using P-splines with dynamically updated boundaries within a unified iterative framework. Deriving the iterative process rather than using optimizers provides control over convergence and produces quantities that, for instance, enable fast computation of the estimated effective degrees of freedom and pointwise confidence bands for single-index terms. 

As empirical results demonstrate, fully joint estimation and preserved smooth basis resolution yield substantial improvements in stability and computational efficiency compared to state-of-the-art competitive methods for moderate sample sizes under non-Gaussian distributions. Regarding the tendency of the \texttt{gamFactory} method to drop the first term in some fitted models, the developers are working to understand and resolve this issue. The modeling flexibility of our method is demonstrated by creating a weather index for each year using meteorological covariates, with the relative importance of each covariate varying by year, offering insights into urban mobility and environmental responsiveness. The implemented smoothed visualisation is indispensable for informative diagnostic analysis when dealing with discrete responses or large datasets.

While the current implementation assumes a constant dispersion parameter, is restricted to the exponential family of distributions, and relies exclusively on P-splines, the proposed methodology lays a solid statistical foundation. Building upon this unified framework, extending the stable method to accommodate dispersion models, more general distributional families, and alternative penalty smoothers represents a natural path for future work.

% The model matrices for single-index effects are defined to implement an estimation method that uses the penalized complete Fisher information matrix and a fast approximation update to optimize smoothing via the generalized Fellner-Schall method in the single-index additive case, recycling the derived matrix decompositions. 
% The model matrix depends continuously on the single-index coefficients. This introduces a profound form of nonlinearity, rendering existing optimal algorithms for GPLAMs that use penalized iterative reweighted least squares (PIRLS)-type updates with locally fixed model matrices unsuitable. 

\section*{Supplementary materials}
The supplementary material includes the following: \suppsection{supp_pscore}{Penalized score}, \suppsection{supp_pfisher}{Penalized Fisher}, \suppsection{supp_simu}{Simulation details} and \suppsection{supp_bike}{Bike-sharing details}.
%\suppsection{supp_df(u)}{Derivative of B-splines}

% \section*{Acknowledgements}
% The authors are grateful to Claudia Collarin, Gustavo Henrique de Araujo Pereira, Michelli Karinne Barros da Silva, and Shuangzhe Liu for their helpful comments and suggestions.
% and all anonymous reviewers

%\section*{Declaration of conflicting interests}
%The authors declared no potential conflicts of interest with respect to the research, authorship and/or publication of this article.

\section*{Funding}
This work was partially supported by Conselho Nacional de Desenvolvimento Científico e Tecnológico (CNPq) - Brazil.

\appendix

\section{B-splines} \label{append_eknots}
Consider $m_j > 2d_j$ knots for the $j$-th term, named $t_1^j < \cdots < t_{m_j}^j$, in which the $d_j$ first and last values are outer knots, and the other $m_j-2d_j$ are internal knots. Crucially, both the outer and internal knots are strictly equally spaced, avoiding multiple (coincident) knots at the boundaries, as required for P-splines with difference penalties \citep{durban_2015}. For example, consider a cubic B-spline ($d_j=4$) with $m_j = 13$ knots for the single-index covariate $u^j = \{{\bf z}^j\}^\T {\boldsymbol \alpha}^j$, then there are $5$ internal knots. In each step of the iterative process the knots $t_1^j < \cdots < t_{13}^j$ are updated such that $t_4^j = \min_i \hat{u}^j_i - \Delta \times \epsilon$, $t_{10}^j =\max_i \hat{u}^j_i + \Delta \times \epsilon$ and the entire sequence $t_{1}^j, \cdots, t_{13}^j$ is equally spaced, where $\Delta=\max_i \hat{u}^j_i - \min_i \hat{u}^j_i$ is the current domain range and $\epsilon=0.001$ is a small numerical tolerance. This boundary extension ensures that all data points fall strictly within the inner intervals, a computational safeguard for numerical stability similar to that adopted in the \texttt{mgcv} package. Because ${\boldsymbol \alpha}^j$ has unitary size, we have that
\[
-||{\bf z}^j|| \leq u^j \leq ||{\bf z}^j||,
\]
for all observations. Using fixed boundaries based on $\min_i ||{\bf z}^j_i||$ and $\max_i ||{\bf z}^j_i||$ would result in a very wide basis domain. Consequently, as the single-index coefficients evolve, the single-index covariates could cover only a small part of their domain, causing most basis functions to be wasted on empty regions. This loss of resolution, combined with the fact that the basis must be recomputed at each step of the iterative process anyway, motivated us to adopt dynamically updated boundaries.

The B-spline basis is computed recursively as
\[
\text{N}_i^{k_j}(u^j) = \frac{u^j - t_i^j}{t_{i+{k_j}}^j - t_i^j}\text{N}_i^{{k_j}-1}(u^j) + \frac{t_{i+{k_j}+1}^j - u^j}{t_{i+{k_j}+1}^j - t_{i+1}^j} \text{N}_{i+1}^{{k_j}-1}(u^j),
\]
for $i = 1, \cdots, q_j+1$, with the recursion basis
\[
\text{N}_i^0(u^j) = \left \{ 
\begin{array}{cc} 1 & t_i^j \le u^j < t_{i+1}^j \\
                  0 & \text{otherwise}.
\end{array} \right.
\]
%with an additional convention to deal with repeated knots from the borders
%\[
%\frac{u^j - t_i^j}{t_{i+{k_j}}^j - t_i^j} = 0 \ \text{if} \ t_i^j = t_{i+{k_j}}^j \quad \text{and} \quad \frac{t_{i+{k_j}+1}^j - u^j}{t_{i+{k_j}+1}^j - t_{i+1}^j} = 0 \ \text{if} \ t_{i+1}^j = t_{i+{k_j}+1}^j,
%\]
%with an additional convention to ensure the partition of unity property
%\[
%    \text{N}_{q_j+1}^{k_j}(u^j) =  1 \ \text{if} \ u^j = t_{m_j}^j.
%\]
In this notation $q_j + 1 = m_j - d_j$, with $d_j=4$ corresponding to a cubic B-spline and ${\boldsymbol \gamma}^j = (\gamma_1^j, \cdots, \gamma_{q_j+1}^j)^\T$ are penalized coefficients to be estimated. To compute the basis, we use the \texttt{splines} R package \citep{splines_1999} by explicitly supplying our uniformly spaced knot sequence. Note that this package cannot be installed from CRAN but is part of base R, with its most recent update from 2019.

\section{Jacobian matrices} \label{append_jacobian}
The Jacobian matrix of ${\boldsymbol \alpha}^j$ with respect to $\tilde{\boldsymbol \alpha}^j$ is given by

\[
{\bf J}^j = \left [\begin{array}{cccc}
\partial \alpha_1^j/\partial \tilde{\alpha}_1^j & \partial \alpha_1^j/\partial \tilde{\alpha}_2^j & \cdots & \partial \alpha_1^j/\partial \tilde{\alpha}_{s_j}^j\\
\partial\alpha_2^j/\partial \tilde{\alpha}_1^j & \partial \alpha_2^j/\partial \tilde{\alpha}_2^j & \cdots & \partial \alpha_2^j/\partial \tilde{\alpha}_{s_j}^j\\
\vdots \\
\partial\alpha_{s_j + 1}^j/\partial \tilde{\alpha}_1^j & \partial \alpha_{s_j + 1}^j/\partial \tilde{\alpha}_2^j & \cdots & \partial \alpha_{s_j + 1}^j/\partial \tilde{\alpha}_{s_j}^j
\end{array} \right], 
\]
with
\begin{align*}
\frac{\partial \alpha_1^j}{\partial \tilde{\alpha}_k^j} % & = \frac{\partial}{\partial \tilde{\alpha}_k^j} \left \{\left [1 + \sum_{i}^{s_j} \{\tilde{\alpha}_i^j\}^2 \right]^{-\frac{1}{2}} \right \} \\
& = -\tilde{\alpha}_k^j \left[1 + \sum_{i}^{s_j}\{\tilde{\alpha}_i^j\}^2 \right ]^{-\frac{3}{2}} \\
% & = \frac{-\tilde{\alpha}_k^j}{\left[1 + \sum_{i}^{s_j}\{\tilde{\alpha}_i^j\}^2 \right ]^{\frac{1}{2}}\left[1 + \sum_{i}^{s_j}\{\tilde{\alpha}_i^j\}^2 \right ]} \\
& = \frac{-{\alpha}_{k+1}^j}{\left[1 + \sum_{i}^{s_j}\{\tilde{\alpha}_i^j\}^2 \right ]}. 
\end{align*}
Also, 
\begin{align*}
\frac{\partial \alpha_{k+1}^j}{\partial \tilde{\alpha}_k^j} % & = \frac{\partial}{\partial \tilde{\alpha}_k^j} \left \{ \tilde{\alpha}_k^j \left [1 + \sum_{i}^{s_j} \{\tilde{\alpha}_i^j\}^2 \right]^{-\frac{1}{2}} \right \} \\
& = -\{\tilde{\alpha}_k^j\}^2 \left[1 + \sum_{i}^{s_j}\{\tilde{\alpha}_i^j\}^2 \right ]^{-\frac{3}{2}} + \alpha_1^j\\
& = \frac{-\tilde{\alpha}_k^j {\alpha}_{k+1}^j} {\left[1 + \sum_{i}^{s_j}\{\tilde{\alpha}_i^j\}^2 \right ]} + \alpha_1^j
\end{align*}
and for $l \neq k + 1 $,
\begin{align*}
\frac{\partial \alpha_l^j}{\partial \tilde{\alpha}_k^j} & = -\tilde{\alpha}_{l-1}^j  \tilde{\alpha}_k^j \left[1 + \sum_{i}^{s_j}\{\tilde{\alpha}_i^j\}^2 \right ]^{-\frac{3}{2}} \\
& = \frac{-\tilde{\alpha}_{l-1}^j {\alpha}_{k+1}^j} {\left[1 + \sum_{i}^{s_j}\{\tilde{\alpha}_i^j\}^2 \right ]}.
\end{align*}
Then,
\[
{\bf J}^j =  \frac{-1}{\left[1 + \sum_{i}^{s_j}\{\tilde{\alpha}_i^j\}^2 \right ]} \left [\begin{array}{cccc}
\alpha_2^j & \alpha_3^j & \cdots & \alpha^j_{s_j + 1} \\
\tilde{\alpha}_{1}^j \alpha_{2}^j & \tilde{\alpha}_{1}^j \alpha_{3}^j & \cdots & \tilde{\alpha}_{1}^j \alpha_{s_j + 1}^j\\
\tilde{\alpha}_{2}^j \alpha_{2}^j & \tilde{\alpha}_{2}^j \alpha_{3}^j & \cdots & \tilde{\alpha}_{2}^j \alpha_{s_j + 1}^j\\
\vdots \\
\tilde{\alpha}_{s_j}^j \alpha_{2}^j & \tilde{\alpha}_{s_j}^j \alpha_{3}^j & \cdots & \tilde{\alpha}_{s_j}^j \alpha_{s_j + 1}^j\\
\end{array} \right] + 
\left [\begin{array}{cccc}
0 & 0 & \cdots & 0\\
\alpha_{1}^j & 0 & \cdots & 0\\
0 & \alpha_{1}^j & \cdots & 0\\
\vdots \\
0 & 0 & \cdots & \alpha_{1}^j 
\end{array} \right] .
\]

\section{B-splines derivative} \label{append_df}
From \citet{deboor_spline}, we have
\begin{align*}
f_j' & = \sum_{i}^{q_j+1} \frac{d}{du^j}\left\{ \text{N}_i^{k_j}(u^j) \right\} \gamma_i^j \\ 
& = \ k_j \sum_{i}^{q_j+1} \left \{ \frac{\text{N}_i^{k_j-1}(u^j)}{t_{i+k_j}^j - t_i^j} - \frac{\text{N}_{i+1}^{k_j-1}(u^j)}{t_{i+k_j+1}^j - t_{i+1}^j} \right \} \gamma_i^j \\
& =  k_j \left \{ \frac{\text{N}_1^{k_j-1}(u^j)}{t_{k_j+1}^j - t_1^j} \gamma_1^j - \frac{\text{N}_{2}^{k_j-1}(u^j)}{t_{k_j+2}^j - t_{2}^j} \gamma_1^j + \frac{\text{N}_2^{k_j-1}(u^j)}{t_{k_j+2}^j - t_2^j} \gamma_2^j - \frac{\text{N}_{3}^{k_j-1}(u^j)}{t_{k_j+3}^j - t_{3}^j} \gamma_2^j + \cdots \right . \\
& \hspace{1.2cm} + \left . \frac{\text{N}_{q_j+1}^{k_j-1}(u^j)}{t_{k_j+q_j+1}^j - t_{q_j+1}^j} \gamma_{q_j}^j - \frac{\text{N}_{q_j+2}^{k_j-1}(u^j)}{t_{k_j+q_j+2}^j - t_{q_j+2}^j} \gamma_{q_j}^j \right \} \\
& = \sum_{i}^{q_j+1} \text{N}_{i+1}^{k_j-1}(u^j)\frac{k_j(\gamma_{i+1}^j-\gamma_{i}^j)}{t_{k_j+i+1}^j - t_{i+1}^j} + k_j \left \{ \frac{\text{N}_1^{k_j-1}(u^j)}{t_{k_j+1}^j - t_1^j} \gamma_1^j  - \frac{\text{N}_{q_j+2}^{k_j-1}(u^j)}{t_{k_j+q_j+2}^j - t_{q_j+2}^j} \gamma_{q_j+1}^j \right \},
\end{align*}
therefore $f_j'$ is a B-spline of degree $k_j-1$ when $\text{N}_1^{k_j-1}(u^j) = \text{N}_{q_j+1}^{k_j-1}(u^j) = 0$. As discussed by \citet[Equation 2.11]{eilers_joys}, this is an elegant formulation for the derivative using B-splines of a lower degree and first-order differences of the coefficients. However, it is more practical to maintain the same coefficients. The \texttt{splines} R package stops at the second step to compute the derivative, preserving the coefficients of the $j$-th original B-spline ${f}_j$ and returning the basis
\begin{align*}
f_j' & = \sum_{i}^{q_j+1} \left \{ \frac{ {\text{N}}_i^{k_j-1}(u^j) k_j}{t_{i+k_j}^j - t_i^j} - \frac{  {\text{N}}_{i+1}^{k_j-1}(u^j) k_j}{t_{i+k_j+1}^j - t_{i+1}^j} \right \} {\gamma}_i^j = \sum_{i}^{q_j+1} \{ {\text{N}}_i^{k_j}(u^j)^{(1)}\} {\gamma}_i^j.
\end{align*}
For the additive functions having zero mean
\begin{align*}
\tilde{f}_j'  & = \frac{d}{du^j} \left\{ {f}_j - \frac{ {\bf 1}^\T {\bf f}_j}{n} \right\} = {f}_j' - \frac{d}{du^j} \left\{ \frac{ {\bf 1}^\T {\bf f}_j}{n} \right\} = {f}_j' - \frac{ {\bf 1}^\T {\bf f}_j'}{n}.
\end{align*}
Then, to obtain the derivative of the reparametrized function, one can simply apply the reparameterization procedure to the derivative basis. The notation is $\tilde{\bf f}_j' = \{ \tilde{\bf N}^j \}^{(1)} \tilde{\boldsymbol \gamma}^j$ in which the rows of matrix $\{ \tilde{\bf N}^j \}^{(1)}$ are $\{ \{\tilde{\bf N}_i^j\}^{(1)} \}^\T$ with $  \{\tilde{\bf N}_i^j\}^{(1)} = [\tilde{\text{N}}_1^{k_j}(u_i^j)^{(1)}, \cdots, \tilde{\text{N}}_{q_j}^{k_j}(u^j_i)^{(1)}]^\T$. 

\section{PGAM-type iterative process} \label{append_pgam}
Note that
$$
{\boldsymbol \psi} = [{\bf M}^\T {\bf W} {\bf M} + \phi^{-1} {\bf P}_\lambda]^{-1} {\bf M}^\T {\bf W} {\bf M} {\boldsymbol \psi} + [{\bf M}^\T {\bf W} {\bf M} + \phi^{-1} {\bf P}_\lambda]^{-1} \phi^{-1} {\bf P}_\lambda {\boldsymbol \psi},
$$
therefore
\begin{align*}
{\boldsymbol \psi} + {\bf K}_{\psi \psi}^{-1} {\bf U}_\psi & = {\boldsymbol \psi} + [{\bf M}^\T {\bf W} {\bf M} + \phi^{-1} {\bf P}_\lambda ]^{-1} \{ {\bf M}^\T {\bf W}^{\frac{1}{2}}{\bf V}^{-\frac{1}{2}}({\bf y} - {\boldsymbol \mu}) - \phi^{-1} {\bf P}_\lambda {\boldsymbol \psi} \} \\
& = [{\bf M}^\T {\bf W} {\bf M} + \phi^{-1} {\bf P}_\lambda ]^{-1} \{ {\bf M}^\T {\bf W} {\bf M} {\boldsymbol \psi} + {\bf M}^\T {\bf W}^{\frac{1}{2}}{\bf V}^{-\frac{1}{2}}({\bf y} - {\boldsymbol \mu}) \} \\
& = [{\bf M}^\T {\bf W} {\bf M} + \phi^{-1} {\bf P}_\lambda]^{-1} {\bf M}^\T {\bf W} \{  {\bf M} {\boldsymbol \psi} + {\bf W}^{-\frac{1}{2}}{\bf V}^{-\frac{1}{2}}({\bf y} - {\boldsymbol \mu}) \}.
\end{align*}
Then, the PGAM-type procedure is given by
\begin{align*}
{\boldsymbol \psi}^{(t+1)} & = {\boldsymbol \psi}^{(t)} + \{ {\bf K}_{\psi \psi}^{(t)} \}^{-1} {\bf U}_\psi^{(t)} \\
& = [\{{\bf M}^{(t)}\}^\T {\bf W}^{(t)} {\bf M}^{(t)} + \phi^{-1} {\bf P}_\lambda]^{-1} \{{\bf M}^{(t)}\}^\T {\bf W}^{(t)} \tilde{\bf y}^{(t)}, 
\end{align*}
with $\tilde{\bf y} = {\bf M}{\boldsymbol \psi} + {\bf W}^{-\frac{1}{2}}{\bf V}^{-\frac{1}{2}}({\bf y} - {\boldsymbol \mu})$.

\section{Smoothing update} \label{append_smooth}
The penalized log-likelihood function \eqref{eq_loglike} under a Bayesian viewpoint \citep[Appendix B]{wood_2025} its recovered, for $\phi$ fixed,  using the prior $\pi({\boldsymbol \psi})$ such that ${\boldsymbol \psi} \stackrel{\rm } {\sim} \text{N}({\bf 0}, {\bf P}_\lambda^{-})$, in which ${\bf P}_\lambda^{-}$ is the Moore-Penrose pseudo-inverse of ${\bf P}_\lambda$. Denoting the Gaussian approximation of the posterior by $\pi_G({\boldsymbol \psi} | {\bf y})$ such that ${\boldsymbol \psi} | {\bf y} \stackrel{\rm } {\sim} \text{N}(\hat{\boldsymbol \psi}, {\bf K}_{\psi \psi}^{-1})$ then Laplace approximation of the marginal log-likelihood is
\begin{align*}
\text{L}_p({\boldsymbol \lambda}) & = \text{L}({\boldsymbol \psi}) + \pi({\boldsymbol \psi}) - \pi_G({\boldsymbol \psi}|{\bf y}) \\
& = \text{L}({\boldsymbol \psi}) - \frac{1}{2}\log{|{\bf P}_\lambda^{-}|} - \frac{1}{2} {\boldsymbol \psi} ^\T {\bf P}_{\lambda} {\boldsymbol \psi} + \frac{1}{2}\log{|{\bf K}_{\psi \psi}^{-1}|} + \text{const},
\end{align*}
then 
\begin{align*}
\frac{\partial \text{L}_p({\boldsymbol \lambda})}{\partial \lambda_1} & = \frac{\partial}{\partial \lambda_1} \left \{ - \frac{1}{2}\log{|{\bf P}_\lambda^{-}|} - \frac{1}{2} {\boldsymbol \psi} ^\T {\bf P}_{\lambda} {\boldsymbol \psi} + \frac{1}{2}\log{|{\bf K}_{\psi \psi}^{-1}|} + \text{const} \right \} \\
& =  \frac{\partial}{\partial \lambda_1} \left \{ \frac{1}{2}\log{|{\bf P}_\lambda|} - \frac{1}{2} {\boldsymbol \psi} ^\T {\bf P}_{\lambda} {\boldsymbol \psi} - \frac{1}{2}\log{|{\bf K}_{\psi \psi}|} + \text{const} \right \} \\
& = \frac{1}{2} \text{tr} \left\{ {\bf P}_\lambda^{-}  \frac{\partial}{\partial \lambda_1} \{ {\bf P}_\lambda \} \right\} - \frac{1}{2} {\boldsymbol \psi} ^\T {\bf P}^1 {\boldsymbol \psi} - \frac{1}{2} \text{tr} \left\{ {\bf K}_{\psi \psi}^{-1}  \frac{\partial}{\partial \lambda_1} \{ {\bf K}_{\psi \psi} \} \right\} \\
& = \frac{1}{2} \text{tr} \left\{ {\bf P}_\lambda^{-} {\bf P}^1 \right\} - \frac{1}{2} {\boldsymbol \psi} ^\T {\bf P}^1 {\boldsymbol \psi} - \frac{1}{2} \text{tr} \left\{ {\bf K}_{\psi \psi}^{-1} \frac{\partial}{\partial \lambda_1} \{ {\bf M}^\T {\bf W} {\bf M} + \phi^{-1} {\bf P}_\lambda \} \right\} \\
& = \frac{1}{2} \text{tr} \left\{ {\bf P}_\lambda^{-} {\bf P}^1 \right\} - \frac{1}{2} {\boldsymbol \psi} ^\T {\bf P}^1 {\boldsymbol \psi} - \frac{1}{2} \text{tr} \left\{ [ \phi {\bf M}^\T {\bf W} {\bf M} + {\bf P}_\lambda ]^{-1}   {\bf P}^1 \right\}.
% & = \frac{1}{2} \text{tr} \left\{ {\bf P}_\lambda^{-} {\bf P}^1 \right\} - \frac{1}{2} {\boldsymbol \psi} ^\T {\bf P}^1 {\boldsymbol \psi} - \frac{1}{2} \text{tr} \left\{ [ \phi {\bf M}^\T {\bf W} {\bf M} + {\bf P}_\lambda ]^{-1} {\bf P}^1 \right\}.
\end{align*}
The first theorem by \citet{wood_2017} results that $\text{tr} \left\{ {\bf P}_\lambda^{-} {\bf P}^1 \right\} - \text{tr} \left\{ [ \phi {\bf M}^\T {\bf W} {\bf M} + {\bf P}_\lambda ]^{-1}   {\bf P}^1 \right\} > 0$. Then, an efficient and simple update for smoothing optimization via the generalized Fellner-Schall method for the additive single-index case is given by
\begin{align*}
\lambda_1^{(t+1)} & = \frac{\text{tr} \left\{ {\bf P}_\lambda^{-} {\bf P}^1 \right\} - \text{tr} \left\{ [ \phi \{{\bf M}^{(t)}\}^\T {\bf W}^{(t)} {\bf M}^{(t)} + {\bf P}_\lambda ]^{-1}  {\bf P}^1 \right\}}{ \{ {\boldsymbol \psi}^{(t+1)} \} ^\T {\bf P}^1 {\boldsymbol \psi}^{(t+1)} } \lambda_1^{(t)} \\
%& = \frac{\text{tr} \left\{ {\bf P}_\lambda^{-} {\bf P}^1 \right\} - \phi^{-1} \text{tr} \left\{ [ \{{\bf M}^{(t)}\}^\T {\bf W}^{(t)} {\bf M}^{(t)} + \phi^{-1} {\bf P}_\lambda ]^{-1}   {\bf P}^1 \right\}}{ \{ {\boldsymbol \psi}^{(t)} \} ^\T {\bf P}^1 {\boldsymbol \psi}^{(t)} } \lambda_1^{(t)} \\
& = \frac{\text{tr} \left\{ {\bf P}_\lambda^{-} {\bf P}^1 \right\} - \phi^{-1} \text{tr} \left\{ \text{cp} \{ \{{\bf L}^{(t)}\}^{-\T} \}  {\bf P}^1 \right\}}{ \{ {\boldsymbol \psi}^{(t+1)} \} ^\T {\bf P}^1 {\boldsymbol \psi}^{(t+1)} } \lambda_1^{(t)},
\end{align*}
denoting ${\bf Q} = {\bf P}_\lambda^{-} - \phi^{-1} \text{cp} \{ {\bf L}^{-\T} \}$, with $\text{cp}$ being the cross-product matrix, we have
\begin{align*}
\lambda_j^{(t+1)} & = \frac{\text{tr} \left\{ {\bf Q}^{(t)}  {\bf P}^j \right\} }{ \{ \tilde{\boldsymbol \gamma}^{j(t+1)}  \} ^\T \tilde{\bf P}^j \tilde{\boldsymbol \gamma}^{j(t+1)} } \lambda_j^{(t)}.
\end{align*}

\section{Algorithm implementation} \label{append_implem}
We propose the following implementation that integrates the smoothing parameter updates for direct estimation of all model coefficients:
\begin{enumerate}
\item[(a)] Get initial model:
\begin{enumerate}[leftmargin=1cm] 
\item[(a.1)] Specify the dimension of the P-spline basis $q_j$, the order of each B-spline basis $d_j$ and the order of difference $\textit{dif}_j$.
\item[(a.2)] Obtain ${\boldsymbol \beta}$ from \texttt{gam(y$\sim$X-1, family=EF("link"))}.
\end{enumerate}
In an m-loop: 
\begin{enumerate}[leftmargin=1cm]
\item[(a.3)] Obtain $\tilde{\boldsymbol \alpha}^j$ from \texttt{runif(s\_j, -1, 1)} ensuring that $\max \{ \alpha_1^j, \cdots, \alpha_{s_j+1}^j \} < 0.8$ and ${\alpha}_1^j > 0.2$ then compute the single-index covariate ${\bf u}^j = {\bf Z}^j {\boldsymbol \alpha}^j$.
\item[(a.4)] Define the $t_1^j < \cdots < t_{m_j}^j$ (as \autoref{append_eknots}) for computing the $d_j$-order basis matrices $\tilde{\bf N}^j$ and $\{ \tilde{\bf N}^j \}^{(1)}$. 
\item[(a.5)] Obtain $\tilde{\boldsymbol \gamma}^j$ from \texttt{gam(y$\sim$offset(off)+N\_til\_j-1, family=EF("link"))} in which $\texttt{off}={\bf X} {\boldsymbol \beta}$ then compute $\tilde{\bf f}_j = \{ \tilde{\bf N}^j \}  \tilde{\boldsymbol \gamma}^j$ and $\tilde{\bf f}_j' = \{ \tilde{\bf N}^j \}^{(1)}  \tilde{\boldsymbol \gamma}^j$ (as \autoref{append_df}).
\item[(a.6)] Compute the Jacobian matrix ${\bf J}^j$ (as \autoref{append_jacobian}) to compute the single-index term model matrix $\tilde{\bf T}^j = \{\tilde{\bf F}^j\}^{(1)} {\bf Z}^j {\bf J}^j$.
\item[(a.7)] Compute the matrix $\tilde{\bf P}^j = \{\tilde{\bf D}_{\textit{dif}_j}^j\}^\T \tilde{\bf D}_{\textit{dif}_j}^j$.
\end{enumerate}
At the end of the m-loop: 
\begin{enumerate}[leftmargin=1cm]
\item[(a.8)] Obtain ${\boldsymbol \lambda}$ from \texttt{runif(m, 1, 1000)} and $\phi$ from \texttt{runif(1, 1, 100)}.
\item[(a.9)] Compute each penalization ${\bf P}^j$ in \eqref{eq_penalty_matrix} through $\tilde{\bf P}^1, \cdots, \tilde{\bf P}^m$ and $s_1, \cdots, s_m$ to compute the model penalization ${\bf P}_{\lambda} = \sum_j \lambda_j {\bf P}^j$.
\item[(a.10)] Obtain ${\boldsymbol \mu} = g^{-1}\{\texttt{off} + \sum_j \tilde{\bf f}_j \}$ then, with weights $\omega_i = \{g'(\mu_i)^{2}\text{V}_i\}^{-1}$, compute the weighted model matrix $\tilde{\bf M} = \{ {\bf W} \}^{\frac{1}{2}} [{\bf X}, \tilde{\bf N}^1, \tilde{\bf T}^1, \cdots, \tilde{\bf N}^m, \tilde{\bf T}^m]$.
\end{enumerate}
\item[(b)] Get ${\boldsymbol \psi}$ update: 
\begin{enumerate}[leftmargin=1cm]
\item[(b.1)] Obtain the Fisher decomposition ${\bf L}$ from \texttt{chol(crossprod(M\_til) + P/phi + diag(1e-7)} and compute the dependent modified response $\tilde{\bf y} = {\bf M}{\boldsymbol \psi} + {\bf W}^{-\frac{1}{2}}{\bf V}^{-\frac{1}{2}}({\bf y} - {\boldsymbol \mu})$.
\item[(b.2)] Perform the iterative process \eqref{eq_chol} updating ${{\boldsymbol \psi}}$ then, recomputing the knots, update ${\boldsymbol \mu}$ and $\tilde{\bf M}$ (as initial m-loop).
\end{enumerate}
\item[(c)] Get ${\boldsymbol \lambda}$ update: 
\begin{enumerate}[leftmargin=1cm]
\item[(c.1)] Obtain the variance decomposition ${\bf B}$ from \texttt{forwardsolve(t(L), diag(ncol(L)))} to compute the matrix ${\bf Q} = {\bf P}_\lambda^{-} - \phi^{-1} \text{cp} \{ {\bf B} \}$.
\item[(c.2)] Perform the smoothing optimization \eqref{eq_lambda_up} updating ${{\boldsymbol \lambda}}$ then update ${\bf P}_{\lambda}$.
\end{enumerate}
\item[(d)] Get $\phi$ update: 
\begin{enumerate}[leftmargin=1cm]
\item[(d.1)] Compute the $\textit{edf} = \text{tr} \left \{ \text{cp} \{ {\bf B} \} \text{cp} \{ \tilde{\bf M} \} \right \}$ updating $\phi$ consistently by $(n - \textit{edf}) / \sum_i (y_i-\mu_i)^2 \text{V}_i^{-1}$ (only when necessary in EF).
\end{enumerate}
\item[(e)] Convergence check: 
\begin{enumerate}[leftmargin=1cm]
\item[(e.1)] Back to step (a.3) if: $\alpha_1^{max} = \max \{\alpha_1^1, \alpha_1^2, \cdots, \alpha_1^m\} < 0.05$ or $\lambda^{min} = \min \{\lambda_1, \cdots, \lambda_m\} < 0$ or $\texttt{met} > 10^{6}$ or model iteration number $\geq 80$.
\end{enumerate}
\item[(f)] Stop criterion check:
\begin{enumerate}[leftmargin=1cm]
\item[(f.1)] Obtain the penalized log-likelihood $\text{L}_p^{(t+1)}$, then compute the iteration metric $\texttt{met} = |\text{L}_p^{(t+1)} - \text{L}_p^{(t)}|/(|\text{L}_p^{(t)}| + 10^{-4})$. 
\item[(f.2)] Save the current model if: \texttt{met} is the smallest metric ever obtained. Stop the procedure if: $\texttt{met} < 10^{-6}$ or total iteration number $\geq 500$. Repeat steps (b)-(f) until the stop condition is met, then return the saved model.
\end{enumerate}
\end{enumerate}
As \citet{collarin_2025b} highlight, dynamically updating the boundaries of the single-index covariates may compromise the differentiability of the penalized log-likelihood function, leading to numerical instability by inducing discontinuity. While they avoid this issue by fixing the boundaries, we maintain dynamic updates and employ the complete penalized Fisher information to guide the joint maximization of all coefficients. Although this formulation significantly improves numerical stability, it cannot completely prevent the algorithm from evaluating non-differentiable regions. Therefore, step (e.1) serves as a numerical safeguard, restarting the algorithm whenever potential irregularities are encountered. The restrictions in step (a.3) prevent problematic initial values. Furthermore, recycling ${\bf L}$ from (b.1) in (c.1) and ${\bf B}$ from (c.1) in (d.1) resembles efficient backfitting and substantially reduces computational cost. The ${\boldsymbol \lambda}$ and $\phi$ starting values have wide ranges to avoid introducing an artificial bias that favors the proposed method, while the other fixed values in some steps come from various stress tests.
%where the Fisher information itself is undefined.
%that may severely limit basis resolution
%A comparative empirical analysis is presented in \autoref{sec_simulation}.
%The correction in (f.1) is to avoid division by zero

\section{Single-index variance} \label{append_vcov_f} 
The $\tilde f_1(u_i^1)$ estimated is $[\tilde{\text{N}}_1^{k_1}(\hat u_i^1), \cdots, \tilde{\text{N}}_{q_1}^{k_1}(\hat u_i^1)] \hat{\tilde{\boldsymbol \gamma}}^1 = h_i(\hat{\tilde{\boldsymbol \gamma}}^1, \hat{\tilde{\boldsymbol \alpha}}^1)$, then asymptotically
\begin{align*}
\text{var} \left\{ h_i(\hat{\tilde{\boldsymbol \gamma}}^1, \hat{\tilde{\boldsymbol \alpha}}^1) \right\} & = 
\left [\begin{array}{c}
\dfrac{\partial h_i(\tilde{\boldsymbol \gamma}^1, \tilde{\boldsymbol \alpha}^1)}{\partial \tilde{\boldsymbol \gamma}^1} \\ 
\dfrac{\partial h_i(\tilde{\boldsymbol \gamma}^1, \tilde{\boldsymbol \alpha}^1)}{\partial \tilde{\boldsymbol \alpha}^1} 
\end{array} \right] ^\T
\text{var} \left [\begin{array}{c}
\hat{\tilde{\boldsymbol \gamma}}^1 \\ 
\hat{\tilde{\boldsymbol \alpha}}^1 
\end{array} \right]
\left [\begin{array}{c}
\dfrac{\partial h_i(\tilde{\boldsymbol \gamma}^1, \tilde{\boldsymbol \alpha}^1)}{\partial \tilde{\boldsymbol \gamma}^1} \\ 
\dfrac{\partial h_i(\tilde{\boldsymbol \gamma}^1, \tilde{\boldsymbol \alpha}^1)}{\partial \tilde{\boldsymbol \alpha}^1} 
\end{array} \right] \\
%& = \left[ \tilde{\text{N}}_1^{k_j}(u_i^j) \quad \cdots \quad \tilde{\text{N}}_{q_j}^{k_j}(u_i^j) \quad \tilde{f}'_j(u_i^j) \left\{ \sum_k^{s_j+1} \frac{d \alpha^j_k}{d \tilde{\alpha}_1^j} z_{ik}^j \right\}  \quad \cdots \quad \tilde{f}'_j(u_i^j) \left\{ \sum_k^{s_j+1} \frac{d \alpha^j_k}{d \tilde{\alpha}_m^j} z_{ik}^j \right\} \right] 
%var \left [\begin{array}{c}
%\hat{\tilde{\boldsymbol \gamma}}^j \\ 
%\hat{\tilde{\boldsymbol \alpha}}^j 
%\end{array} \right]
%\left [\begin{array}{c}
%\dfrac{\partial h_i(\tilde{\boldsymbol \gamma}^j, \tilde{\boldsymbol \alpha}^j)}{\partial \tilde{\boldsymbol \gamma}^j} \\ 
%\dfrac{\partial h_i(\tilde{\boldsymbol \gamma}^j, \tilde{\boldsymbol \alpha}^j)}{\partial \tilde{\boldsymbol \alpha}^j} 
%\end{array} \right] \\
%& =  \left [\begin{array}{cc}
%\{\tilde{\bf N}_i^j\}^\T & \{\tilde{\bf T}_i^j\}^\T\\  
%\end{array} \right]
%var \left [\begin{array}{c}
%\hat{\tilde{\boldsymbol \gamma}}^j \\ 
%\hat{\tilde{\boldsymbol \alpha}}^j 
%\end{array} \right]
%\left [\begin{array}{c}
%\tilde{\bf N}_i^j \\ 
%\tilde{\bf T}_i^j
%\end{array} \right], \\
& = \dfrac{1}{\phi} \left [\begin{array}{cc}
\{\tilde{\bf N}_i^1\}^\T & \{\tilde{\bf T}_i^1\}^\T\\  
\end{array} \right]
\text{cp} \{ {\bf B} \}_{[p+1:p+1+q_1+s_1]}
\left [\begin{array}{c}
\tilde{\bf N}_i^1 \\ 
\tilde{\bf T}_i^1
\end{array} \right], 
\end{align*}
where ${\bf B}$ denotes the solution of the triangular system ${\bf L}^\T {\bf B} = {\bf I}_{p + \sum_j q_j + s_j}$, $\text{cp}$ denote the cross-product matrix and the subscript $[p+1:p+1+q_1+s_1]$ indicates the block of the covariance matrix corresponding to parameters $\tilde{\boldsymbol \gamma}^1$ and $\tilde{\boldsymbol \alpha}^1$. So that
\begin{align*}
\left [\begin{array}{c}
\text{var} \{ \tilde f_1(u_1^1) \} \\
\vdots \\
\text{var} \{ \tilde f_1(u_n^1) \}
\end{array} \right] & = \dfrac{1}{\phi} \text{diag} \left\{ \left [\begin{array}{cc}
\{ \tilde{\bf N}_1^1 \}^\T & \{ \tilde{\bf T}_1^1 \}^\T \\
\vdots & \vdots \\
\{ \tilde{\bf N}_n^1 \}^\T & \{ \tilde{\bf T}_n^1 \}^\T
\end{array} \right]
\text{cp} \{ {\bf B} \}_{[p+1:p+q_1+s_1+1]}
\left [\begin{array}{c}
\tilde{\bf N}_1^1 \cdots \tilde{\bf N}_n^1\\ 
\tilde{\bf T}_1^1 \cdots \tilde{\bf T}_n^1
\end{array} \right] \right\} \\
& = \dfrac{1}{\phi} \text{diag} \left\{  \left [\begin{array}{cc}
\tilde{\bf N}^1 & \tilde{\bf T}^1 \\
\end{array} \right]
\text{cp} \{ {\bf B} \}_{[p+1:p+q_1+s_1+1]}
\left [\begin{array}{c}
\{ \tilde{\bf N}^1 \}^\T \\ 
\{ \tilde{\bf T}^1 \}^\T 
\end{array} \right]  \right\}\\
& = \dfrac{1}{\phi} \text{diag} \left\{ \left [\begin{array}{cc}
\tilde{\bf N}^1 & \tilde{\bf T}^1 \\
\end{array} \right]
\text{cp} \{ {\bf B} \}_{[p+1:p+q_1+s_1+1]}
\left [\begin{array}{cc}
\tilde{\bf N}^1 & \tilde{\bf T}^1\\  
\end{array} \right]^\T \right\}. 
\end{align*}
Note that computing this matrix and then extracting the diagonal to obtain the variances is $O[n^2(q_1 + s_1)]$ flops. But computing the interest variance vector, using the Hadamard matrix product ($\circ$), is approximately $O[n(q_1 + s_1)^2]$ flops, with the number of operations reduced by exploiting the triangular structure of {\bf B}:
\begin{align*}
\left [\begin{array}{c}
\text{var} \{ \tilde f_1(u_1^1) \} \\
\vdots \\
\text{var} \{ \tilde f_1(u_n^1) \}
\end{array} \right]
& = \dfrac{1}{\phi} \text{rowSums} \left\{
\left [\begin{array}{cc}
\tilde{\bf N}^1 & \tilde{\bf T}^1 \\
\end{array} \right]
{\bf B}^\T_{[p+1:p+q_1+s_1]} 
\circ
\left [\begin{array}{cc}
\tilde{\bf N}^1 & \tilde{\bf T}^1 \\
\end{array} \right]
{\bf B}^\T_{[p+1:p+q_1+s_1]}
\right\}.
\end{align*}

\bibliography{sn-biblio}



\section*{Supplementary material}

# Supplementary Material to: Stable direct estimation for GPLSIAMs using P-splines with dynamically updated boundaries

Danilo V. Silva[1] and Gilberto A. Paula[1*]

[1]Department of Statistics, Institute of Mathematics, Statistics and Computer Science, Universidade de São Paulo, São Paulo, Brazil.

*Corresponding author(s). E-mail(s): giapaula@ime.usp.br;
Contributing authors: danilo.silva@ime.usp.br;

## 1 Penalized score

The penalized score of $\beta_1$ is

$$\begin{aligned}
\mathrm{U}_{\beta_1} &= \frac{\partial}{\partial \beta_1}\left\{\sum_i^n \phi\{y_i\tau_i - b(\tau_i)\} + c(y_i,\phi) - \frac{1}{2}\sum_j^m \lambda_j \{\tilde{\boldsymbol{\gamma}}^j\}^{\mathrm{T}}\tilde{\mathbf{P}}^j\tilde{\boldsymbol{\gamma}}^j\right\}\\
&= \phi\sum_i^n y_i \frac{d\tau_i}{d\mu_i}\frac{d\mu_i}{d\eta_i}x_{i1} - b'(\tau_i)\frac{d\tau_i}{d\mu_i}\frac{d\mu_i}{d\eta_i}x_{i1}\\
&= \phi\sum_i^n x_{i1}\frac{1}{\mathrm{V}_i}\frac{d\mu_i}{d\eta_i}(y_i-\mu_i)\\
&= \phi\sum_i^n x_{i1}\sqrt{\frac{\omega_i}{\mathrm{V}_i}}(y_i-\mu_i),
\end{aligned}$$

with $\omega_i = \{g'(\mu_i)\}^{-2}\mathrm{V}_i^{-1}$, $\mathrm{V}_i = d\mu_i/d\tau_i$. Let $\mathbf{X}$ be the linear term model matrix with rows $\mathbf{x}_i^{\mathrm{T}}$, $\mathbf{W} = \mathrm{diag}\{\omega_1,\cdots,\omega_n\}$, $\mathbf{V} = \mathrm{diag}\{\mathrm{V}_1,\cdots,\mathrm{V}_n\}$, $\mathbf{y} = (y_1,\cdots,y_n)^{\mathrm{T}}$ and $\boldsymbol{\mu} = (\mu_1,\cdots,\mu_n)^{\mathrm{T}}$. Then, the penalized score of $\boldsymbol{\beta}$ is

$$\mathrm{U}_\beta = \frac{\partial \mathrm{L}_p(\boldsymbol{\psi},\phi,\boldsymbol{\lambda})}{\partial \boldsymbol{\beta}} = \phi\mathbf{X}^{\mathrm{T}}\mathbf{W}^{\frac{1}{2}}\mathbf{V}^{-\frac{1}{2}}(\mathbf{y}-\boldsymbol{\mu}).$$

Let $\tilde{\mathbf{P}}^j$ be the matrix with rows $\{\tilde{\mathbf{P}}_i^j\}^{\mathrm{T}}$ in which $\tilde{\mathbf{P}}_i^j = [\tilde{\mathrm{P}}_{i1}^j, \cdots, \tilde{\mathrm{P}}_{iq_j}^j]^{\mathrm{T}}$ for each penalized coefficient from respective additive term. The penalized score of $\tilde{\gamma}_1^j$ ($\tilde{\mathbf{P}}^j$ symmetry) is

$$\begin{aligned}
\mathrm{U}_{\tilde{\gamma}_1^j} &= \frac{\partial}{\partial \tilde{\gamma}_1^j} \left\{ \sum_i^n \phi\{y_i \tau_i - b(\tau_i)\} + c(y_i, \phi) - \frac{1}{2} \sum_j^m \lambda_j \{\tilde{\boldsymbol{\gamma}}^j\}^{\mathrm{T}} \tilde{\mathbf{P}}^j \tilde{\boldsymbol{\gamma}}^j \right\} \\
&= \phi \sum_i^n y_i \frac{d\tau_i}{d\mu_i} \frac{d\mu_i}{d\eta_i} \tilde{\mathrm{N}}_1^{k_j}(u_i^j) - b'(\tau_i) \frac{d\tau_i}{d\mu_i} \frac{d\mu_i}{d\eta_i} \tilde{\mathrm{N}}_1^{k_j}(u_i^j) - \lambda_j \sum_i^{q_j} \tilde{\mathrm{P}}_{1i}^j \tilde{\gamma}_i^j \\
&= \phi \sum_i^n \tilde{\mathrm{N}}_1^{k_j}(u_i^j) \frac{1}{\mathrm{V}_i} \frac{d\mu_i}{d\eta_i} (y_i - \mu_i) - \lambda_j \sum_i^{q_j} \tilde{\mathrm{P}}_{1i}^j \tilde{\gamma}_i^j \\
&= \phi \sum_i^n \tilde{\mathrm{N}}_1^{k_j}(u_i^j) \sqrt{\frac{\omega_i}{\mathrm{V}_i}} (y_i - \mu_i) - \lambda_j \sum_i^{q_j} \tilde{\mathrm{P}}_{1i}^j \tilde{\gamma}_i^j,
\end{aligned}$$

with $\omega_i = \{g'(\mu_i)\}^{-2} \mathrm{V}_i^{-1}$, $\mathrm{V}_i = d\mu_i / d\tau_i$. Let $\tilde{\mathbf{N}}^j$ be the $j$-th reparametrized B-spline basis with rows $\{\tilde{\mathbf{N}}_i^j\}^{\mathrm{T}}$, $\mathbf{W} = \mathrm{diag}\{\omega_1, \cdots, \omega_n\}$, $\mathbf{V} = \mathrm{diag}\{\mathrm{V}_1, \cdots, \mathrm{V}_n\}$, $\mathbf{y} = (y_1, \cdots, y_n)^{\mathrm{T}}$ and $\boldsymbol{\mu} = (\mu_1, \cdots, \mu_n)^{\mathrm{T}}$. Then, the penalized score of $\tilde{\boldsymbol{\gamma}}^j$ is

$$\mathrm{U}_{\tilde{\gamma}^j} = \frac{\partial \mathrm{L}_p(\boldsymbol{\psi}, \phi, \boldsymbol{\lambda})}{\partial \tilde{\boldsymbol{\gamma}}^j} = \phi \{\tilde{\mathbf{N}}^j\}^{\mathrm{T}} \mathbf{W}^{\frac{1}{2}} \mathbf{V}^{-\frac{1}{2}} (\mathbf{y} - \boldsymbol{\mu}) - \lambda_j \tilde{\mathbf{P}}^j \tilde{\boldsymbol{\gamma}}^j.$$

The penalized score of $\phi$ is

$$\begin{aligned}
\mathrm{U}_{\phi} &= \frac{\partial}{\partial \phi} \left\{ \sum_i^n \phi\{y_i \tau_i - b(\tau_i)\} + c(y_i, \phi) - \frac{1}{2} \sum_j^m \lambda_j \{\tilde{\boldsymbol{\gamma}}^j\}^{\mathrm{T}} \tilde{\mathbf{P}}^j \tilde{\boldsymbol{\gamma}}^j \right\} \\
&= \sum_i^n y_i \tau_i - b(\tau_i) + c'(y_i, \phi).
\end{aligned}$$

# 2 Penalized Fisher

The penalized Fisher information of $(\beta_1, \beta_2)$ is

$$\begin{aligned}
\mathrm{K}_{\beta_1 \beta_2} &= -\mathrm{E}\left[ \frac{\partial}{\partial \beta_2} \left\{ \phi \sum_i^n x_{i1} \sqrt{\frac{\omega_i}{\mathrm{V}_i}} (y_i - \mu_i) \right\} \right] \\
&= -\phi \sum_i^n \mathrm{E}\left[ \frac{\partial}{\partial \beta_2} \left\{ x_{i1} \sqrt{\frac{\omega_i}{\mathrm{V}_i}} \right\} (y_i - \mu_i) - x_{i1} \sqrt{\frac{\omega_i}{\mathrm{V}_i}} \frac{\mathrm{V}_i}{\mathrm{V}_i} \frac{d\mu_i}{d\eta_i} x_{i2} \right] \\
&= \phi \sum_i^n x_{i1} \omega_i x_{i2},
\end{aligned}$$

with $\omega_i = \{g'(\mu_i)\}^{-2}\mathrm{V}_i^{-1}$, $\mathrm{V}_i = d\mu_i/d\tau_i$. Let $\mathbf{X}$ be the linear term model matrix with rows $\mathbf{x}_i^{\mathrm{T}}$ and $\mathbf{W} = \mathrm{diag}\{\omega_1, \cdots, \omega_n\}$. Then, the penalized Fisher information of $(\boldsymbol{\beta}, \boldsymbol{\beta})$ is

$$\mathbf{K}_{\beta\beta} = \mathrm{E}\left[-\frac{\partial^2 \mathrm{L}_p(\boldsymbol{\psi}, \phi, \boldsymbol{\lambda})}{\partial\boldsymbol{\beta}\partial\boldsymbol{\beta}^{\mathrm{T}}}\right] = \phi\mathbf{X}^{\mathrm{T}}\mathbf{W}\mathbf{X}.$$

Let $\tilde{\mathbf{P}}^j$ be the matrix with rows $\{\tilde{\mathbf{P}}_i^j\}^{\mathrm{T}}$ in which $\tilde{\mathbf{P}}_i^j = [\tilde{\mathrm{P}}_{i1}^j, \cdots, \tilde{\mathrm{P}}_{iq_j}^j]^{\mathrm{T}}$ for each penalized coefficient from respective additive term. The penalized Fisher information of $(\tilde{\gamma}_1^j, \tilde{\gamma}_2^j)$ is

$$\begin{aligned}
\mathrm{K}_{\tilde{\gamma}_1^j\tilde{\gamma}_2^j} &= -\mathrm{E}\left[\frac{\partial}{\partial\tilde{\gamma}_2^j}\left\{\phi\sum_i^n \tilde{\mathrm{N}}_1^{k_j}(u_i^j)\sqrt{\frac{\omega_i}{\mathrm{V}_i}}(y_i - \mu_i) - \lambda_j\sum_i^{q_j}\tilde{\mathrm{P}}_{1i}^j\tilde{\gamma}_i^j\right\}\right] \\
&= \phi\sum_i^n \tilde{\mathrm{N}}_1^{k_j}(u_i^j)\sqrt{\frac{\omega_i}{\mathrm{V}_i}}\frac{\mathrm{V}_i}{\mathrm{V}_i}\frac{d\mu_i}{d\eta_i}\tilde{\mathrm{N}}_2^{k_j}(u_i^j) + \lambda_j\tilde{\mathrm{P}}_{12}^j \\
&= \phi\sum_i^n \tilde{\mathrm{N}}_1^{k_j}(u_i^j)\omega_i\tilde{\mathrm{N}}_2^{k_j}(u_i^j) + \lambda_j\tilde{\mathrm{P}}_{12}^j,
\end{aligned}$$

with $\omega_i = \{g'(\mu_i)\}^{-2}\mathrm{V}_i^{-1}$, $\mathrm{V}_i = d\mu_i/d\tau_i$. Let $\tilde{\mathbf{N}}^j$ be the $j$-th reparametrized B-spline basis with rows $\{\tilde{\mathbf{N}}_i^j\}^{\mathrm{T}}$ and $\mathbf{W} = \mathrm{diag}\{\omega_1, \cdots, \omega_n\}$. Then, the penalized Fisher information of $(\tilde{\boldsymbol{\gamma}}^j, \tilde{\boldsymbol{\gamma}}^j)$ is

$$\mathbf{K}_{\tilde{\gamma}^j\tilde{\gamma}^j} = \mathrm{E}\left[-\frac{\partial^2 \mathrm{L}_p(\boldsymbol{\psi}, \phi, \boldsymbol{\lambda})}{\partial\tilde{\boldsymbol{\gamma}}^j\partial\{\tilde{\boldsymbol{\gamma}}^j\}^{\mathrm{T}}}\right] = \phi\{\tilde{\mathbf{N}}^j\}^{\mathrm{T}}\mathbf{W}\tilde{\mathbf{N}}^j + \lambda_j\tilde{\mathbf{P}}_{dif_j}^j.$$

Let $\tilde{\mathbf{P}}^j$ be the matrix with rows $\{\tilde{\mathbf{P}}_i^j\}^{\mathrm{T}}$ in which $\tilde{\mathbf{P}}_i^j = [\tilde{\mathrm{P}}_{i1}^j, \cdots, \tilde{\mathrm{P}}_{iq_j}^j]^{\mathrm{T}}$ for each penalized coefficient from respective additive term. The penalized Fisher information of $(\tilde{\gamma}_1^j, \tilde{\gamma}_2^{j'})$ for all $j \neq j'$ is

$$\begin{aligned}
\mathrm{K}_{\tilde{\gamma}_1^j\tilde{\gamma}_2^{j'}} &= -\mathrm{E}\left[\frac{\partial}{\partial\tilde{\gamma}_2^{j'}}\left\{\phi\sum_i^n \tilde{\mathrm{N}}_1^{k_j}(u_i^j)\sqrt{\frac{\omega_i}{\mathrm{V}_i}}(y_i - \mu_i) - \lambda_j\sum_i^{q_j}\tilde{\mathrm{P}}_{1i}^j\tilde{\gamma}_i^j\right\}\right] \\
&= \phi\sum_i^n \tilde{\mathrm{N}}_1^{k_j}(u_i^j)\sqrt{\frac{\omega_i}{\mathrm{V}_i}}\frac{\mathrm{V}_i}{\mathrm{V}_i}\frac{d\mu_i}{d\eta_i}\tilde{\mathrm{N}}_2^{k_{j'}}(u_i^{j'}) \\
&= \phi\sum_i^n \tilde{\mathrm{N}}_1^{k_j}(u_i^j)\omega_i\tilde{\mathrm{N}}_2^{k_{j'}}(u_i^{j'}),
\end{aligned}$$

with $\omega_i = \{g'(\mu_i)\}^{-2}\mathrm{V}_i^{-1}$, $\mathrm{V}_i = d\mu_i/d\tau_i$. Let $\tilde{\mathbf{N}}^j$ be the $j$-th reparametrized B-spline basis with rows $\{\tilde{\mathbf{N}}_i^j\}^{\mathrm{T}}$, $\tilde{\mathbf{N}}^{j'}$ the $j'$-th basis with rows $\{\tilde{\mathbf{N}}_i^{j'}\}^{\mathrm{T}}$ and $\mathbf{W} =$

diag$\{\omega_1, \cdots, \omega_n\}$. Then, the penalized Fisher information of $(\tilde{\gamma}^j, \tilde{\gamma}^{j'})$ is

$$\mathbf{K}_{\tilde{\gamma}^j\tilde{\gamma}^{j'}} = \mathrm{E}\left[-\frac{\partial^2 \mathrm{L}_p(\boldsymbol{\psi}, \phi, \boldsymbol{\lambda})}{\partial \tilde{\boldsymbol{\gamma}}^j \partial \{\tilde{\boldsymbol{\gamma}}^{j'}\}^{\mathrm{T}}}\right] = \phi\{\tilde{\mathbf{N}}^j\}^{\mathrm{T}}\mathbf{W}\tilde{\mathbf{N}}^{j'}.$$

The penalized Fisher information of $(\beta_1, \tilde{\gamma}_2^j)$ is

$$\begin{aligned}
\mathrm{K}_{\beta_1\tilde{\gamma}_2^j} &= -\mathrm{E}\left[\frac{\partial}{\partial \tilde{\gamma}_2^j}\left\{\phi\sum_i^n x_{i1}\sqrt{\frac{\omega_i}{\mathrm{V}_i}}(y_i - \mu_i)\right\}\right] \\
&= -\phi\sum_i^n \mathrm{E}\left[\frac{\partial}{\partial \tilde{\gamma}_2^j}\left\{x_{i1}\sqrt{\frac{\omega_i}{\mathrm{V}_i}}\right\}(y_i - \mu_i) - x_{i1}\sqrt{\frac{\omega_i}{\mathrm{V}_i}}\frac{\mathrm{V}_i}{\mathrm{V}_i}\frac{d\mu_i}{d\eta_i}\tilde{\mathrm{N}}_2^{k_j}(u_i^j)\right] \\
&= \phi\sum_i^n x_{i1}\omega_i\tilde{\mathrm{N}}_2^{k_j}(u_i^j),
\end{aligned}$$

with $\omega_i = \{g'(\mu_i)\}^{-2}\mathrm{V}_i^{-1}$, $\mathrm{V}_i = d\mu_i/d\tau_i$. Let $\mathbf{X}$ be the linear term model matrix with rows $\mathbf{x}_i^{\mathrm{T}}$, $\tilde{\mathbf{N}}^j$ the $j$-th reparametrized B-spline basis with rows $\{\tilde{\mathbf{N}}_i^j\}^{\mathrm{T}}$ and $\mathbf{W} =$ diag$\{\omega_1, \cdots, \omega_n\}$. Then, the penalized Fisher information of $(\boldsymbol{\beta}, \tilde{\boldsymbol{\gamma}}^j)$ is

$$\mathbf{K}_{\beta\tilde{\gamma}^j} = \mathrm{E}\left[-\frac{\partial^2 \mathrm{L}_p(\boldsymbol{\psi}, \phi, \boldsymbol{\lambda})}{\partial \boldsymbol{\beta} \partial \{\tilde{\boldsymbol{\gamma}}^j\}^{\mathrm{T}}}\right] = \phi\mathbf{X}^{\mathrm{T}}\mathbf{W}\tilde{\mathbf{N}}^j.$$

The penalized Fisher information of $(\tilde{\alpha}_1^j, \tilde{\alpha}_2^{j'})$ for all $j \neq j'$ is

$$\begin{aligned}
\mathrm{K}_{\tilde{\alpha}_1^j\tilde{\alpha}_2^{j'}} &= -\mathrm{E}\left[\frac{\partial}{\partial \tilde{\alpha}_2^{j'}}\left\{\phi\sum_i^n\left\{\frac{d\alpha_1^j}{d\tilde{\alpha}_1^j}z_{i1}^j + \cdots + \frac{d\alpha_{s_j+1}^j}{d\tilde{\alpha}_1^j}z_{is_j+1}^j\right\}\tilde{f}_j'(u_i^j)\sqrt{\frac{\omega_i}{\mathrm{V}_i}}(y_i - \mu_i)\right\}\right] \\
&= \phi\sum_i^n\left\{\sum_k^{s_j+1}\frac{d\alpha_k^j}{d\tilde{\alpha}_1^j}z_{ik}^j\right\}\tilde{f}_j'(u_i^j)\sqrt{\frac{\omega_i}{\mathrm{V}_i}}\frac{\mathrm{V}_i}{\mathrm{V}_i}\frac{d\mu_i}{d\eta_i}\tilde{f}_{j'}'(u_i^{j'})\left\{\sum_k^{s_{j'}+1}\frac{d\alpha_k^{j'}}{d\tilde{\alpha}_2^{j'}}z_{ik}^{j'}\right\} \\
&= \phi\sum_i^n\left\{\sum_k^{s_j+1}\frac{d\alpha_k^j}{d\tilde{\alpha}_1^j}z_{ik}^j\right\}\tilde{f}_j'(u_i^j)\omega_i\tilde{f}_{j'}'(u_i^{j'})\left\{\sum_k^{s_{j'}+1}\frac{d\alpha_k^{j'}}{d\tilde{\alpha}_2^{j'}}z_{ik}^{j'}\right\},
\end{aligned}$$

with $\omega_i = \{g'(\mu_i)\}^{-2}\mathrm{V}_i^{-1}$, $\mathrm{V}_i = d\mu_i/d\tau_i$. Let $\mathbf{Z}^j$ be the matrix with rows $\{\mathbf{z}_i^j\}^{\mathrm{T}}$, $\mathbf{Z}^{j'}$ the matrix with rows $\{\mathbf{z}_i^{j'}\}^{\mathrm{T}}$ and $\mathbf{W} =$ diag$\{\omega_1, \cdots, \omega_n\}$. Also, let $\mathbf{J}^j$ be the $j$-th Jacobian matrix of $\boldsymbol{\alpha}^j$ with respect to $\tilde{\boldsymbol{\alpha}}^j$, $\mathbf{J}^{j'}$ the $j'$-th Jacobian matrix, $\{\tilde{\mathbf{F}}^j\}^{(1)} =$ diag$\{\tilde{\mathbf{f}}_j'\}$ in which $\tilde{\mathbf{f}}_j' = \{\tilde{\mathbf{N}}^j\}^{(1)}\tilde{\boldsymbol{\gamma}}^j$ and $\{\tilde{\mathbf{F}}^{j'}\}^{(1)} =$ diag$\{\tilde{\mathbf{f}}_{j'}'\}$. Then, the penalized Fisher information of $(\tilde{\boldsymbol{\alpha}}^j, \tilde{\boldsymbol{\alpha}}^{j'})$ is

$$\mathbf{K}_{\tilde{\alpha}^j\tilde{\alpha}^{j'}} = \mathrm{E}\left[-\frac{\partial^2 \mathrm{L}_p(\boldsymbol{\psi}, \phi, \boldsymbol{\lambda})}{\partial \tilde{\boldsymbol{\alpha}}^j \partial \{\tilde{\boldsymbol{\alpha}}^{j'}\}^{\mathrm{T}}}\right] = \phi\{\{\tilde{\mathbf{F}}^j\}^{(1)}\mathbf{Z}^j\mathbf{J}^j\}^{\mathrm{T}}\mathbf{W}\{\tilde{\mathbf{F}}^{j'}\}^{(1)}\mathbf{Z}^{j'}\mathbf{J}^{j'}.$$

The penalized Fisher information of $(\beta_1, \tilde{\alpha}_2^j)$ is

$$
\begin{aligned}
\mathrm{K}_{\beta_1\tilde{\alpha}_2^j} &= -\mathrm{E}\left[\frac{\partial}{\partial\tilde{\alpha}_2^j}\left\{\phi\sum_i^n x_{i1}\sqrt{\frac{\omega_i}{\mathrm{V}_i}}(y_i-\mu_i)\right\}\right] \\
&= -\phi\sum_i^n \mathrm{E}\left[\frac{\partial}{\partial\tilde{\alpha}_2^j}\left\{x_{i1}\sqrt{\frac{\omega_i}{\mathrm{V}_i}}\right\}(y_i-\mu_i) - x_{i1}\sqrt{\frac{\omega_i}{\mathrm{V}_i}}\frac{\mathrm{V}_i}{\mathrm{V}_i}\frac{d\mu_i}{d\eta_i}\tilde{f}_j'(u_i^j)\left\{\sum_k^{s_j+1}\frac{d\alpha_k^j}{d\tilde{\alpha}_2^j}z_{ik}^j\right\}\right] \\
&= \phi\sum_i^n x_{i1}\omega_i\tilde{f}_j'(u_i^j)\left\{\sum_k^{s_j+1}\frac{d\alpha_k^j}{d\tilde{\alpha}_2^j}z_{ik}^j\right\},
\end{aligned}
$$

with $\omega_i = \{g'(\mu_i)\}^{-2}\mathrm{V}_i^{-1}$, $\mathrm{V}_i = d\mu_i/d\tau_i$. Let $\mathbf{X}$ be the linear term model matrix with rows $\mathbf{x}_i^{\mathrm{T}}$, $\mathbf{Z}^j$ the matrix with rows $\{\mathbf{z}_i^j\}^{\mathrm{T}}$ and $\mathbf{W} = \mathrm{diag}\{\omega_1, \cdots, \omega_n\}$. Also, let $\mathbf{J}^j$ be the $j$-th Jacobian matrix of $\boldsymbol{\alpha}^j$ with respect to $\tilde{\boldsymbol{\alpha}}^j$ and $\{\tilde{\mathbf{F}}^j\}^{(1)} = \mathrm{diag}\{\tilde{\mathbf{f}}_j'\}$ in which $\tilde{\mathbf{f}}_j' = \{\tilde{\mathbf{N}}^j\}^{(1)}\tilde{\boldsymbol{\gamma}}^j$. Then, the penalized Fisher information of $(\boldsymbol{\beta}, \tilde{\boldsymbol{\alpha}}^j)$ is

$$
\mathbf{K}_{\beta\tilde{\alpha}^j} = \mathrm{E}\left[-\frac{\partial^2 \mathrm{L}_p(\boldsymbol{\psi}, \phi, \boldsymbol{\lambda})}{\partial\boldsymbol{\beta}\partial\{\tilde{\boldsymbol{\alpha}}^j\}^{\mathrm{T}}}\right] = \phi\mathbf{X}^{\mathrm{T}}\mathbf{W}\{\tilde{\mathbf{F}}^j\}^{(1)}\mathbf{Z}^j\mathbf{J}^j.
$$

The penalized Fisher information of $(\tilde{\alpha}_1^j, \tilde{\gamma}_2^j)$ is

$$
\begin{aligned}
\mathrm{K}_{\tilde{\alpha}_1^j\tilde{\gamma}_2^j} &= -\mathrm{E}\left[\frac{\partial}{\partial\tilde{\gamma}_2^j}\left\{\phi\sum_i^n\left\{\frac{d\alpha_1^j}{d\tilde{\alpha}_1^j}z_{i1}^j + \cdots + \frac{d\alpha_{s_j+1}^j}{d\tilde{\alpha}_1^j}z_{is_j+1}^j\right\}\tilde{f}_j'(u_i^j)\sqrt{\frac{\omega_i}{\mathrm{V}_i}}(y_i-\mu_i)\right\}\right] \\
&= \phi\sum_i^n\left\{\sum_k^{s_j+1}\frac{d\alpha_k^j}{d\tilde{\alpha}_1^j}z_{ik}^j\right\}\tilde{f}_j'(u_i^j)\sqrt{\frac{\omega_i}{\mathrm{V}_i}}\frac{\mathrm{V}_i}{\mathrm{V}_i}\frac{d\mu_i}{d\eta_i}\tilde{\mathrm{N}}_2^{k_j}(u_i^j) \\
&= \phi\sum_i^n\left\{\sum_k^{s_j+1}\frac{d\alpha_k^j}{d\tilde{\alpha}_1^j}z_{ik}^j\right\}\tilde{f}_j'(u_i^j)\omega_i\tilde{\mathrm{N}}_2^{k_j}(u_i^j),
\end{aligned}
$$

with $\omega_i = \{g'(\mu_i)\}^{-2}\mathrm{V}_i^{-1}$, $\mathrm{V}_i = d\mu_i/d\tau_i$. Let $\mathbf{Z}^j$ be the matrix with rows $\{\mathbf{z}_i^j\}^{\mathrm{T}}$, $\tilde{\mathbf{N}}^j$ the $j$-th reparametrized B-spline basis with rows $\{\tilde{\mathbf{N}}_i^j\}^{\mathrm{T}}$ and $\mathbf{W} = \mathrm{diag}\{\omega_1, \cdots, \omega_n\}$. Also, let $\mathbf{J}^j$ be the $j$-th Jacobian matrix of $\boldsymbol{\alpha}^j$ with respect to $\tilde{\boldsymbol{\alpha}}^j$ and $\{\tilde{\mathbf{F}}^j\}^{(1)} = \mathrm{diag}\{\tilde{\mathbf{f}}_j'\}$ in which $\tilde{\mathbf{f}}_j' = \{\tilde{\mathbf{N}}^j\}^{(1)}\tilde{\boldsymbol{\gamma}}^j$. Then, the penalized Fisher information of $(\tilde{\boldsymbol{\alpha}}^j, \tilde{\boldsymbol{\gamma}}^j)$ is

$$
\mathbf{K}_{\tilde{\alpha}^j\tilde{\gamma}^j} = \mathrm{E}\left[-\frac{\partial^2 \mathrm{L}_p(\boldsymbol{\psi}, \phi, \boldsymbol{\lambda})}{\partial\tilde{\boldsymbol{\alpha}}^j\partial\{\tilde{\boldsymbol{\gamma}}^j\}^{\mathrm{T}}}\right] = \phi\{\{\tilde{\mathbf{F}}^j\}^{(1)}\mathbf{Z}^j\mathbf{J}^j\}^{\mathrm{T}}\mathbf{W}\tilde{\mathbf{N}}^j.
$$

The penalized Fisher information of $(\tilde{\alpha}_1^j, \tilde{\gamma}_2^{j'})$ is

$$
\mathrm{K}_{\tilde{\alpha}_1^j\tilde{\gamma}_2^{j'}} = -\mathrm{E}\left[\frac{\partial}{\partial\tilde{\gamma}_2^{j'}}\left\{\phi\sum_i^n\left\{\frac{d\alpha_1^j}{d\tilde{\alpha}_1^j}z_{i1}^j + \cdots + \frac{d\alpha_{s_j+1}^j}{d\tilde{\alpha}_1^j}z_{is_j+1}^j\right\}\tilde{f}_j'(u_i^j)\sqrt{\frac{\omega_i}{\mathrm{V}_i}}(y_i-\mu_i)\right\}\right]
$$

$$= \phi \sum_i^n \left\{ \sum_k^{s_j+1} \frac{d\alpha_k^j}{d\tilde{\alpha}_1^j} z_{ik}^j \right\} \tilde{f}_j'(u_i^j) \sqrt{\frac{\omega_i}{\mathrm{V}_i} \frac{\mathrm{V}_i}{\mathrm{V}_i} \frac{d\mu_i}{d\eta_i}} \tilde{\mathrm{N}}_2^{k_{j'}}(u_i^{j'})$$

$$= \phi \sum_i^n \left\{ \sum_k^{s_j+1} \frac{d\alpha_k^j}{d\tilde{\alpha}_1^j} z_{ik}^j \right\} \tilde{f}_j'(u_i^j) \omega_i \tilde{\mathrm{N}}_2^{k_{j'}}(u_i^{j'}),$$

with $\omega_i = \{g'(\mu_i)\}^{-2}\mathrm{V}_i^{-1}$, $\mathrm{V}_i = d\mu_i/d\tau_i$. Let $\mathbf{Z}^j$ be the matrix with rows $\{\mathbf{z}_i^j\}^{\mathrm{T}}$, $\tilde{\mathbf{N}}^{j'}$ the $j'$-th reparametrized B-spline basis with rows $\{\tilde{\mathbf{N}}_i^{j'}\}^{\mathrm{T}}$ and $\mathbf{W} = \mathrm{diag}\{\omega_1, \cdots, \omega_n\}$. Also, let $\mathbf{J}^j$ be the $j$-th Jacobian matrix of $\boldsymbol{\alpha}^j$ with respect to $\tilde{\boldsymbol{\alpha}}^j$ and $\{\tilde{\mathbf{F}}^j\}^{(1)} = \mathrm{diag}\{\tilde{\mathbf{f}}_j'\}$ in which $\tilde{\mathbf{f}}_j' = \{\tilde{\mathbf{N}}^j\}^{(1)}\tilde{\boldsymbol{\gamma}}^j$. Then, the penalized Fisher information of $(\tilde{\boldsymbol{\alpha}}^j, \tilde{\boldsymbol{\gamma}}^{j'})$ is

$$\mathbf{K}_{\tilde{\alpha}^j \tilde{\gamma}^{j'}} = \mathrm{E}\left[ -\frac{\partial^2 \mathrm{L}_p(\boldsymbol{\psi}, \phi, \boldsymbol{\lambda})}{\partial \tilde{\boldsymbol{\alpha}}^j \partial \{\tilde{\boldsymbol{\gamma}}^{j'}\}^{\mathrm{T}}} \right] = \phi \{\{\tilde{\mathbf{F}}^j\}^{(1)} \mathbf{Z}^j \mathbf{J}^j\}^{\mathrm{T}} \mathbf{W} \tilde{\mathbf{N}}^{j'}.$$

The penalized Fisher information of $(\phi, \beta_2)$ is

$$\mathrm{K}_{\phi\beta_2} = -\mathrm{E}\left[ \frac{\partial}{\partial \beta_2} \left\{ \sum_i^n y_i \tau_i - b(\tau_i) + c'(y_i, \phi) \right\} \right] = 0,$$

Then, the penalized Fisher information of $(\phi, \boldsymbol{\beta})$ is

$$\mathbf{K}_{\phi\beta} = \mathrm{E}\left[ -\frac{\partial^2 \mathrm{L}_p(\boldsymbol{\psi}, \phi, \boldsymbol{\lambda})}{\partial \phi \partial \boldsymbol{\beta}^{\mathrm{T}}} \right] = \mathbf{0}_{1\times p}.$$

Analogously to the last $\mathbf{K}_{\phi\tilde{\gamma}^j} = \mathbf{0}_{1\times q_j}$, $\mathbf{K}_{\phi\tilde{\alpha}^j} = \mathbf{0}_{1\times s_j}$ and $\mathrm{K}_{\phi\phi} = -\sum_i^n \mathrm{E}\left[c''(y_i, \phi)\right]$.

# 3 Simulation details

In this section, we present additional details on the simulation study, which is organized into three scenarios: Poisson I, Gamma, and Poisson II. Across all scenarios, the fitted coefficients for the proposed method are shown after removing a few unstable fitted GPLSIAMs to ensure clear visualization. To avoid redundancy, the complete R functions used for the simulations are provided exclusively in the Poisson I scenario. For the Gamma and Poisson II scenarios, only the function for generating the dataset is presented, as the remaining functions require either minor (Gamma) or no adaptations (Poisson II). Note that the observed increase in effective degrees of freedom with sample size is expected. As the sample size grows, the data provide sufficient information to support more complex curves. This is consistent with fitted GPLAMs via the `mgcv` package when smoothing parameters are estimated.

## 3.1 Poisson I

Figure 1 (all top) presents the nonlinear functions from the postulated model, which have a similar variation range. Figure 1 (all bottom) exhibits, for illustration, a

response simulation of $n = 200$ observations in the Poisson I scenario. This is a simple problem, as evident from the dispersion of observations.

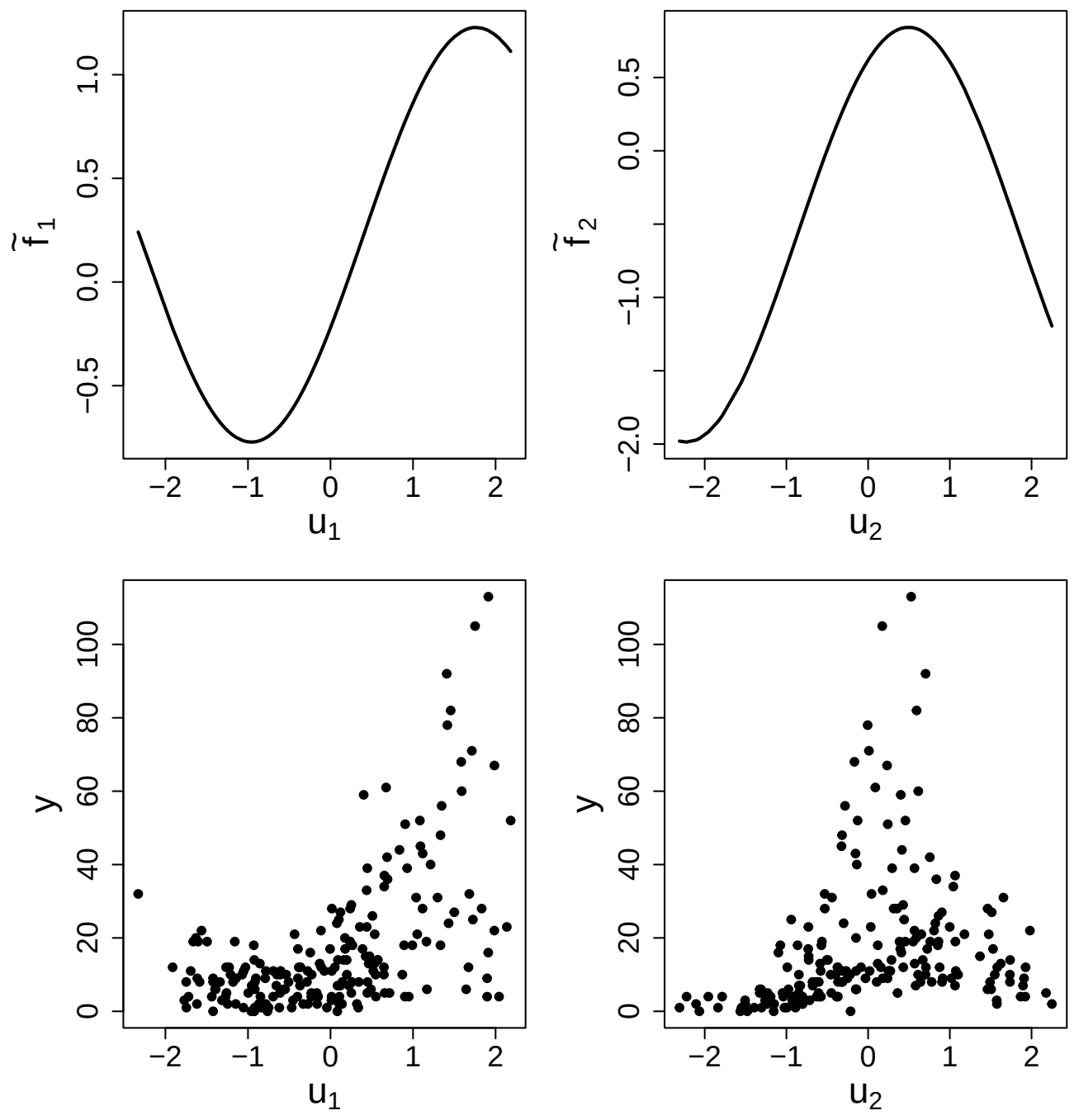


**Fig. 1** True $\tilde{f}_1$ (top-left), and $\tilde{f}_2$ (top-right) nonlinear functions versus their respective single-index covariates. The $n = 200$ response simulation versus $u_1$ (bottom-left), and $u_2$ (bottom-right) single-index covariate in Poisson I scenario.

Under the proposed method, for $n = 200$, only 0.4% of the fitted models are removed. Figure 2 (top-left) displays the elapsed time in seconds, Figure 2 (top-middle) the number of iterations of the returned model, and Figure 2 (top-right) its fitted effective degrees of freedom. Figure 2 (bottom-left and bottom-middle) presents the fitted linear coefficients $\beta_1$ and $\beta_2$, and the single-index coefficient $\alpha_1^1$ is in Figure 2 (bottom-right). The remaining single-index coefficients are in Figure 3 (all top and bottom-left), and the smoothing parameters are in Figure 3 (bottom-middle and bottom-right).

The `data_gen()` function used in the simulation to generate the dataset with `n` rows and `R` different responses from the postulated GPLSIAM in the Poisson I scenario is given by

```
> # data gen
> data_gen <- function(n, R, seed=13){
+   # prep
+   set.seed(seed)
+   dat <- list()
+   dat$X <- cbind(1, matrix(runif(n), nrow=n))
```

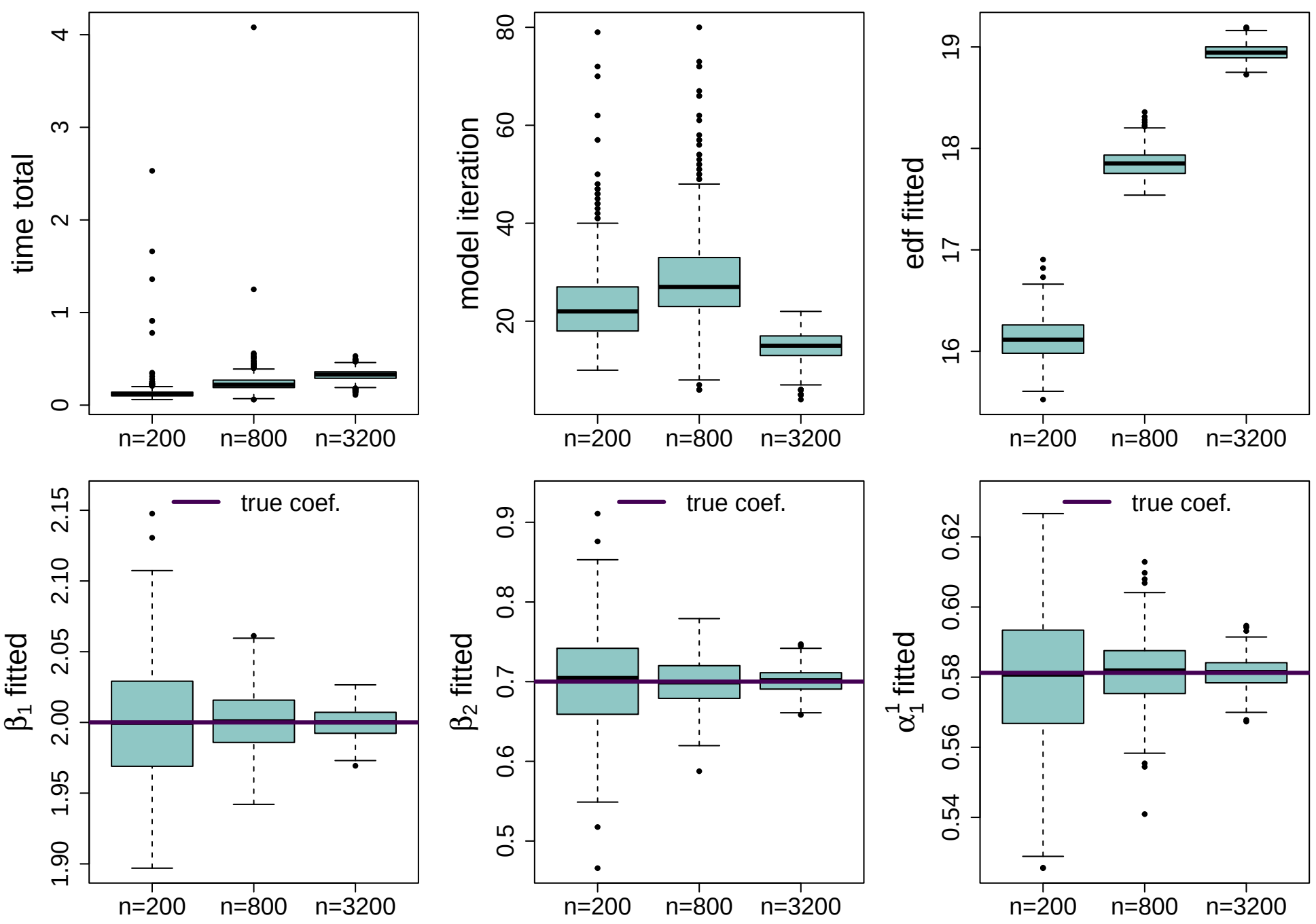


**Fig. 2** Elapsed total time (top-left), final iteration (top-middle), and fitted effective degrees of freedom (top-right). Fitted linear coefficients $\beta_1$ (bottom-left) and $\beta_2$ (bottom-middle). Fitted single-index coefficient $\alpha_1^1$ (bottom-left) from stable fitted GPLSIAMs with the proposed method in the Poisson I scenario.

```
+   dat$Z1 <- matrix(runif(2*n), nrow=n)
+   dat$Z2 <- matrix(runif(3*n), nrow=n)
+
+   # scale
+   dat$Z1 <- scale(dat$Z1)
+   dat$Z2 <- scale(dat$Z2)
+
+   # prep f1
+   alpha1 <- c(1, -1.4)
+   kk1 <- sqrt(sum(alpha1^2))
+   alpha1 <- alpha1/kk1
+   dat$u1 <- dat$Z1%*%alpha1
+   t1 <- dat$u1/sqrt(12) + sum(0.5*alpha1)
+   dat$f1 <- sin(4*t1)
+   dat$f1 <- dat$f1 - mean(dat$f1)
+
+   # prep f2
+   alpha2 <- c(1, 1.7, -0.8)
+   kk2 <- sqrt(sum(alpha2^2))
```

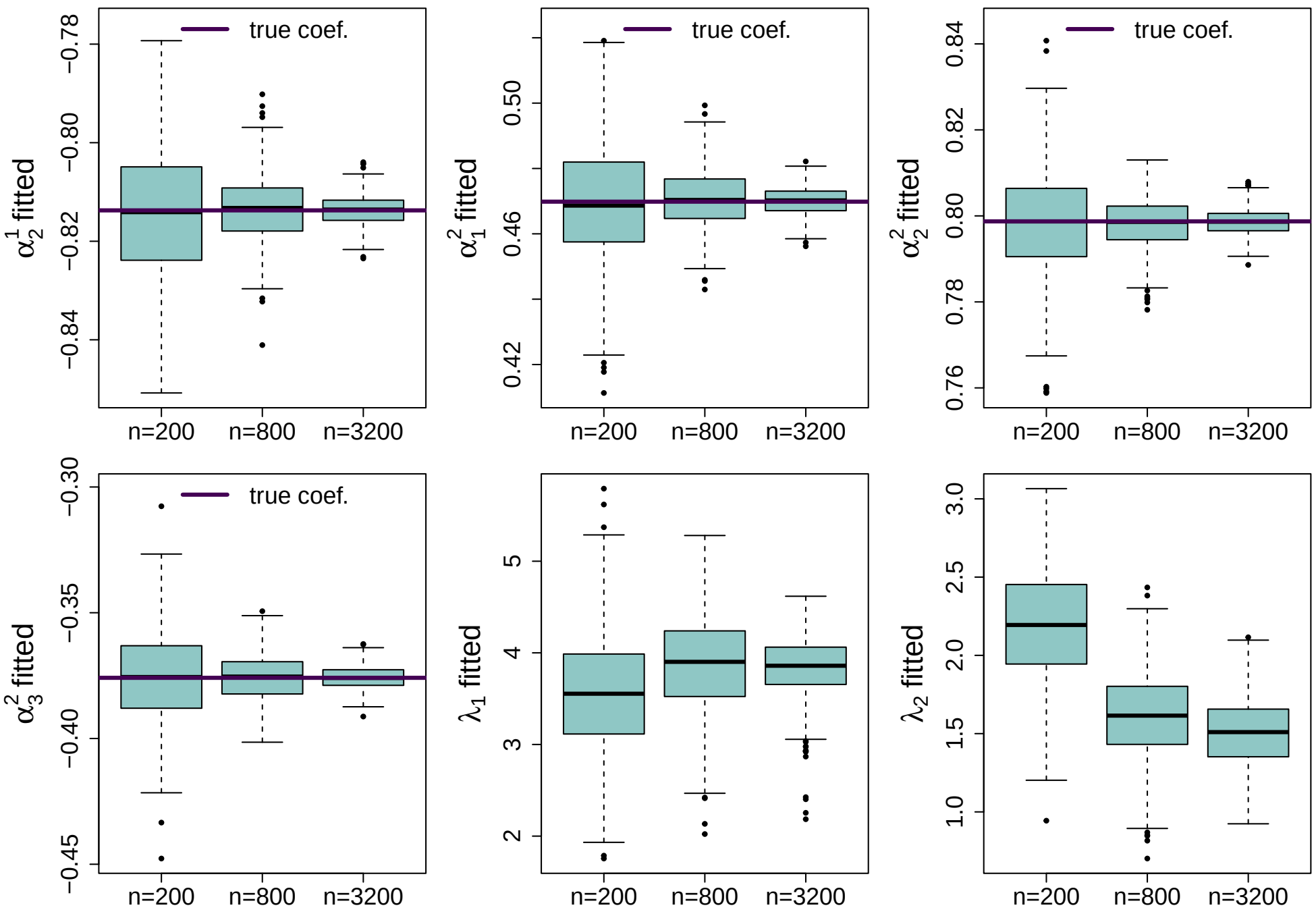


**Fig. 3** Fitted single-index coefficient $\alpha_2^1$ (top-left), $\alpha_1^2$ (top-middle), $\alpha_2^2$ (top-right), and $\alpha_3^2$ (bottom-left). Fitted smoothing parameter $\lambda_1$ (bottom-middle), and $\lambda_2$ (bottom-right) from stable fitted GPLSIAMs with the proposed method in the Poisson I scenario.

```
+   alpha2 <- alpha2/kk2
+   dat$u2 <- dat$Z2%*%alpha2
+   t2 <- dat$u2/sqrt(12) + sum(0.5*alpha2)
+   dat$f2 <- sin(4*t2) - cos(4*t2)
+   dat$f2 <- dat$f2 - mean(dat$f2)
+
+   # response simu
+   beta <- c(2, 0.7)
+   dat$mu <- exp(dat$X%*%beta + dat$f1 + dat$f2)
+   y_list <- list()
+   for(j in 1:R){
+     yj <- rpois(nrow(dat$mu), lambda=dat$mu)
+     y_list <- c(y_list, list(yj))
+   }
+   names(y_list) <- paste0("y", 1:R)
+
+   # return
+   dat <- c(dat, y_list)
+   dat
+ }
```

The auxiliary functions for the Poisson I scenario are given by

```
> # link
> gmu <- function(mu) log(mu)
>
> # inverse link
> inv_gmu <- function(eta){
+   # return
+   thresh <- -log(.Machine$double.eps)
+   eta <- pmin(thresh, pmax(eta, -thresh))
+   exp(eta)
+ }
>
> # variance function
> vmu <- function(mu) mu
>
> # weight
> wmu <- function(mu){
+   # return
+   dgmu_dmu <- 1/mu
+   dgmu_dmu^(-2)/vmu(mu)
+ }
>
> # ploglik
> ploglik <- function(fit){
+   # return
+   tmu <- log(fit$mu)
+   bmu <- fit$mu
+   cy <- sapply(y, function(j) -sum(log(1:j)))
+   cy[y==0] <- 0
+   ploglik <- sum(fit$phi*(y*tmu - bmu) + cy)
+   ploglik <- as.numeric(ploglik - t(fit$psi)%*%fit$P%*%fit$psi)
+   ploglik
+ }
>
> # ralpha_til
> ralpha_til <- function(s){
+   # prep
+   bad <- T
+   while(bad==T){
+     # simu
+     temp_til <- runif(s, -1, 1)
+     temp <- c(1, temp_til)
+     kk <- sqrt(sum(temp^2))
+     temp <- temp/kk
+
```

```
+     # check
+     max_temp <- max(abs(temp))
+     max_temp <- max_temp < 0.8
+     pri_temp <- head(temp, 1)
+     pri_temp <- pri_temp > 0.2
+     if(max_temp & pri_temp) bad <- F
+   }
+
+   # return
+   alpha_til <- temp_til
+   alpha_til
+ }
```

The start_fit() function, which sets the initial values according to the proposed method for the Poisson distribution, is shown below. The R packages Matrix (Bates et al. 2025), splines (Bates and Venables 1999) and mgcv (Wood 2025) (for generalized linear model fit) are needed.

```
> # fit start
> start_fit <- function(){
+   # prep
+   n <- nrow(X)
+   p <- ncol(X)
+   m <- length(Z)
+   qj_vec <- rep(9, m)
+   mj_vec <- qj_vec + 1 + 4 - 8
+   fit <- list()
+   P_big <- list()
+
+   # beta
+   beta <- coef(gam(y~X-1, family=poisson("log")))
+   M <- X
+   psi <- beta
+   psi_ini <- 1
+   psi_fin <- p
+   eta <- X%*%beta
+
+   # m-loop
+   for(j in 1:m){
+     # prep
+     Zj <- Z[[j]]
+     sj <- ncol(Zj) - 1
+
+     # alpha_tilj
+     alpha_tilj <- ralpha_til(sj)
+
```

```
+     # alphaj
+     alphaj <- c(1, alpha_tilj)
+     kkj <- sqrt(sum(alphaj^2))
+     alphaj <- alphaj/kkj
+
+     # uj
+     uj <- Zj %*% alphaj
+
+     # N_tilj
+     minj <- min(uj)
+     maxj <- max(uj)
+     delta <- maxj-minj
+     minj <- minj - delta*0.001
+     maxj <- maxj + delta*0.001
+     h <- (maxj - minj)/(mj_vec[j]+1)
+     tj <- seq(minj-3*h, maxj+3*h, length.out=mj_vec[j]+8)
+     N_tilj <- splineDesign(uj, knots=tj, ord=4, outer.ok=T)
+     N_tilj <- scale(N_tilj, scale=F)
+     N_tilj <- N_tilj[,-ncol(N_tilj)]
+
+     # gamma_tilj
+     qj <- ncol(N_tilj)
+     gamma_tilj <- coef(gam(y~N_tilj, family=poisson("log")))[-1]
+
+     # add N_tilj
+     M <- cbind(M, N_tilj)
+     psi <- c(psi, gamma_tilj)
+     psi_ini <- c(psi_ini, tail(psi_fin,1) + 1)
+     psi_fin <- c(psi_fin, tail(psi_fin,1) + qj)
+
+     # dN_tilj
+     dN_tilj <- splineDesign(uj, knots=tj, ord=4, outer.ok=T, derivs=1)
+     dN_tilj <- scale(dN_tilj, scale=F)
+     dN_tilj <- dN_tilj[,-ncol(dN_tilj)]
+
+     # df_tilj
+     df_tilj <- dN_tilj%*%gamma_tilj
+
+     # Jj
+     Jj <- outer(c(1,alpha_tilj), -alphaj[-1]/kkj^2)
+     for (i in 1:sj) Jj[i+1,i] <- Jj[i+1,i] + 1/kkj
+
+     # T_tilj
+     T_tilj <- as.vector(df_tilj)*Zj%*%Jj
+
```

```
+     # add T_tilj
+     M <- cbind(M, T_tilj)
+     psi <- c(psi, alpha_tilj)
+     psi_ini <- c(psi_ini, tail(psi_fin,1) + 1)
+     psi_fin <- c(psi_fin, tail(psi_fin,1) + sj)
+
+     # f_tilj
+     f_tilj <- N_tilj%*%gamma_tilj
+
+     # eta
+     eta <- eta + f_tilj
+
+     # P_tilj
+     D_tilj <- diff(diag(qj+1), differences=2)
+     D_tilj <- D_tilj[,-(qj+1)]
+     P_tilj <- Matrix(crossprod(D_tilj))
+     P_big <- c(P_big, list(bdiag(P_tilj, diag(sj)*0)))
+   }
+
+   # save
+   fit$n <- n
+   fit$m <- m
+   fit$mj_vec <- mj_vec
+   fit$psi <- as.matrix(psi)
+   fit$psi_pos <- list(ini=psi_ini, fin=psi_fin)
+   fit$mu <- inv_gmu(eta)
+   fit$M_til <- sqrt(wmu(fit$mu))*M
+
+   # lambda + phi
+   lambda <- runif(m, 1, 1000)
+   phi <- 1
+
+   # save
+   fit$lambda <- lambda
+   fit$phi <- phi
+
+   # penalization
+   P <- 0
+   P_big2 <- P_big
+   for(j in 1:m){
+     # bdiag
+     indic <- numeric(m)
+     indic[j] <- 1
+     P_big2[[j]] <- diag(p)*0
+     for(i in 1:m) P_big2[[j]] <- bdiag(P_big2[[j]], P_big[[i]]*indic[i])
```

```
+     P <- P + P_big2[[j]]*lambda[j]
+   }
+
+   # save
+   fit$P_big2 <- P_big2
+   fit$P <- P
+   fit$metric <- 10
+   fit
+ }
```

The update_fit() function to update the model estimates according to the proposed method is given by

```
> # fit update
> update_fit <- function(fit){
+   # prep
+   m <- fit$m
+
+   # psi
+   L <- chol(crossprod(fit$M_til) + fit$P/fit$phi + diag(1e-7, nrow(fit$P)))
+   y_til2 <- fit$M_til%*%fit$psi + (y-fit$mu)/sqrt(vmu(fit$mu))
+   b <- forwardsolve(t(L), t(fit$M_til)%*%y_til2)
+   psi_new <- backsolve(L, b)
+
+   # lambda
+   B <- forwardsolve(t(L), diag(ncol(L)))
+   Q <- MASS::ginv(as.matrix(fit$P)) - crossprod(B)/fit$phi
+   P <- 0
+   for(j in 1:(fit$m)){
+     # lambda + P
+     lambdaj <- fit$lambda[j]/(t(psi_new)%*%fit$P_big2[[j]]%*%psi_new)
+     fit$lambda[j] <- sum(diag(Q%*%fit$P_big2[[j]]))*lambdaj
+     P <- P + fit$P_big2[[j]]*fit$lambda[j]
+   }
+
+   # phi
+   vcov <- crossprod(B)
+   edf <- diag(vcov %*% crossprod(fit$M_til))
+
+   # re-compute fit
+   beta <- psi_new[fit$psi_pos$ini[1]:fit$psi_pos$fin[1]]
+   M <- X
+   eta <- X%*%beta
+   fit$beta <- beta
+
```

```
+   # m-loop
+   for(j in 1:m){
+     # prep
+     Zj <- Z[[j]]
+
+     # gamma_tilj
+     gamma_tilj <- psi_new[fit$psi_pos$ini[2*j]:fit$psi_pos$fin[2*j]]
+
+     # alpha_tilj
+     alpha_tilj <- psi_new[fit$psi_pos$ini[2*j+1]:fit$psi_pos$fin[2*j+1]]
+
+     # alphaj
+     alphaj <- c(1, alpha_tilj)
+     kkj <- sqrt(sum(alphaj^2))
+     alphaj <- alphaj/kkj
+
+     # uj
+     uj <- Zj %*% alphaj
+
+     # N_tilj
+     minj <- min(uj)
+     maxj <- max(uj)
+     delta <- maxj-minj
+     minj <- minj - delta*0.001
+     maxj <- maxj + delta*0.001
+     h <- (maxj - minj)/(fit$mj_vec[j]+1)
+     tj <- seq(minj-3*h, maxj+3*h, length.out=fit$mj_vec[j]+8)
+     N_tilj <- splineDesign(uj, knots=tj, ord=4, outer.ok=T)
+     N_tilj <- scale(N_tilj, scale=F)
+     N_tilj <- N_tilj[,-ncol(N_tilj)]
+
+     # add N_tilj
+     M <- cbind(M, N_tilj)
+
+     # dN_tilj
+     dN_tilj <- splineDesign(uj, knots=tj, ord=4, outer.ok=T, derivs=1)
+     dN_tilj <- scale(dN_tilj, scale=F)
+     dN_tilj <- dN_tilj[,-ncol(dN_tilj)]
+
+     # df_tilj
+     df_tilj <- dN_tilj%*%gamma_tilj
+
+     # Jj
+     Jj <- outer(c(1,alpha_tilj), -alphaj[-1]/kkj^2)
+     for (i in 1:ncol(Jj)) Jj[i+1,i] <- Jj[i+1,i] + 1/kkj
+
```

```
+     # T_tilj
+     T_tilj <- as.vector(df_tilj)*Zj%*%Jj
+
+     # add T_tilj
+     M <- cbind(M, T_tilj)
+
+     # f_tilj
+     f_tilj <- N_tilj%*%gamma_tilj
+
+     # eta
+     eta <- eta + f_tilj
+
+     # save
+     fit$gamma_til[[j]] <- gamma_tilj
+     fit$alpha[[j]] <- alphaj
+     fit$alpha_til[[j]] <- alpha_tilj
+     fit$f[[j]] <- f_tilj
+     fit$u[[j]] <- uj
+   }
+
+   # save
+   ploglik1 <- ploglik(fit)
+   fit$psi <- psi_new
+   fit$mu <- inv_gmu(eta)
+   fit$M_til <- sqrt(wmu(fit$mu))*M
+   fit$P <- P
+   fit$B <- B
+   fit$edf <- edf
+   fit$psi_var <- diag(vcov)/fit$phi
+   ploglik2 <- ploglik(fit)
+   fit$metric <- abs(ploglik2-ploglik1)/(abs(ploglik1) + 1e-4)
+   fit
+ }
```

The fit_gplsiam() function and its seeded version gplsiam() to fit GPLSIAM in Poisson I scenario using the proposed method are given by

```
> # fit
> fit_gplsiam <- function(X, Z, y, best_fit=F){
+   # initial step
+   fit <- start_fit()
+
+   # prep
+   if(!is.list(best_fit)){
+     # create
```

```
+     fit$tot_ite <- 0
+     best_fit <- fit
+   }
+
+   # prep
+   tot_ite <- best_fit$tot_ite
+   ite <- 0
+
+   # iterative process
+   while(fit$metric > 1e-6 & tot_ite < 499){
+     # update
+     fit <- update_fit(fit)
+
+     # bad convergence
+     pri_alpha <- min(sapply(fit$alpha, head, n=1)) < 0.05
+     min_lambda <- min(fit$lambda) < 0
+     big_metric <- fit$metric > 1e+6
+     if(ite >= 80 | pri_alpha | min_lambda | big_metric){
+       # restart
+       best_fit$tot_ite <- tot_ite
+       return(fit_gplsiam(X, Z, y, best_fit))
+     }
+
+     # cat
+     tot_ite <- tot_ite + 1
+     ite <- ite + 1
+     msg <- paste("\r", "Iteration actual:", ite, "- Iteration total:", tot_ite)
+     cat(msg, "- The metric:", fit$metric, strrep(" ", 20))
+
+     # save best_fit
+     if(fit$metric < best_fit$metric){
+       # best_fit
+       fit$ite <- ite
+       best_fit <- fit
+     }
+   }
+
+   # pointwise band
+   for(j in 1:best_fit$m){
+     # prep
+     ini <- best_fit$psi_pos$ini[2*j]
+     fin <- best_fit$psi_pos$fin[2*j+1]
+
+     # fj_upp + fj_low
+     B_fj <- best_fit$B[ini:fin, ini:fin]
```

```
+     vcov_fj <- best_fit$M_til[,ini:fin] %*% t(B_fj)
+     vcov_fj <- rowSums(vcov_fj*vcov_fj)/best_fit$phi
+     best_fit$f_upp[[j]] <- best_fit$f[[j]] + qnorm(0.975)*sqrt(vcov_fj)
+     best_fit$f_low[[j]] <- best_fit$f[[j]] + qnorm(0.025)*sqrt(vcov_fj)
+   }
+
+   # prep
+   best_fit$B <- NULL
+   best_fit$M_til <- NULL
+   best_fit$P_big2 <- NULL
+   best_fit$P <- NULL
+   best_fit$alpha_til <- NULL
+
+   # return
+   best_fit$tot_ite <- tot_ite
+   term_name <- paste0("f", 1:best_fit$m)
+   names(best_fit$gamma_til) <- term_name
+   names(best_fit$alpha) <- term_name
+   names(best_fit$f) <- term_name
+   names(best_fit$f_upp) <- term_name
+   names(best_fit$f_low) <- term_name
+   names(best_fit$u) <- term_name
+   best_fit
+ }
>
> # fit with seed
> gplsiam <- function(X, Z, y){
+   # set
+   set.seed(13)
+   fit <- fit_gplsiam(X, Z, y, best_fit=F)
+   fit
+ }
```

The `si()` auxiliary function and `gplsiam_2step()` to fit GPLSIAM in the Poisson I scenario using the two-step method, extending Li et al. (2025) for multiple single-index effects, are given by

```
> # si
> si <- function(alpha_til, y, X, Z, opt=T, qj=9, fx=F){
+   # global iter count
+   tot_ite <<- tot_ite + 1
+
+   # alpha1 and u1
+   s1 <- ncol(Z[[1]])-1
+   alpha_til1 <- alpha_til[1:s1]
```

```
+   alpha1 <- c(1, alpha_til1)
+   kk1 <- sqrt(sum(alpha1^2))
+   alpha1 <- alpha1/kk1
+   u1 <- Z[[1]]%*%alpha1
+
+   # alpha2 and u2
+   s2 <- ncol(Z[[2]])-1
+   alpha_til2 <- alpha_til[(1:s2)+s1]
+   alpha2 <- c(1, alpha_til2)
+   kk2 <- sqrt(sum(alpha2^2))
+   alpha2 <- alpha2/kk2
+   u2 <- Z[[2]]%*%alpha2
+
+   # model
+   b <- gam(y~X +
+              s(u1, fx=fx, k=qj+1, bs="ps", m=c(2,2)) +
+              s(u2, fx=fx, k=qj+1, bs="ps", m=c(2,2)) -
+              1, family="poisson", method="ML")
+
+   # return
+   if(opt) b$gcv.ubre else{
+     # alpha1 and J1
+     b$alpha[[1]] <- alpha1
+
+     # alpha2 and J2
+     b$alpha[[2]] <- alpha2
+     b
+   }
+ }
>
> # fit
> gplsiam_2step <- function(X, Z, y){
+   # tot_ite count
+   tot_ite <<- 0
+
+   # initial alpha_til
+   s1 <- ncol(Z[[1]])-1
+   s2 <- ncol(Z[[2]])-1
+   alpha_til <- rep(0, s1+s2)
+
+   # fit
+   f0 <- optim(alpha_til, si, y=y, X=X, Z=Z, fx=T, qj=5)
+   f1 <- optim(f0$par, si, y=y, X=X, Z=Z, hessian=T)
+   b <- si(f1$par, y, X=X, Z=Z, opt=F)
+
+   # save
```

```
+   fit <- list()
+   fit$m <- length(Z)
+   fit$mu <- predict(b, type="response")
+   fit$lambda <- b$sp/sapply(b$smooth, "[[", "S.scale")
+   fit$phi <- 1/b$sig2
+   fit$tot_ite <- tot_ite
+   fit$ite <- b$outer.info$iter
+   fit$beta <- b$coefficients[1:ncol(X)]
+   for(j in 1:fit$m){
+     fit$alpha[[j]] <- b$alpha[[j]]
+     fit$f[[j]] <- predict(b, type="terms")[,j+1]
+     fit$u[[j]] <- b$model[,j+2]
+   }
+   fit$edf <- b$edf
+
+   # return
+   term_name <- paste0("f", 1:fit$m)
+   names(fit$alpha) <- term_name
+   names(fit$f) <- term_name
+   names(fit$u) <- term_name
+   fit
+ }
```

The gplsiam_gamfactory() to fit GPLSIAM in Poisson I scenario using gamfactory method, following a recommendation from the package developers that use random initialization of the single-index coefficients and the generalized Fellner-Schall iteration as the optimizer is given by

```
> # fit
> gplsiam_gamfactory <- function(X, Z, y, qj=9){
+   # model
+   a1 <- runif(ncol(Z$Z1),-1,1)
+   a2 <- runif(ncol(Z$Z2),-1,1)
+   b <- gam_nl(y ~ X +
+                 s_nest(Z$Z1, trans=trans_linear(alpha=a1), k=qj, m=c(4,2)) +
+                 s_nest(Z$Z2, trans=trans_linear(alpha=a2), k=qj, m=c(4,2)) -
+                 1, family=fam_poisson(), method="efs")
+
+   # save
+   fit <- list()
+   fit$m <- length(Z)
+   fit$mu <- predict(b, type="response")
+   fit$lambda <- b$sp/sapply(b$smooth, "[[", "S.scale")
+   fit$phi <- 1/b$sig2
+   fit$tot_ite <- b$outer.info$iter
```

```
+   fit$ite <- fit$tot_ite
+   fit$beta <- b$coefficients[1:ncol(X)]
+   md <- 0
+   for(j in 1:fit$m){
+     fit$alpha[[j]] <- b$smooth[[j]]$xt$si$alpha
+     fit$alpha[[j]] <- fit$alpha[[j]]/sqrt(sum(fit$alpha[[j]]^2))
+     fit$alpha[[j]] <- fit$alpha[[j]]/sign(fit$alpha[[j]][1])
+     fit$f[[j]] <- predict(b, type="terms")[,j+1]
+     md_j <- mean(fit$f[[j]])
+     md <- md + md_j
+     fit$f[[j]] <- fit$f[[j]] - md_j
+     fit$u[[j]] <- b$model[,j+2] %*% fit$alpha[[j]]
+   }
+   fit$beta[1] <- fit$beta[1] + md
+   fit$edf <- b$edf
+
+   # return
+   term_name <- paste0("f", 1:fit$m)
+   names(fit$alpha) <- term_name
+   names(fit$f) <- term_name
+   names(fit$u) <- term_name
+   fit
+ }
```

The simulation in the Poisson I scenario using the proposed method is performed via

```
> # load gplsiam
> source("R/gplsiam_poisson1.R")
>
> # prep
> R <- 500
>
> # simu n1
> n <- 200
> dat_n1 <- data_gen(n, R)
> X <- dat_n1$X
> Z <- list(Z1=dat_n1$Z1, Z2=dat_n1$Z2)
> time_n1 <- numeric(R)
> list_n1 <- list()
> for(j in 1:R){
+   name <- paste0("y", j)
+   y <- unlist(dat_n1[name])
+   time_n1[j] <- system.time(b <- gplsiam(X, Z, y), gcFirst=T)[3]
+   list_n1[[name]] <- b
```

```
+ }
 Iteration actual: 14 - Iteration total: 14 - The metric: 7.852271e-08
>
> # simu n2
> n <- 800
> dat_n2 <- data_gen(n, R)
> X <- dat_n2$X
> Z <- list(Z1=dat_n2$Z1, Z2=dat_n2$Z2)
> time_n2 <- numeric(R)
> list_n2 <- list()
> for(j in 1:R){
+   name <- paste0("y", j)
+   y <- unlist(dat_n2[name])
+   time_n2[j] <- system.time(b <- gplsiam(X, Z, y), gcFirst=T)[3]
+   list_n2[[name]] <- b
+ }
 Iteration actual: 25 - Iteration total: 25 - The metric: 9.791735e-07
>
> # simu n3
> n <- 3200
> dat_n3 <- data_gen(n, R)
> X <- dat_n3$X
> Z <- list(Z1=dat_n3$Z1, Z2=dat_n3$Z2)
> time_n3 <- numeric(R)
> list_n3 <- list()
> for(j in 1:R){
+   name <- paste0("y", j)
+   y <- unlist(dat_n3[name])
+   time_n3[j] <- system.time(b <- gplsiam(X, Z, y), gcFirst=T)[3]
+   list_n3[[name]] <- b
+ }
 Iteration actual: 17 - Iteration total: 17 - The metric: 9.040281e-07
>
> # time all
> sum(time_n1,time_n2,time_n3)/60
[1] 5.909333
```

The simulation in the Poisson I scenario using the two-step method is performed via

```
> # load gplsiam
> source("R/gplsiam_poisson1.R")
>
> # prep
> R <- 500
>
```

```
> # simu n1
> n <- 200
> dat_n1 <- data_gen(n, R)
> X <- dat_n1$X
> Z <- list(Z1=dat_n1$Z1, Z2=dat_n1$Z2)
> time_n1 <- numeric(R)
> list_n1 <- list()
> for(j in 1:R){
+   name <- paste0("y", j)
+   y <- unlist(dat_n1[name])
+   time_n1[j] <- system.time(b <- gplsiam_2step(X, Z, y), gcFirst=T)[3]
+   list_n1[[name]] <- b
+   cat("\r", "Model actual:", j, strrep(" ", 20))
+ }
 Model actual: 500
>
> # simu n2
> n <- 800
> dat_n2 <- data_gen(n, R)
> X <- dat_n2$X
> Z <- list(Z1=dat_n2$Z1, Z2=dat_n2$Z2)
> time_n2 <- numeric(R)
> list_n2 <- list()
> for(j in 1:R){
+   name <- paste0("y", j)
+   y <- unlist(dat_n2[name])
+   time_n2[j] <- system.time(b <- gplsiam_2step(X, Z, y), gcFirst=T)[3]
+   list_n2[[name]] <- b
+   cat("\r", "Model actual:", j, strrep(" ", 20))
+ }
 Model actual: 500
>
> # simu n3
> n <- 3200
> dat_n3 <- data_gen(n, R)
> X <- dat_n3$X
> Z <- list(Z1=dat_n3$Z1, Z2=dat_n3$Z2)
> time_n3 <- numeric(R)
> list_n3 <- list()
> for(j in 1:R){
+   name <- paste0("y", j)
+   y <- unlist(dat_n3[name])
+   time_n3[j] <- system.time(b <- gplsiam_2step(X, Z, y), gcFirst=T)[3]
+   list_n3[[name]] <- b
+   cat("\r", "Model actual:", j, strrep(" ", 20))
```

```
+ }
 Model actual: 500
>
> # time all
> sum(time_n1,time_n2,time_n3)/60
[1] 166.0063
```

The simulation in the Poisson I scenario using the `gamfactory` method is performed via

```
> # load gplsiam
> source("R/gplsiam_poisson1.R")
>
> # prep
> R <- 500
>
> # simu n1
> n <- 200
> dat_n1 <- data_gen(n, R)
> X <- dat_n1$X
> Z <- list(Z1=dat_n1$Z1, Z2=dat_n1$Z2)
> time_n1 <- numeric(R)
> list_n1 <- list()
> for(j in 1:R){
+   name <- paste0("y", j)
+   y <- unlist(dat_n1[name])
+   time_n1[j] <- system.time(b <- gplsiam_gamfactory(X, Z, y), gcFirst=T)[3]
+   list_n1[[name]] <- b
+   cat("\r", "Model actual:", j, strrep(" ", 20))
+ }
 Model actual: 500
>
> # simu n2
> n <- 800
> dat_n2 <- data_gen(n, R)
> X <- dat_n2$X
> Z <- list(Z1=dat_n2$Z1, Z2=dat_n2$Z2)
> time_n2 <- numeric(R)
> list_n2 <- list()
> for(j in 1:R){
+   name <- paste0("y", j)
+   y <- unlist(dat_n2[name])
+   time_n2[j] <- system.time(b <- gplsiam_gamfactory(X, Z, y), gcFirst=T)[3]
+   list_n2[[name]] <- b
+   cat("\r", "Model actual:", j, strrep(" ", 20))
```

```
+ }
 Model actual: 500
>
> # simu n3
> n <- 3200
> dat_n3 <- data_gen(n, R)
> X <- dat_n3$X
> Z <- list(Z1=dat_n3$Z1, Z2=dat_n3$Z2)
> time_n3 <- numeric(R)
> list_n3 <- list()
> for(j in 1:R){
+   name <- paste0("y", j)
+   y <- unlist(dat_n3[name])
+   time_n3[j] <- system.time(b <- gplsiam_gamfactory(X, Z, y), gcFirst=T)[3]
+   list_n3[[name]] <- b
+   cat("\r", "Model actual:", j, strrep(" ", 20))
+ }
 Model actual: 500
>
> # time all
> sum(time_n1,time_n2,time_n3)/60
[1] 41.743
```

### 3.2 Gamma

Figure 4 (all top) presents the nonlinear functions from the postulated model, which have a similar variation range. Figure 4 (all bottom) exhibits, for illustration, a response simulation of $n = 200$ observations in the Gamma scenario. This is a complex problem, as evident from the dispersion of observations.

Under the proposed method, for $n = 200$, only 4.4% of the fitted models are removed, and 0.2% for $n = 3200$. Figure 5 (top-left) displays the elapsed times in seconds, Figure 5 (top-middle) the number of iterations of the returned model, and Figure 5 (top-right) its fitted effective degrees of freedom. Figure 5 (center-left) shows the fitted precision parameter, the fitted linear coefficients $\beta_1$ and $\beta_2$ are in Figure 5 (center-middle and center-right), all single-index coefficients are in Figure 5 (all bottom) and Figure 6 (all top and all center), and the smoothing parameters are in Figure 6 (all bottom).

The `data_gen()` function used in the simulation to generate the dataset with `n` rows and `R` different responses from the postulated GPLSIAM in the Gamma scenario is given by

```
> # data gen
> data_gen <- function(n, R, seed=13){
+   # prep
+   set.seed(seed)
+   dat <- list()
```

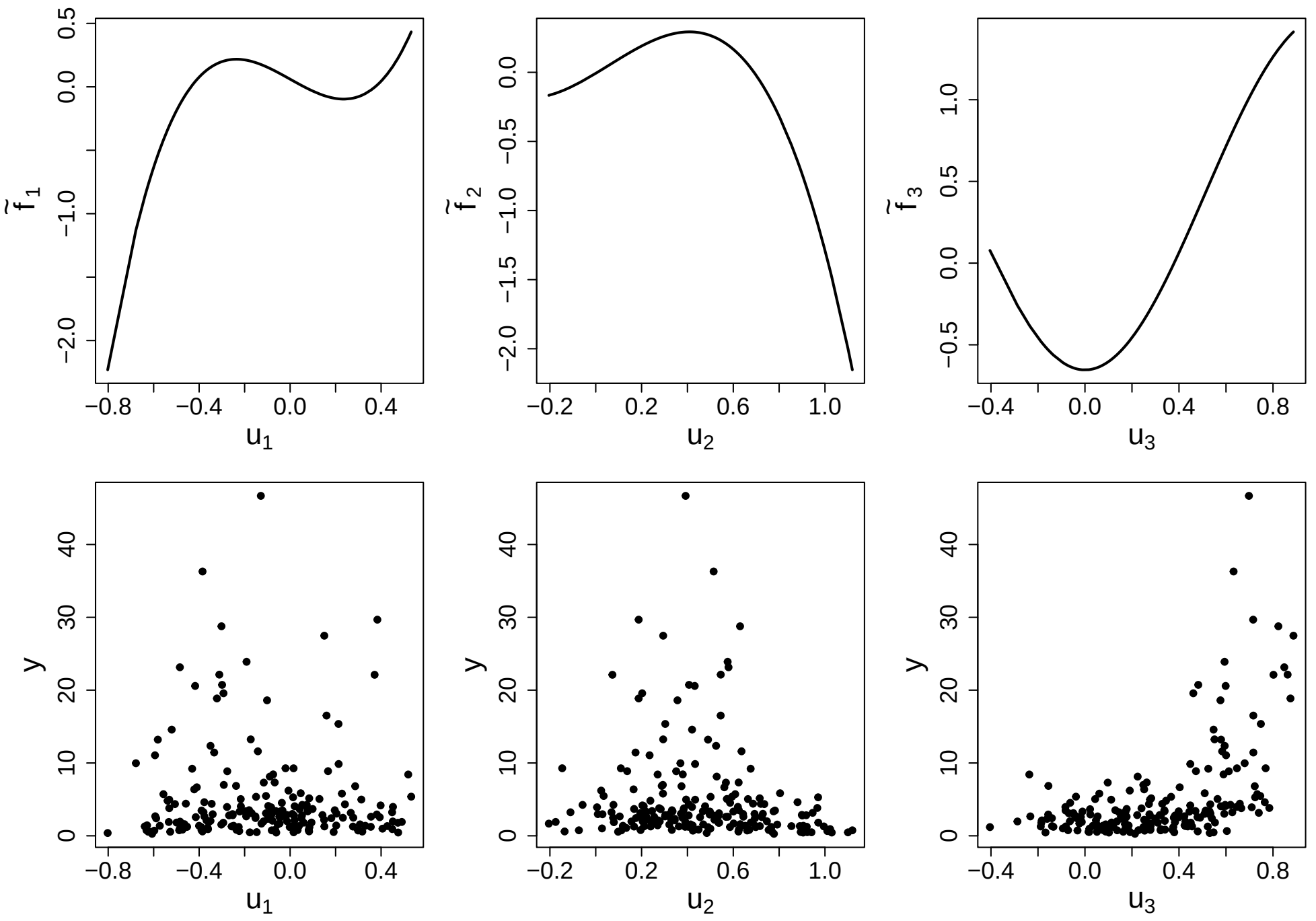


**Fig. 4** True $\tilde{f}_1$ (top-left), $\tilde{f}_2$ (top-middle), and $\tilde{f}_3$ (top-right) additive functions versus their respective single-index covariates. The $n = 200$ response simulation versus $u_1$ (bottom-left), $u_2$ (bottom-middle) and $u_3$ (bottom-right) single-index covariate in Poisson II scenario.

```
+   dat$X <- cbind(1, matrix(runif(n), nrow=n))
+   dat$Z1 <- matrix(runif(2*n), nrow=n)
+   dat$Z2 <- matrix(runif(3*n), nrow=n)
+   dat$Z3 <- matrix(runif(4*n), nrow=n)
+
+   # prep f1
+   alpha1 <- c(1, -1.4)
+   kk1 <- sqrt(sum(alpha1^2))
+   alpha1 <- alpha1/kk1
+   dat$u1 <- dat$Z1%*%alpha1
+   dat$f1 <- -sin(dat$u1) + (1.8*dat$u1)^3
+   dat$f1 <- dat$f1 - mean(dat$f1)
+
+   # prep f2
+   alpha2 <- c(1, 1.7, -0.8)
+   kk2 <- sqrt(sum(alpha2^2))
+   alpha2 <- alpha2/kk2
+   dat$u2 <- dat$Z2%*%alpha2
+   dat$f2 <- -3*(dat$u2)^3 + exp(dat$u2)
+   dat$f2 <- dat$f2 - mean(dat$f2)
```

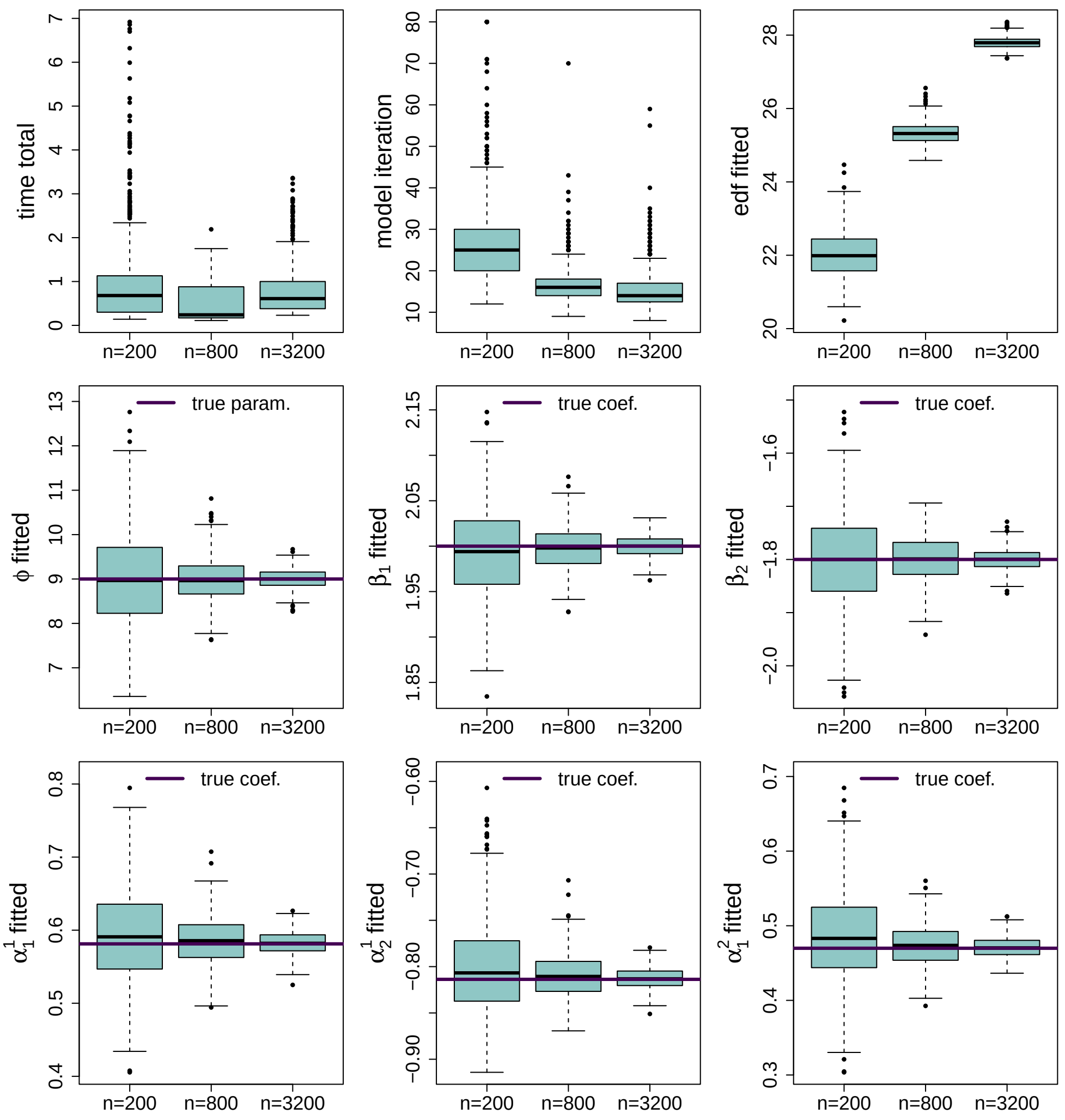


**Fig. 5** Elapsed total time (top-left), final iteration (top-middle), and fitted effective degrees of freedom (top-right). Fitted precision parameter $\phi$ (center-left), and linear coefficients $\beta_1$ (center-middle) and $\beta_2$ (center-right). Fitted single-index coefficient $\alpha_1^1$ (bottom-left), $\alpha_2^1$ (bottom-middle), and $\alpha_1^2$ (bottom-right) from stable fitted GPLSIAMs with the proposed method in the Gamma scenario.

```
+
+    # prep f3
+    alpha3 <- c(1, 3.4, -0.5, -1.6)
+    kk3 <- sqrt(sum(alpha3^2))
+    alpha3 <- alpha3/kk3
+    dat$u3 <- dat$Z3%*%alpha3
+    dat$f3 <- (dat$u3)^2/6 - cos(pi*dat$u3)
+    dat$f3 <- dat$f3 - mean(dat$f3)
```

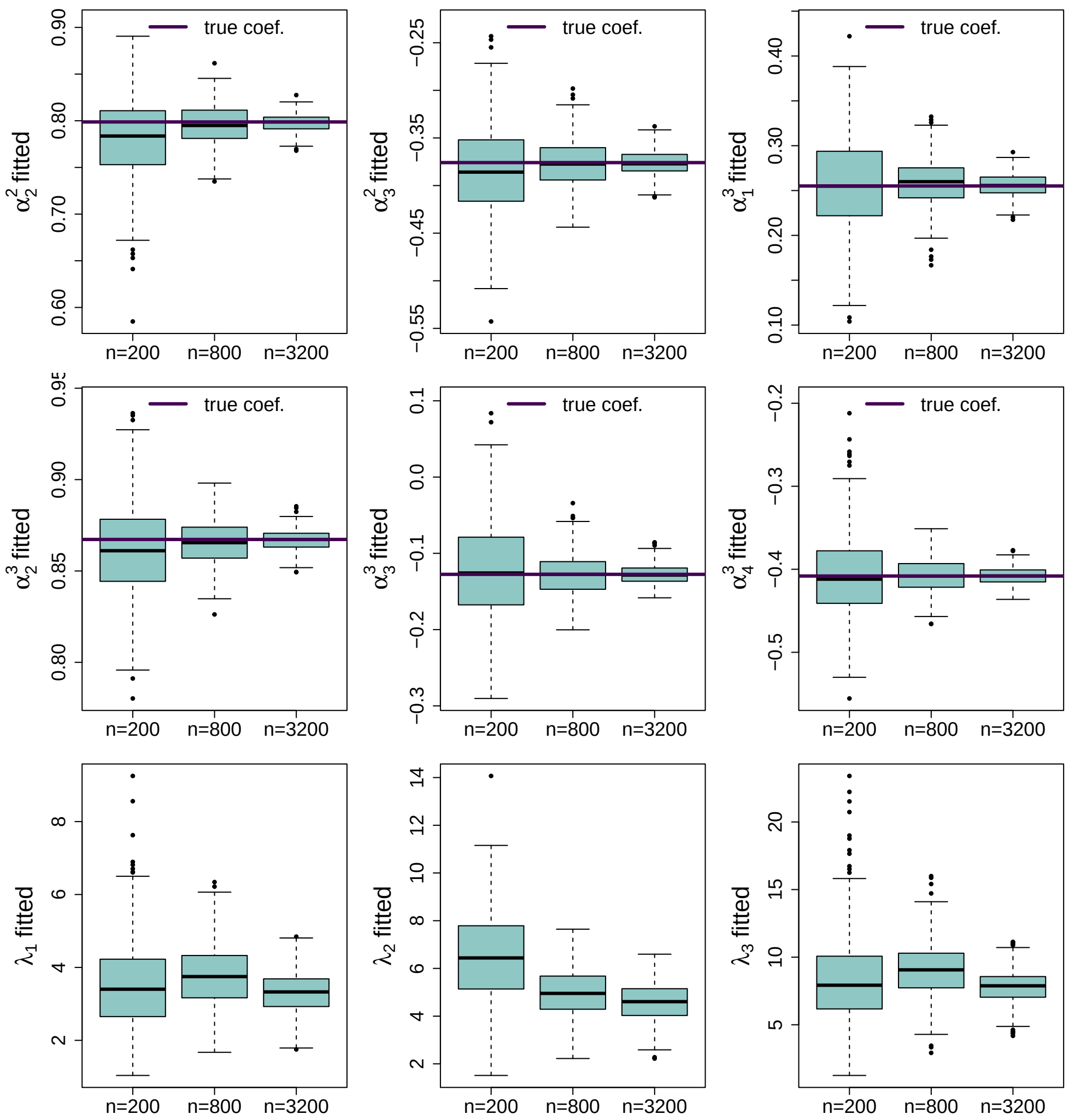


**Fig. 6** Fitted single-index coefficient $\alpha_2^2$ (top-left), $\alpha_3^2$ (top-middle) and $\alpha_1^3$ (top-right). Fitted single-index coefficient $\alpha_2^3$ (center-left), $\alpha_3^3$ (center-middle) and $\alpha_4^3$ (center-right). Fitted smoothing parameter $\lambda_1$ (bottom-left), $\lambda_2$ (bottom-middle), and $\lambda_3$ (bottom-right) from stable fitted GPLSIAMs with the proposed method in the Gamma scenario.

```
+
+   # response simu
+   beta <- c(2, -1.8)
+   dat$mu <- exp(dat$X%*%beta + dat$f1 + dat$f2 + dat$f3)
+   dat$phi <- 9
+   y_list <- list()
+   for(j in 1:R){
+     y_j <- rgamma(nrow(dat$mu), shape=dat$phi, scale=dat$mu/dat$phi)
```

```
+     y_list <- c(y_list, list(y_j))
+   }
+   names(y_list) <- paste0("y", 1:R)
+
+   # return
+   dat <- c(dat, y_list)
+   dat
+ }
```

### 3.3 Poisson II

Figure 7 (all top) presents the nonlinear functions from the postulated model, which have a similar variation range. Figure 7 (all bottom) exhibits, for illustration, a response simulation of $n = 200$ observations in the Poisson II scenario. This is a complex problem, as evident from the dispersion of observations.

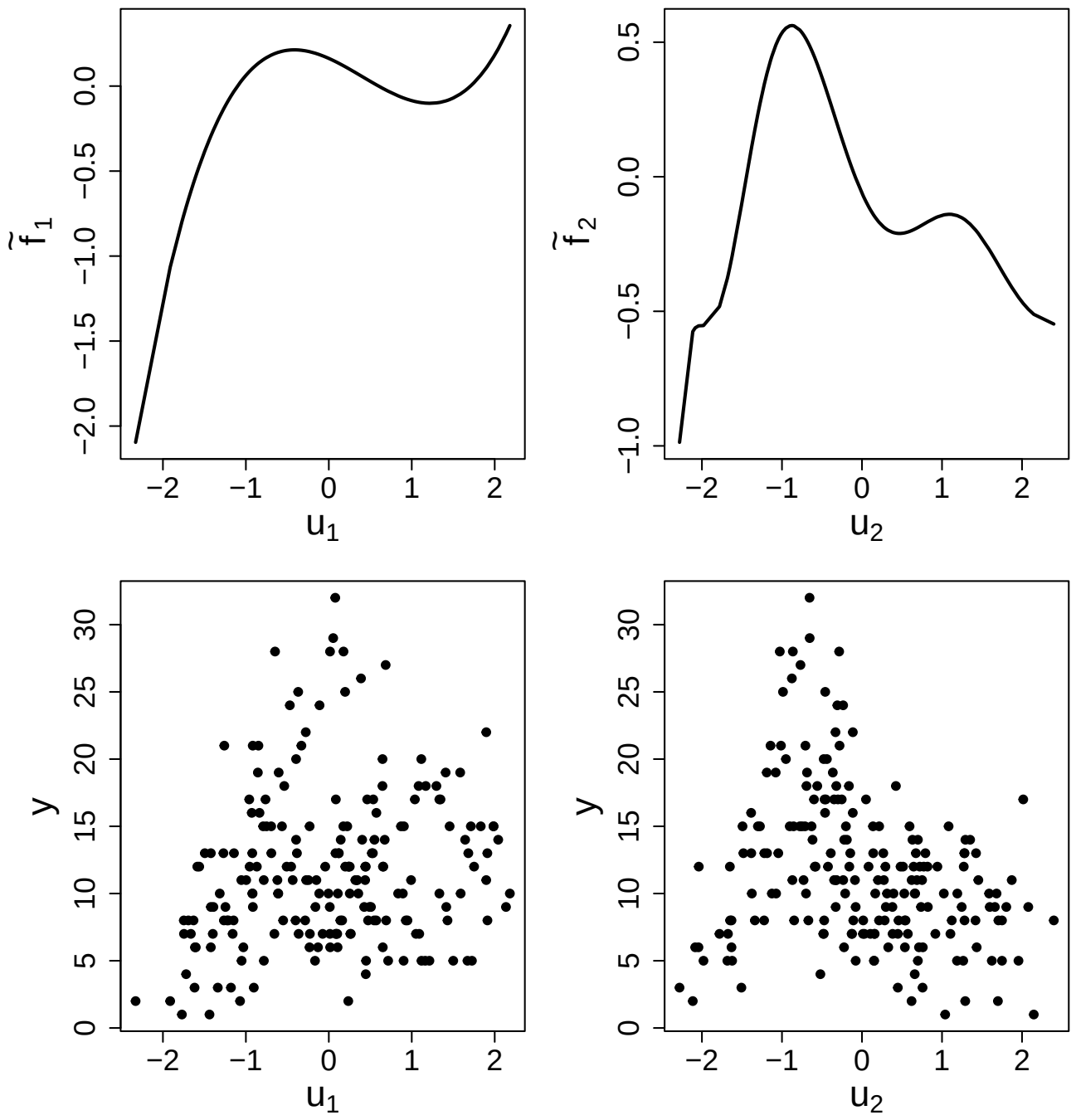


**Fig. 7** True $\tilde{f}_1$ (top-left), and $\tilde{f}_2$ (top-right) nonlinear functions versus their respective single-index covariates. The $n = 200$ response simulation versus $u_1$ (bottom-left), and $u_2$ (bottom-right) single-index covariate in Poisson II scenario.

Under the proposed method, for $n = 200$, only 3.6% of the fitted models are removed, and 0.8% for $n = 800$. Figure 8 (top-left) displays the elapsed time in seconds, Figure 8 (top-middle) the number of iterations of the returned model, and Figure 8 (top-right) its fitted effective degrees of freedom. Figure 2 (bottom-left and

bottom-middle) presents the fitted linear coefficients $\beta_1$ and $\beta_2$, and the single-index coefficient $\alpha_1^1$ is in Figure 8 (bottom-right). The remaining single-index coefficients are in Figure 9 (all top and bottom-left), and the smoothing parameters are in Figure 9 (bottom-middle and bottom-right).

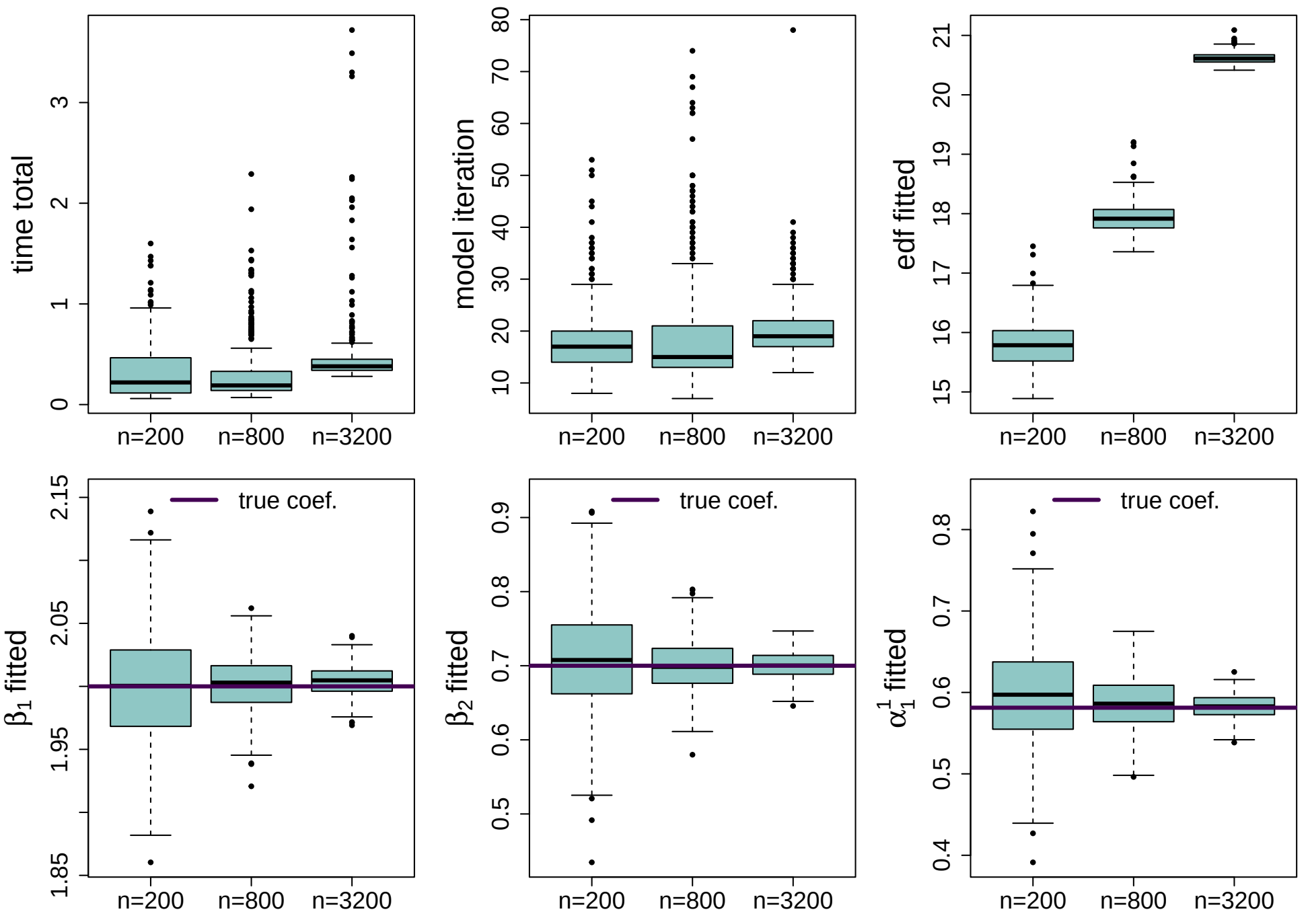


**Fig. 8** Elapsed total time (top-left), final iteration (top-middle), and fitted effective degrees of freedom (top-right). Fitted linear coefficients $\beta_1$ (bottom-left) and $\beta_2$ (bottom-middle). Fitted single-index coefficient $\alpha_1^1$ (bottom-left) from stable fitted GPLSIAMs with the proposed method in the Poisson II scenario.

The `data_gen()` function used in the simulation to generate the dataset with `n` rows and `R` different responses from the postulated GPLSIAM in the Poisson II scenario is given by

```
> # data gen
> data_gen <- function(n, R, seed=13){
+   # prep
+   set.seed(seed)
+   dat <- list()
+   dat$X <- cbind(1, matrix(runif(n), nrow=n))
+   dat$Z1 <- matrix(runif(2*n), nrow=n)
+   dat$Z2 <- matrix(runif(3*n), nrow=n)
+
+   # scale
```

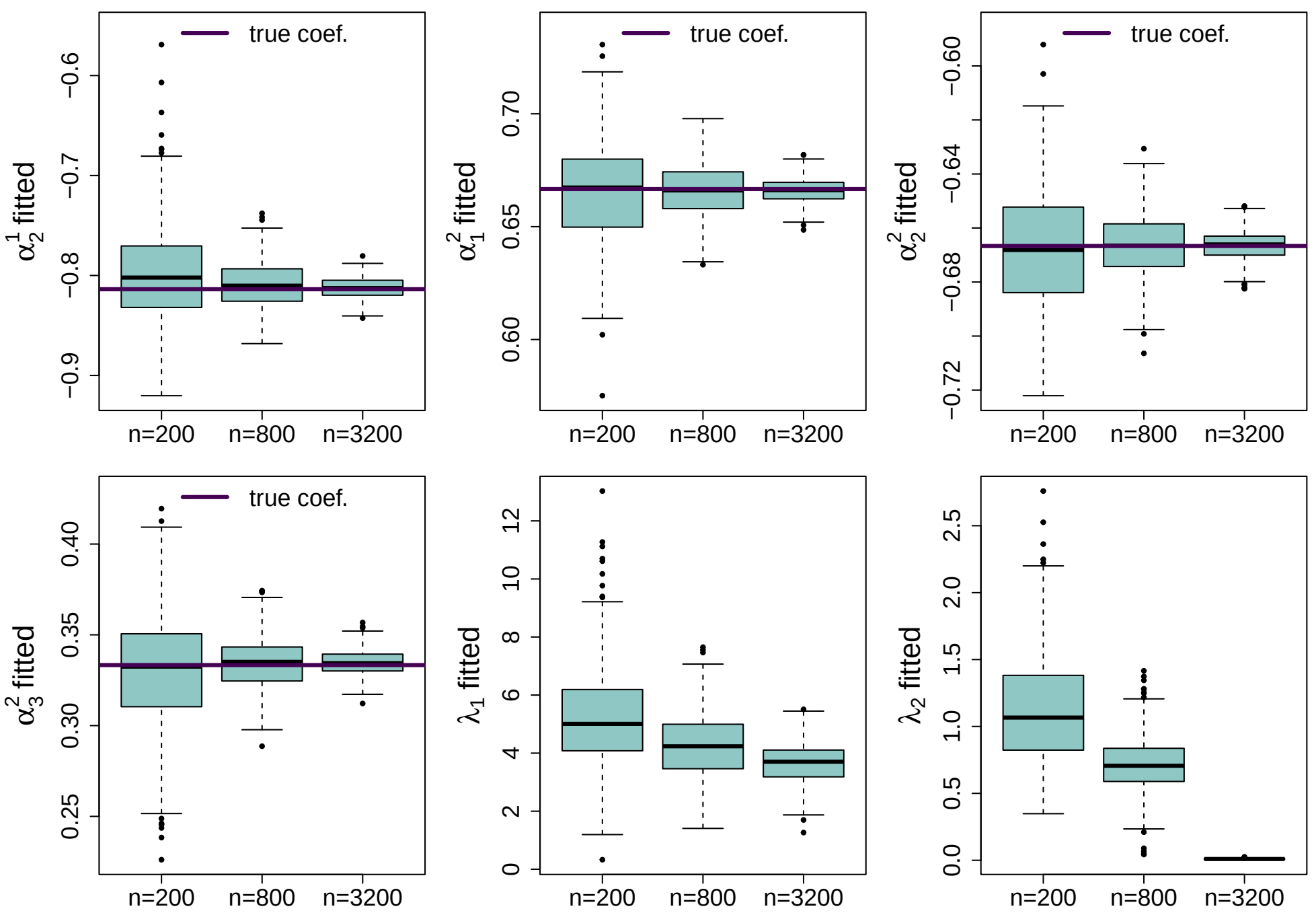


**Fig. 9** Fitted single-index coefficient $\alpha_2^1$ (top-left), $\alpha_1^2$ (top-middle), $\alpha_2^2$ (top-right), and $\alpha_3^2$ (bottom-left). Fitted smoothing parameter $\lambda_1$ (bottom-middle), and $\lambda_2$ (bottom-right) from stable fitted GPLSIAMs with the proposed method in the Poisson II scenario.

```
+   dat$Z1 <- scale(dat$Z1)
+   dat$Z2 <- scale(dat$Z2)
+
+   # prep f1
+   alpha1 <- c(1, -1.4)
+   kk1 <- sqrt(sum(alpha1^2))
+   alpha1 <- alpha1/kk1
+   dat$u1 <- dat$Z1%*%alpha1
+   t1 <- dat$u1/sqrt(12) + sum(0.5*alpha1)
+   dat$f1 <- -sin(t1) + (1.8*t1)^3
+   dat$f1 <- dat$f1 - mean(dat$f1)
+
+   # prep f2
+   alpha2 <- c(1, -1, 1/2)
+   kk2 <- sqrt(sum(alpha2^2))
+   alpha2 <- alpha2/kk2
+   dat$u2 <- dat$Z2%*%alpha2
+   t2 <- (dat$u2/sqrt(12) + sum(0.5*alpha2) + 0.41)/1.4
+   dat$f2 <- 0.2*t2^11 * (10*(1-t2))^6 + 10*(10*t2)^3 * (1-t2)^10
+   dat$f2 <- dat$f2/8
```

```
+   dat$f2 <- dat$f2 - mean(dat$f2)
+
+   # response simu
+   beta <- c(2, 0.7)
+   dat$mu <- exp(dat$X%*%beta + dat$f1 + dat$f2)
+   y_list <- list()
+   for(j in 1:R){
+     yj <- rpois(nrow(dat$mu), lambda=dat$mu)
+     y_list <- c(y_list, list(yj))
+   }
+   names(y_list) <- paste0("y", 1:R)
+
+   # return
+   dat <- c(dat, y_list)
+   dat
+ }
```

# 4 Bike-sharing demand details

Figure 10 and Figure 11 present the nonlinear relation between the count of total rental bikes and the environmental covariates. We can see a different form in each year, suggesting interaction. The seasonal covariates do not appear to interact with year.

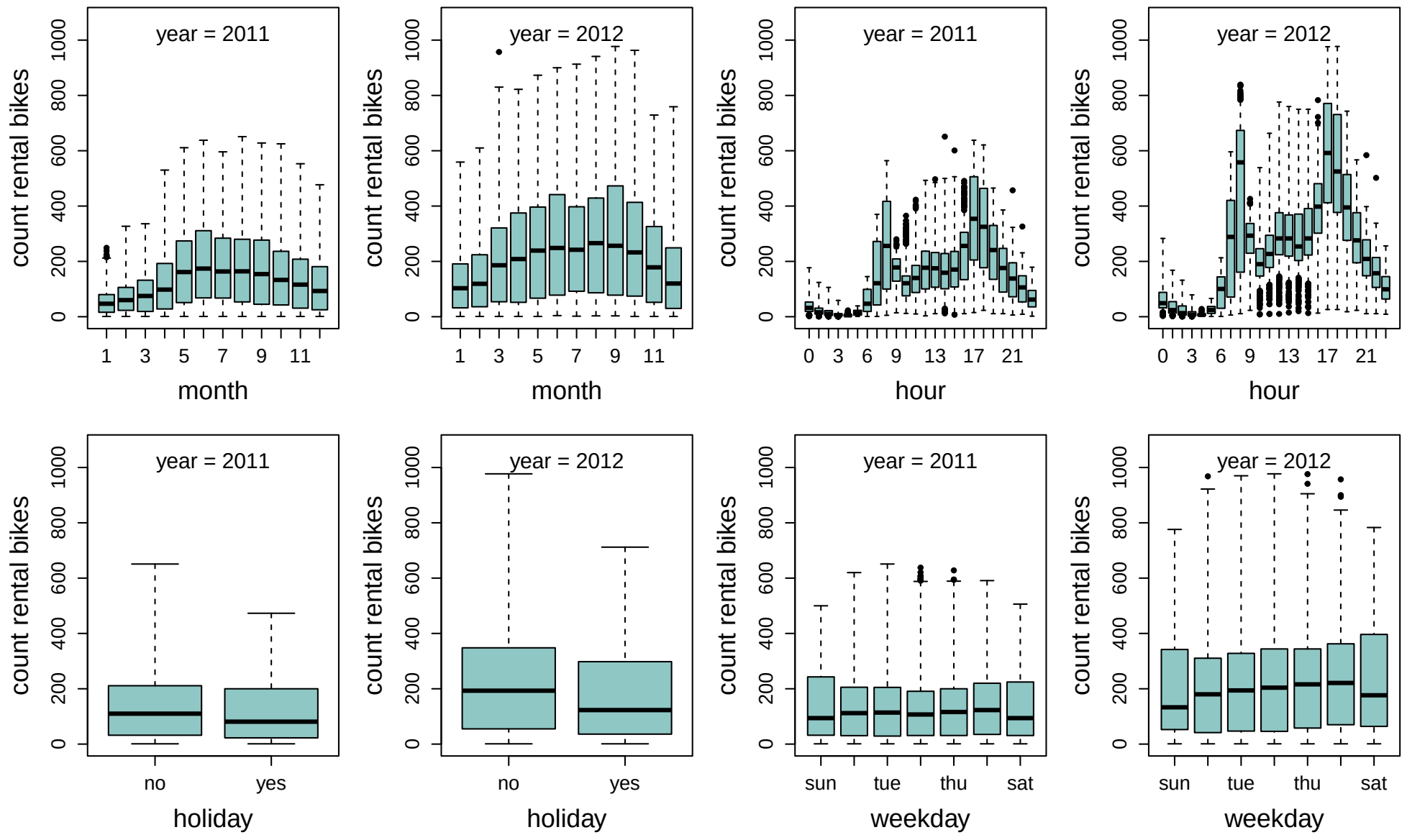


**Fig. 10** Adjusted boxplots (Hubert and Vandervieren 2008) for count of rental bikes versus month (two top-left), hour (two top-right), holiday (two bottom-left), and weekday (two bottom-right) for each year.

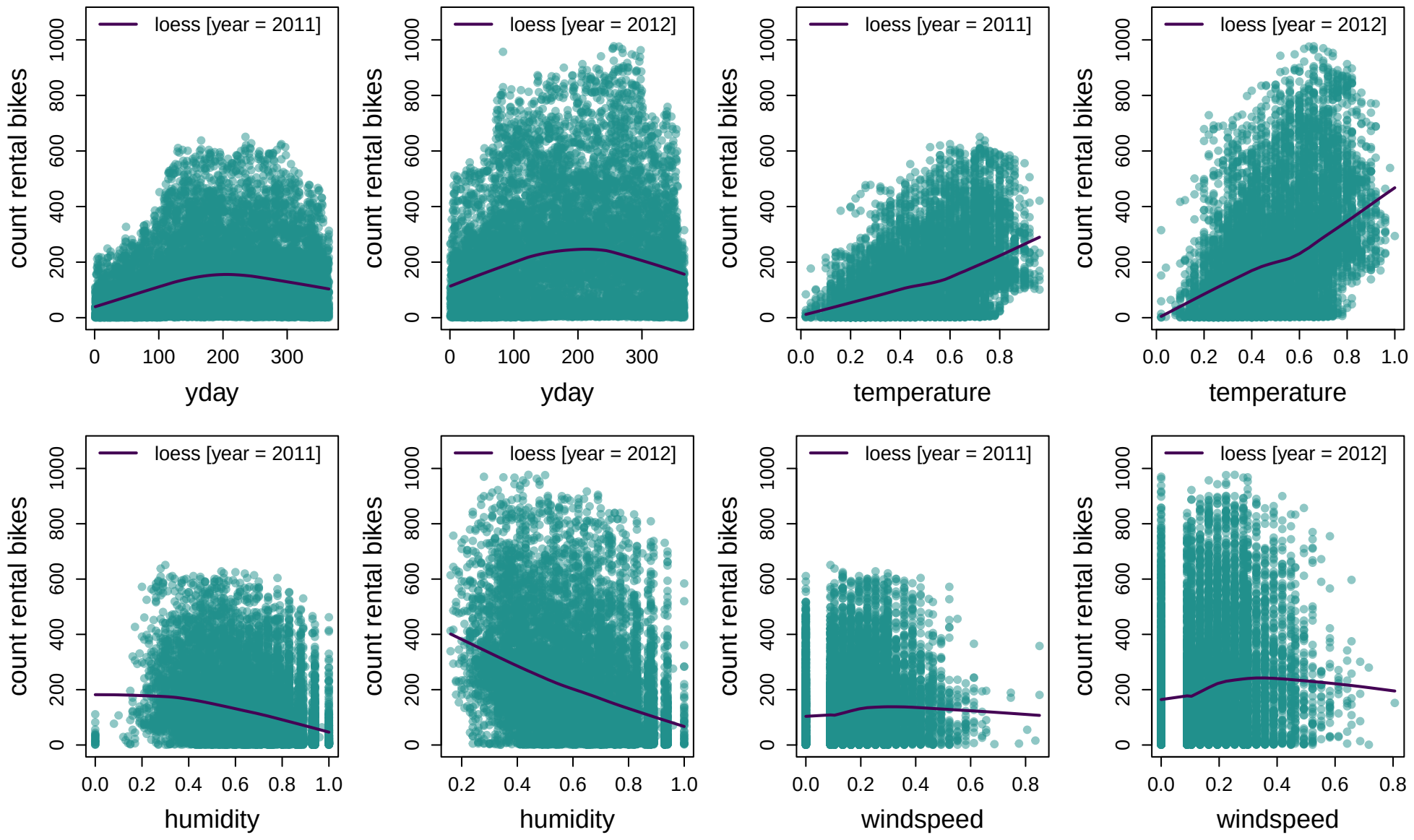


**Fig. 11** Dispersion plot for count of rental bikes versus yday (two top-left), temperature (two top-right), humidity (two bottom-left), and wind speed (two bottom-right) for each year.

The R functions used in the application are minor adaptations of the functions presented in the simulation. The adaptations include specialization for the Bernoulli distribution, an additional loop for nonlinear functions that are not single-index, and an incidence-matrix organization for the single-index interaction effect. The `res_gplsiam()` function to perform basic diagnostics of the fitted GPLASIM according to the proposed method for the Bernoulli is given by

```
> # residual function
> res_gplsiam <- function(fit, R=1){
+   # prep
+   n <- fit$n
+   mu <- fit$mu
+   y <- fit$y
+
+   # quantile residual
+   a <- pbinom(y-1, 1, mu)
+   b <- pbinom(y, 1, mu)
+   rq_list <- list()
+   for(j in 1:R) rq_list[[j]] <- list(rq=qnorm(runif(n, min=a, max=b)))
+
+   # prep
+   rq_matrix <- sapply(rq_list, "[[", "rq")
+   rq <- c(rq_matrix)
+   rq_theor <- qnorm(ppoints(n))
```

```
+   rq_cente <- apply(rq_matrix, 2, sort) - rq_theor
+
+   # prep
+   par(mar=c(4.5,5,1,1), mfrow=c(1,3), pch=19, cex.lab=1.8, cex.axis=1.5)
+   col <- c("white", turbo(11))
+   smoothScatter(rq~rep(mu, R), xlab="fitted value", ylab="quantile residual",
+                 colramp=colorRampPalette(col),
+                 nbin=200, nrpoints=0, ylim=c(-4.5,4.5))
+   abline(h=0, lwd=2, lty=2)
+
+   # rq vs index
+   smoothScatter(rq~rep(1:n, R), xlab="index", ylab="quantile residual",
+                 colramp=colorRampPalette(col),
+                 nbin=200, nrpoints=0, ylim=c(-4.5,4.5))
+   abline(h=0, lwd=2, lty=2)
+
+   # rq wormplot
+   p <- pnorm(rq_theor)
+   se <- (1/dnorm(rq_theor)) * sqrt(p*(1-p)/n)
+   conf_upp <- qnorm(0.975)*se
+   conf_low <- qnorm(0.025)*se
+   smoothScatter(rq_cente~rep(rq_theor, R), ylab="centered quantile residual",
+                 xlab="theoretical quantile", colramp=colorRampPalette(col),
+                 nbin=200, nrpoints=0, ylim=c(-0.2,0.2))
+   abline(h=0, lwd=2, lty=2)
+   lines(conf_upp~rq_theor, lwd=2, lty=2)
+   lines(conf_low~rq_theor, lwd=2, lty=2)
+ }
```

The application using the proposed method is performed below. The R package pROC (Robin et al. 2025) is needed.

```
> # load data
> dat <- readRDS("data/bike_hour_casa.rds")
> by <- as.numeric(dat$yr)
>
> # model
> X <- model.matrix(~holiday+weekday+yr, data=dat)
> U <- list(dat$yday, dat$hr)
> Z1 <- model.matrix(~hum+windspeed-1, data=dat)
> Z <- list(Z1, Z1)
> y <- dat$hdemand
> time <- system.time(b <- gplsiam(X, U, Z, y, by), gcFirst=T)[3]
 Iteration actual: 18 - Iteration total: 18 - The metric: 5.711917e-07
> time
```

```
elapsed
   2.92
>
> # summary beta
> ini <- b$psi_pos$ini[1]
> fin <- b$psi_pos$fin[1]
> beta <- b$beta
> round(beta, 2)
[1] -3.05 -0.89  0.61  0.75  0.70  0.86  1.13  0.44  2.14
> sd <- sqrt(b$psi_var[ini:fin])
> round(sd, 2)
[1] 0.14 0.15 0.10 0.09 0.09 0.09 0.10 0.09 0.06
> w <- abs(beta/sd)
> round(1 - pnorm(w), 3)
[1] 0 0 0 0 0 0 0 0 0
>
> # summary alpha_til_1
> ini <- b$psi_pos$ini[5]
> fin <- b$psi_pos$fin[5]
> alpha_1 <- b$alpha$f3
> round(alpha_1, 2)
[1] 0.96 0.29
> alpha_til_1 <- alpha_1[-1]/alpha_1[1]
> round(alpha_til_1, 2)
[1] 0.31
> sd <- sqrt(b$psi_var[ini:fin])
> round(sd, 2)
[1] 0.06
> w <- abs(alpha_til_1/sd)
> round(1 - pnorm(w), 3)
[1] 0
>
> # summary alpha_til_2
> ini <- b$psi_pos$ini[7]
> fin <- b$psi_pos$fin[7]
> alpha_2 <- b$alpha$f4
> round(alpha_2, 2)
[1] 0.75 0.67
> alpha_til_2 <- alpha_2[-1]/alpha_2[1]
> round(alpha_til_2, 2)
[1] 0.89
> sd <- sqrt(b$psi_var[ini:fin])
> round(sd, 2)
[1] 0.1
> w <- abs(alpha_til_2/sd)
```

```
> round(1 - pnorm(w), 3)
[1] 0
>
> # summary lambda
> lambda <- b$lambda
> round(lambda, 3)
[1]  0.013  0.010 37.191 45.971
>
> # summary edf
> edf <- sum(b$edf)
> round(edf, 3)
[1] 44.476
> edf_1 <- sum(b$edf[b$psi_pos$ini[2]:b$psi_pos$fin[2]])
> round(edf_1, 3)
[1] 11.604
> edf_2 <- sum(b$edf[b$psi_pos$ini[3]:b$psi_pos$fin[3]])
> round(edf_2, 3)
[1] 10.893
> edf_3 <- sum(b$edf[b$psi_pos$ini[4]:b$psi_pos$fin[4]])
> round(edf_3, 3)
[1] 6.057
> edf_4 <- sum(b$edf[b$psi_pos$ini[6]:b$psi_pos$fin[6]])
> round(edf_4, 3)
[1] 4.922
>
> # plot diag
> b$y <- dat$hdemand
> res_gplsiam(b, R=40)
>
> # summary auc
> roc <- roc(dat$hdemand, b$mu, levels=0:1, direction="<")
> roc$auc
Area under the curve: 0.9489
> coords(roc, "best")
  threshold specificity sensitivity
1 0.4571647   0.8405556   0.9060747
```



\end{document}